\documentclass[a4paper,11pt]{article}
\usepackage{jheppub} % for details on the use of the package, please see the JINST-author-manual
\DeclareMathAlphabet\bfcal{OMS}{cmsy}{b}{n}

\usepackage{dcolumn}% Align table columns on decimal point
\usepackage{bm}% bold math
\usepackage{epsfig}
\usepackage{graphicx}% Include figure files
\usepackage{amsmath}
\usepackage{amssymb}
\usepackage{mathrsfs}
\usepackage{color}
\usepackage[usenames,dvipsnames]{xcolor}
\usepackage{hyperref}
\usepackage{feynmp-auto}
\usepackage{tikz}
\usepackage{braket}
\usepackage[caption=false]{subfig}
\usepackage{multirow}
\usepackage{array}
\usepackage{slashed}
\usepackage{rotating}
\usepackage{caption}
\usepackage{subcaption}
\usepackage{bbold}

\usepackage[normalem]{ulem}
\usepackage{marginnote}
\newcommand\identity{1\kern-0.25em\text{l}}
\newcommand\vv{\text{v}}
\newcommand\Pv{{\bf P}}
\newcommand\pv{{\bf p}}
\newcommand\qv{{\bf q}}
\newcommand\kv{{\bf k}}
\newcommand\Dv{{\bf D}}
\newcommand\bv{{\bf b}}

\newcommand\Pc{{\mathcal P}}

\newcommand{\Otsosing}{\braket{\mathcal{O}^{J/\psi}(^3S_1^{[1]})}}
\newcommand{\Otsooct}{\braket{\mathcal{O}^{J/\psi}(^3S_1^{[8]})}}
\newcommand{\Oosz}{\braket{\mathcal{O}^{J/\psi}(^1S_0^{[8]})}}
\newcommand{\Otpz}{\braket{\mathcal{O}^{J/\psi}(^3P_0^{[8]})}}

\DeclareMathOperator*{\sumint}{%
\mathchoice%
  {\ooalign{$\displaystyle\sum$\cr\hidewidth$\displaystyle\int$\hidewidth\cr}}
  {\ooalign{\raisebox{.14\height}{\scalebox{.7}{$\textstyle\sum$}}\cr\hidewidth$\textstyle\int$\hidewidth\cr}}
  {\ooalign{\raisebox{.2\height}{\scalebox{.6}{$\scriptstyle\sum$}}\cr$\scriptstyle\int$\cr}}
  {\ooalign{\raisebox{.2\height}{\scalebox{.6}{$\scriptstyle\sum$}}\cr$\scriptstyle\int$\cr}}
}

\title{\boldmath The role of the soft scale for $J/\psi$ production in the transverse momentum dependent framework}

\author[a]{Marston Copeland,}
\author[b]{Sean Fleming,}
\author[a]{Reed Hodges,}

\affiliation[a]{Department of Physics, Duke University, Durham, NC 27708, USA}
\affiliation[b]{Department of Physics, University of Arizona, Tucson, AZ 85721, USA}

\emailAdd{paul.copeland@duke.edu}
\emailAdd{spf@arizona.edu}
\emailAdd{reed.hodges@duke.edu}

\abstract{We use vNRQCD to study power corrections in the $\vv$ expansion due to soft gluon radiation during $J/\psi$ production at small transverse momentum. We categorize four new $J/\psi$ production operators that mediate the transition of perturbatively produced color-octet charm quark/anti-quark pairs to charm quarks in a $^3S_1^{[1]}$ state via soft gluon emission. We then use Soft Collinear Effective Theory and vNRQCD to derive a factorization theorem for $J/\psi$ production in SIDIS in terms of the gluon transverse momentum dependent (TMD) PDFs in the proton and new objects which we call TMD soft transition functions. We show that the TMD soft transition function leads in the $\vv$ power-counting with respect to the color-octet TMD shape functions that have been used in previous studies of $J/\psi$ production at small transverse momentum.  }

\begin{document}
\maketitle
\flushbottom

\section{Introduction}
A central goal of the long range plan for nuclear science~\cite{LRP2023} is to determine how quarks and gluons (partons) are distributed in a nucleon. 
The dynamics of partons are encoded in theoretically defined distribution functions that can be extracted from high energy collision data. For example, data from deep inelastic scattering experiments can be used to ``measure'' so-called parton distribution functions (PDFs) which describes the distribution of the fraction of the nucleon lightcone momentum carried by the partons. Any experiment that contains a nucleon in the initial state provides an opportunity to extract various PDFs. 
%Of course there are number of collider experiments that have been and are still being carried out with different initial states, but if these initial states involve a nucleon then these experiments provide an opportunity to extract various PDFs. 
For a comprehensive review of PDFs and their measurements see ref.~\cite{Butterworth:2015oua}. 

While information on the distribution of lightcone momentum among the constituent partons deepens our understanding of the structure of the nucleon, it also begs for additional information - such as the distribution of momentum in the nucleon that is transverse to it's direction of motion. This information is encoded in a broader version of the PDFs known as transverse momentum dependent PDFs (TMDPDFs)~\cite{Boussarie:2023izj}. These functions can be extracted from experimental observables that are sensitive to the partonic transverse momentum in the initial state.

Bound states of heavy quarks ($Q\bar{Q}$), known as quarkonium, have recently emerged as a promising probe of nucleon structure at small transverse momentum scales. This is due to two features: the lowest lying vector meson states of quarkonium ($J/\psi, \Upsilon$) have extremely clean decays to $e^+ e^-$ and $\mu^+ \mu^-$, and are massive enough so that their decays products are visible even when the states have almost no transverse momentum. Furthermore, many of the dominant quarkonium production mechanisms require a gluon from the initial state to form a $Q\bar{Q}$ pair. This means that quarkonium production can access the gluon TMDPDFs in the proton. 

%The $J/\psi$ meson, a $c\bar{c}$ bound state in a $^3S_1^{[1]}$ configuration, is particularly attractive to study because of its clear decay signal to lepton pairs, making it easy to measure experimentally. 

Theoretically, quarkonium production is attractive because it can be  described using non-relativistic quantum chromodynamics (NRQCD) \cite{Bodwin:1994jh}. NRQCD is an effective field theory of QCD that exploits the large physical scale of the heavy quark masses by systematically expanding quarkonium observables in powers of the strong coupling, $\alpha_s$, and the small relative velocity of the heavy quark pair, $\vv$. For the $J/\psi$ meson, $\vv^2 \sim \alpha_s(2m_c) \sim 0.3$. In the collinear factorization framework, NRQCD has been applied to describe quarkonium production for a wide range of processes, such as quarkonium production in SIDIS \cite{Fleming:1997fq, Yuan:2000cn, Beneke:1998re, Chu:2024fpo, Maxia:2024cjh}, $e^+e^-$ annihilation \cite{Bodwin:2010fi, Braaten:2002fi, Hagiwara:2003cw, Bodwin:2008nf}, proton-proton collisions  \cite{Braaten:1994xb, Braaten:1994vv}, jets \cite{Baumgart:2014upa,Bain:2016clc,Bain:2017wvk,Kang:2017yde,Dai:2017cjq,Wang:2025drz, Copeland:2025osx}, exclusive processes \cite{Flett:2024htj,Ivanov:2004vd,Chen:2019uit,Blask:2025jua}, and heavy-ion collisions \cite{Sharma:2012dy,Yao:2018nmy, Yao:2020kqy, Yang:2024ejk}. The full list is too expansive to capture, see refs. \cite{Hoang:2002ae, Brambilla:2010cs} for a review. 

NRQCD is useful for studying for the production of quarkonium because of the NRQCD factorization conjecture \cite{Bodwin:1994jh,Nayak:2005rw,Nayak:2005rt,Nayak:2006fm}. This expresses the production cross section as a sum of products of perturbatively calculable partonic cross sections with vacuum production matrix elements, known as the ``Long Distance Matrix Elements" (LDMEs), each of which scales as some power of $\vv$. Thus the production cross section is a double series in $\alpha_s$ and $\vv$. An important aspect of NRQCD factorization is that heavy quark pairs can be produced in the hard process in quantum number configurations that are different from the final state quarkonium. The heavy quarks then undergo non-perturbative transitions via the radiation of ultrasoft gluons (gluons with momentum on the order of ${\cal O}(m_Q \vv^2) \sim  300 - 500$  MeV $\sim \Lambda_{QCD}$) in order to flip the $Q\bar{Q}$ pair to the correct quantum numbers to form the quarkonium boundstate. This process is encoded by the LDMEs which are thought to be universal constants that can be extracted from experiment. For $J/\psi$ production in particular, the most important LDMEs are the color-singlet LDME, which describes a color-singlet $c\bar{c}$ pair in a ${}^3S_1$ spin-angular momentum configuration hadronizing to $J/\psi$, and three color-octet LDMEs with a color-octet $c\bar{c}$ pair in either a ${}^3S_1$, ${}^1S_0$ or ${}^3P_J$ configuration hadronizing to $J/\psi$. The $^3S_1$ color-singlet LDME is the leading operator in the NRQCD expansion and scales as $\vv^3$ while the color-octet LDMEs are thought to scale like $\vv^7.$

NRQCD factorization has had success in predicting $J/\psi$ production cross sections, though a  number of discrepancies exist between theory and experimental observations. One of these discrepancies appears when studying the color-octet LDMEs. Though they each scale as the same power of $\vv$, various extractions of the $J/\psi$'s LDMEs from fits to the worlds $J/\psi$ production data \cite{Butenschoen:2011yh,Butenschoen:2012qr,Chao:2012iv,Bodwin:2014gia} show that the $^1S_0^{[8]}$ LDME is an order of magnitude larger than the other octet LDMEs. This disagrees with theoretical expectations. Interestingly, a new analysis was performed \cite{Brambilla:2024iqg} that used heavy quark spin symmetry to relate the LDMEs of the $\eta_c$ to those of the $J/\psi$. Using this constraint, the authors found that the $^1S_0^{[8]}$ LDME was suddenly one to two orders of magnitude smaller than its peers, the $^3S_1^{[8]}$ and $^3P_0^{[8]}$ LDMEs. This is in sharp contrast with more traditional extractions, making the narrative even more confusing. In addition, the color-octet mechanisms appear to be poorly constrained in general, with some of the parameters having uncertainties as large as $100-200\%$. Another, perhaps not unrelated, problem is the famous ``$J/\psi$ polarization puzzle". $J/\psi$ mesons produced at large $p_T$ in hadron colliders were predicted to be 100\% transversely polarized thirty years ago based on fundamental NRQCD assumptions. The argument goes as follows. At hadron colliders, charm quarks will be produced primarily via gluon fragmentation at large $p_T$ \cite{Braaten:1994xb} putting them in a $^3S_1^{[8]}$ state at leading order in $\alpha_s$ \cite{Braaten:1994vv}. The fragmenting gluon will be transversely polarized and the produced charm quarks will inherit this polarization. During the non-perturbative evolution of the $^3S_1^{[8]}$ charm quarks into a $J/\psi$, NRQCD spin symmetry guarantees that transitions which flip the $c\bar{c}$'s spin are suppressed by additional powers of $\vv$ so that the $J/\psi$ inherits the transverse polarization of the $c\bar{c}$ pair. If the earliest global extractions of the LDMEs are used \cite{Butenschoen:2011yh,Butenschoen:2012qr}, this prediction catastrophically disagrees with the measured polarization of the $J/\psi$ at the Tevatron and at the LHC \cite{LHCb:2013izl}. Some attempts have been made to resolve this puzzle by restricting the range of transverse momentum data included in global fits \cite{Chao:2012iv, Brambilla:2024iqg} and by resumming large logs of $p_T$ \cite{Bodwin:2014gia}, however the results are still unsatisfying, in part due to the aforementioned large error bars on the color-octet LDMEs that result from these fits. 

A confusing aspect of traditional NRQCD and the NRQCD factorization conjecture is the role that soft gluons with momentum $q^\mu \sim m_Q \vv$ play in the theoretical framework, as opposed to ultra-soft gluons with momentum $k^\mu \sim m\vv^2$. To address this short-coming two new approaches were developed: pNRQCD~\cite{Brambilla:1999xf,Brambilla:2004jw} and vNRQCD~\cite{Luke:1999kz,Rothstein:2018dzq}. In the pNRQCD approach the paradigm is the traditional effective field theory (EFT) picture where successive degrees of freedom are integrated out. Since NRQCD contains both soft and ultra soft degrees of freedom (dof), gluon fields contain momentum fluctuations on the order of $m\vv$ or less and the charm quarks can be off-shell by order $m\vv$ or less. The soft dof are then integrated out by matching onto pNRQCD which only contains dynamics from ultra-soft degrees of freedom. The vNRQCD picture is different as it explicitly contains two different gluon fields, one soft and the other ultra-soft, and interactions in the EFT are multipole expanded so that the heavy quarks only have off-shell fluctuations of order $m\vv^2$ or less. Recently, the production of the low lying vector meson charmonium and bottomonium states was investigated in pNRQCD~\cite{Brambilla:2022rjd,Brambilla:2022ayc}, where the matching of the most important LDMEs onto pNRQCD was performed. However, no work on LDMEs has been done in the vNRQCD framework.

Effective field theories have proven to be excellent tools to study TMD physics \cite{Fleming:2019pzj,Echevarria:2011epo,Echevarria:2012js,vonKuk:2023jfd, vonKuk:2024uxe, vonKuk:2025hdv,Dai:2023rvd,Copeland:2024wwm,Copeland:2024cgq,Ke:2024ytw} and, because quarkonium is a promising probe of nucleon structure at small transverse momentum scales, there has recently been a concentrated effort to describe direct quarkonium production in the small transverse momentum regime using the NRQCD and the TMD framework \cite{Catani:2014qha,Kang:2014tta,Sun:2012vc,Catani:2010pd,Mukherjee:2016cjw,Mukherjee:2015smo,Boer:2012bt,Echevarria:2019ynx,Fleming:2019pzj,DAlesio:2021yws,Boer:2020bbd,Bor:2022fga,Kishore:2021vsm,Scarpa:2019fol,DAlesio:2019qpk,Bacchetta:2018ivt,Mukherjee:2016qxa,Rajesh:2018qks,Godbole:2013bca,Godbole:2012bx,denDunnen:2014kjo,Kang:2014pya,Zhu:2013yxa, Copeland:2023qed, Copeland:2023wbu, Echevarria:2023dme, Echevarria:2024idp, Maxia:2025zee,Boer:2023zit}. This, however, is a complicated problem in the NRQCD formalism as the hadronization process is sensitive to soft gluon radiation, which as we discussed, is not well formulated in the EFT. This is because processes involving quarkonium with total transverse momentum on order of the soft scale, ${\cal O}(m_Q \vv) \sim  0.75 - 1.5 $ GeV, can involve a $Q\bar{Q}$ pair that can radiate an arbitrary number of soft gluons. This radiation can change the final state quarkonium's transverse momentum by a factor of order one \cite{Fleming:2019pzj, Echevarria:2019ynx, Echevarria:2024idp, Copeland:2023wbu}. In principle, this means that the NRQCD factorization conjecture must be modified for the TMD framework since the non-perturbative transition of the $Q\bar{Q}$ pair to quarkonium is no longer given by a constant quantity like the LDME. Instead, this object describing the non-perturbative transition should be replaced by a $\pv_T$ dependent quantity which some have dubbed ``TMD shape functions" (TMDShFs) \cite{Echevarria:2024idp, Fleming:2019pzj, Boer:2023zit}. This can naturally be accomplished in either the pNRQCD or vNRQCD framework. 

In this paper we use vNRQCD to extend the work of refs. \cite{Fleming:2019pzj} and \cite{Echevarria:2024idp} by deriving new TMD $S$-wave production operators that arise due to soft gluon radiation in the limit that the $J/\psi$'s transverse momentum is small. The vacuum production matrix elements of these operators create new $\pv_T$ dependent functions, which we dub TMD soft transition functions (TMDSTFs). The TMDSTFs describe the transition of color-octet $c\bar{c}$ pairs to charm quarks in a $^3S_1^{[1]}$ configuration via soft gluon emission, instead of ultrasoft gluon emission. These operators will be leading in the $\vv$ power-counting with respect to the color-octet TMDShFs of refs. \cite{Boer:2023zit, Echevarria:2024idp}.

%These operators are distinct from the usual color-octet operators, which are also subleading in the $\vv$ power-counting still need to undergo further transitions to reach a $^3S_1^{[1]}$ state. 

%We note that while we have derived our operators by expanding soft gluon emission diagrams to a subleading order, we find that some of our operators are actually leading over all in the $\vv$ power-counting when compared with the color-octet operators from ref. \cite{Echevarria:2024idp}. This is because, at ${\cal O}(\vv^5)$, soft gluon radiation can flip charm quarks in a $^1S_0^{[8]}$ configuration to a $^3S_1^{[1]}$ state in the TMD framework. This means that further ultrasoft gluon radiation is not required and the final operator can be counted as an ${\cal O}(\vv^5)$ object. This is to be contrasted with the $^1S_0^{[8]}$ TMDShF \cite{Echevarria:2024idp} which scales like $\vv^7$ due to the need to radiate off additional ultrasoft gluons in a magnetic dipole transition. 

We begin our work with section \ref{sec: background} by reviewing traditional NRQCD theory and the LDMEs commonly used for $J/\psi$ production. In this section, we also review vNRQCD, which will be the main theoretical framework we use for our analysis. In section \ref{sec: subleading} we derive new $J/\psi$ production operators by expanding the tree-level QCD diagrams for color-octet $c\bar{c}$ pairs radiating an arbitrary number of soft gluons to next-to-leading order in the $\vv$ expansion. We then match onto the corresponding operators in vNRQCD that reproduce our expressions. In section \ref{sec: projection}, we study the implications of these new operators by projecting out the leading order contributions, which place the fields in a $^3S_1^{[1]}$ configuration. In section \ref{sec: SIDIS}, we consider $J/\psi$ produced in SIDIS with small transverse momentum at leading order in $\alpha_s(2m_c)$. We match the hadronic currents onto our new subleading power operators and show that the hadronic tensor factorizes in terms of the gluon TMDPDFs in the nucleon and the new TMDSTFs. In the limit that the soft scale is perturbative, we show that the TMDSTF can be matched onto the $^3S_1^{[1]}$ LDME and that the TMDSTF is leading in the vNRQCD power-counting with respect to the so-called color-octet TMDShFs used in previous studies \cite{Echevarria:2024idp, Boer:2023zit}. We also find the TMDSTF is subleading in the TMD power-counting, which creates an interesting interplay between the TMDSTF and TMDShFs. Finally, we conclude and summarize future steps for this work, of which there are many.

%%%%%%%%%%%%%%%%%%%%%%%%%%%%%%%%%%%%%%
\section{Background}
\label{sec: background}
%%%%%%%%%%%%%%%%%%%%%%%%%%%%%%%%%%%%%%
%
We begin this work by describing the scales involved in the process of quarkonium production, reviewing traditional NRQCD, and discussing vNRQCD, which will be the main effective theory we use for our analysis. 

\subsection{NRQCD}

Systems involving a bound heavy quark-antiquark ($Q\bar{Q}$) pair have a clear separation of three important energy scales, which can be identified as follows:
\begin{itemize}
    \item $m \vv^0$ : \quad hard scale, the mass of the heavy quark;
    \item $m \vv^1$ : \quad soft scale, the relative three-momentum of the $Q\bar{Q}$ pair;
    \item $m \vv^2$ : \quad ultrasoft scale, the binding energy of the heavy $Q\bar{Q}$ pair.
\end{itemize}
Here $\vv \ll 1$ is the relative velocity of the $Q\bar{Q}$ pair and for charmonium, $\vv^2 \approx 0.3$. In NRQCD, $\vv$ is the power-counting parameter of the theory and operators are systematically organized according to their $\vv$ suppression. The NRQCD Lagrangian can be obtained by performing a large $m$ expansion on the full QCD Lagrangian, with the transformation $\Psi \rightarrow e^{-imt}(\psi \quad \chi)^T$. This allows one to write the full QCD field $\Psi$ in terms of nonrelativistic quark fields $\psi$ and antiquark fields $\chi$. The NRQCD lagrangian to order $1/m^3$ is  
\begin{equation}
    \begin{aligned}
        {\mathcal L}_{\text{NRQCD}} = & \; \psi^\dagger\bigg(iD_t + \frac{{\bf D}^2}{2m}\bigg)\psi + \chi^\dagger \bigg(iD_t - \frac{{\bf D}^2}{2m}\bigg)\chi\\
        &\, + \frac{c_F}{2m}[\psi^\dagger(g{\bf B}\cdot \boldsymbol{\sigma})\psi - (\psi \rightarrow \chi)] \\
        & \; + \frac{c_2}{8m^2}[\psi^\dagger(\Dv \cdot g{\bf E}-g{\bf E}\cdot \Dv)\psi + (\psi \rightarrow \chi)] \\
        & \; + \frac{c_3}{8m^2}[\psi^\dagger(i\Dv\times g{\bf E} - g{\bf E}\times i\Dv) \cdot \boldsymbol{\sigma}\psi + (\psi \rightarrow \chi)] \\
        & \; + \frac{c_1}{8m^3}[\psi^\dagger(\Dv^2)^2\psi - (\psi \rightarrow \chi)]\, .
    \end{aligned}
\end{equation}
In NRQCD, the quark and antiquark fields scale like $m\vv^{3/2}$ and their momentum has ``potential" scaling, $p_Q^\mu \sim m(\vv^2, \vv,\vv,\vv) $. The gluons are typically thought to have ultrasoft scaling, that is each of the components of the gluon field scales uniformly like $m\vv^2$, so as not to put the heavy quark fields far off-shell during interactions. However, gluons can in principle have soft scaling as well because the distinction between the soft and ultrasoft fields is not explicitly addressed in the traditional formulation. One of the useful applications of NRQCD is studying $J/\psi$ production in the collinear factorization framework. The production of a $c\bar{c}$ pair occurs at the hard scale, $m$, while the $c\bar{c}$'s hadronization to a $J/\psi$ is a softer process.  This leads to NRQCD factorization, which states that the cross section for $J/\psi$ production can be factorized into perturbative coefficients and non-perturbative LDMEs,
\begin{equation}
\label{eq: NRQCD fact}
    {\rm d}\sigma_{A+B\to J/\psi+X}\ = \sum_n {\rm d}\hat{\sigma}_{A+B\to c\bar{c}(n)+X} \braket{\mathcal{O}^{J/\psi}(n)} \, .
\end{equation}
The sum on $n \equiv {}^{2S+1}L_J^{[c]}$ is over the possible orbital angular momentum, spin, and color configurations for the produced $c\bar{c}$ pair.  The coefficients, ${\rm d}\hat{\sigma}_{A+B\to c\bar{c}(n)+X}$, are partonic cross sections describing the perturbative production of a $c\bar{c}$ pair in the quantum number configuration $n$ at scales $\mu_h$ greater than $2m$. Hence, they are a series in the strong coupling $\alpha_s(\mu_h)$. The LDMEs $\braket{\mathcal{O}^{J/\psi}(n)}$ are 
%the scalar parts of 
vacuum matrix elements of NRQCD operators which describe the hadronization of a $c\bar{c}$ to a $J/\psi$, and are often specified using spectroscopic notation $\braket{\mathcal{O}^{J/\psi}(^{2s+1}L_J^{[c]})}$.  For different $n$, the LDMEs have different scalings in $\vv$ that correspond to how many operator insertions are required to bring the $c\bar{c}$ quantum numbers to those of a $J/\psi$.  As such, eq.~(\ref{eq: NRQCD fact}) is a double expansion in $\alpha_s$ and $\vv$.  Here we list the four LDMEs most relevant to $J/\psi$ production, along with their traditionally-held NRQCD based $\vv$ scaling.
\begin{itemize}
    \item $\Otsosing$: scales as $\vv^3$, requires no operator insertions;
    \item $\Otsooct$: scales as $\vv^7$, requires two chromoelectric transitions, i.e.~insertions of ${\bf A}\cdot \boldsymbol{\nabla}$;
    \item $\Oosz$: scales as $\vv^7$, requires a single chromomagnetic transition, i.e.~an insertion of $\boldsymbol{\sigma} \cdot {\bf B}$;
    \item $\Otpz$: scales as $\vv^7$, requires a single chromoelectric transition and has further suppression due to its initial angular momentum quantum number.
\end{itemize}

NRQCD factorization allows for analytic expressions of the production cross sections to be obtained. This occurs through a matching calculation to determine the ${\rm d}\hat{\sigma}_{A+B\to c\bar{c}(n)+X}$.  The LDMEs are thought to be universal, but are non-perturbative parameters that must be fit to experiment.  Table \ref{tab: LDMEs} shows four different fits to determine these parameters. It is clear that there is notable disagreement between the fits and there are significant uncertainties.  Additionally, in the first three rows, $\Oosz$ is an order of magnitude larger than $\Otsooct$ and $\Otpz$, which is unexpected because all three LDMEs have been traditionally understood to scale as $\vv^7$.  In the last row, ref. \cite{Brambilla:2024iqg} found $\Oosz$ to be over an order of magnitude smaller than the other color-octet LDMEs. This is because they constrain this parameter using $\eta_c$ production data and the heavy quark spin symmetry relation, $\braket{{\cal O}^\eta_c(^3S_1^{[8]})} \approx \Oosz$.

\begin{table}[htbp]
\centering
\begin{tabular}{c|c|c|c|c}
\hline
& $\begin{aligned}\Otsosing \\ \times \, \text{GeV}^3\end{aligned}$ & $\begin{aligned}\Otsooct \\ \times 10^{-2} \, \text{GeV}^3\end{aligned}$ & $\begin{aligned}\Oosz \\ \times 10^{-2} \, \text{GeV}^3\end{aligned}$ & $\begin{aligned}\Otpz/m_c^2 \\ \times 10^{-2} \, \text{GeV}^3\end{aligned}$\\
\hline
B \& K \cite{Butenschoen:2011yh,Butenschoen:2012qr} & $1.32 \pm 0.20$ & $0.224 \pm 0.59$ & $4.97 \pm 0.44$ & $-0.72 \pm 0.88$\\
Chao et al.~\cite{Chao:2012iv} & $1.16\pm 0.20$ & $0.30 \pm 0.12$ & $8.9 \pm 0.98$ & $0.56 \pm 0.21$ \\
Bodwin et al.~\cite{Bodwin:2014gia} & $1.32 \pm 0.20$ & $1.1 \pm 1.0$ & $9.9 \pm 2.2$ & $0.49 \pm 0.44$ \\
Brambilla et al.~\cite{Brambilla:2024iqg} & $1.16 \pm 0.20$ & $1.05 \pm 0.12$ & $0.07 \pm 0.25$ & $1.88 \pm 0.26$ \\
\hline
\end{tabular}
\caption{Different fits for the NRQCD LDMEs.\label{tab: LDMEs}}
\end{table}

\subsection{vNRQCD}

As we have already mentioned numerous times, the original formulation of NRQCD does not distinguish well between gluons with soft and ultrasoft scaling. vNRQCD \cite{Luke:1999kz} overcomes this problem by presenting a useful delineation of the degrees of freedom, ensuring that the soft and ultrasoft modes are handled carefully.  In a manner similar to that of heavy quark effective theory \cite{Georgi:1990um}, the four-momentum of the heavy quark is separated into
\begin{equation}
    P = (E, \Pv) \equiv (k^0, \pv + \kv) \, ,
\end{equation}
where $\pv \sim m\vv$ is a soft momentum and $k\sim m\vv^2$ is an ultrasoft momentum.  The momentum $\pv$ is treated as a discrete label on the nonrelativistic quark and antiquark fields, $\psi_\pv(x)$ and $\chi_\pv(x)$, where the Fourier transform of the position argument $x$ is $\mathcal{O}(m\vv^2)$. Again, on-shell non-relativistic quarks must have potential scaling so that their energy and three-momentum scale as $m\vv^2$ and $m\vv$, respectively.  However, gluons can have their four-momenta scale either as soft or ultrasoft.  vNRQCD splits the gluons into two distinct degrees of freedom, one for each scaling.  The soft gluons, with momenta $q \sim m\vv$, utilize the label momentum notation $A_q$, while the ultrasoft gluons are denoted by $A$ and have momenta of order $m\vv^2$.  The effective Lagrangian describing the dynamics of these degrees of freedom is:
\begin{equation}
    \begin{aligned}
        {\mathcal L}_{\text{vNRQCD}} = & \; -\frac14 G_{us}^{\mu \nu}G_{us, \mu \nu} + \sum_p | p^\mu A_p^\nu - p^\nu A_p^\mu|^2 \\
        & \; + \sum_\pv \psi^\dagger_\pv \bigg( iD^0 - \frac{(\boldsymbol{\mathcal P}-i\Dv)^2}{2m} + \frac{\boldsymbol{\mathcal{P}}^4}{8m^3} + \frac{c_Fg}{2m}{\bf B} _{us}\cdot \boldsymbol{\sigma}\bigg) \psi_\pv + (\psi \rightarrow \chi) \\
        & \; -g^2 \sum_{q,q^\prime,\pv,\pv^\prime} \frac12 U_{\mu \nu}(\pv, \pv^\prime, q, q^\prime) \psi^\dagger_{\pv^\prime}[A_{q^\prime}^\mu, A_q^\nu] \psi_\pv  + (\psi \leftrightarrow \chi, \, T \leftrightarrow \bar{T}) \\
        & \; -g^2 \sum_{q,q^\prime,\pv,\pv^\prime} \frac12 W_{\mu \nu}(\pv, \pv^\prime, q, q^\prime) \psi^\dagger_{\pv^\prime}\{A_{q^\prime}^\mu, A_q^\nu\} \psi_\pv  + (\psi \leftrightarrow \chi, \, T \leftrightarrow \bar{T}) \\
        & \; - \sum_{\pv, \qv} V(\pv, \qv) \psi^\dagger_\qv \psi_\pv \chi^\dagger_{-\qv} \chi_{-\pv} \, .
    \end{aligned}
\end{equation}
The covariant derivative $D$ contains only ultrasoft gluon fields, $iD=i\partial -gA$, from which one forms $G_{us}^{\mu\nu}=\frac{1}{ig}[D^\mu,D^\nu]$ and ${\bf B}_{us}^k = -\frac12\epsilon^{ijk}G_{us}^{ij}$.  The label momentum operator $\mathcal{P}$ projects out the soft momentum of a field:
\begin{equation}
    \boldsymbol{\mathcal P}^i \psi_{\pv} = {\bf p}^i \psi_\pv \, , \qquad \mathcal{P}^\mu A_q ^\nu = q^\mu A_q^\nu \; .
\label{eq: projector}
\end{equation}
The matching to obtain the heavy quark potential $V$ and the coefficients $U_{\mu\nu}$ and $W_{\mu\nu}$ for the interaction between a heavy quark and two soft gluons was first performed in ref.~\cite{Manohar:1999xd}.  A manifestly gauge invariant form of the vNRQCD lagrangian involving soft Wilson lines is presented in ref.~\cite{Rothstein:2018dzq}.

%%%%%%%%%%%%%%%%%%%%%%%%%%%%%%%%
\section{Soft gluon radiation off a $c\bar{c}$ pair}
\label{sec: subleading}
%%%%%%%%%%%%%%%%%%%%%%%%%%%%%%%%

Our aim is to determine the $J/\psi$ production operators arising from the radiation of soft gluons by a $c\bar{c}$ pair. We consider a $c\bar{c}$ pair in some generic quantum number configuration denoted by a vertex structure, $\Gamma$. The vertex structure is placed between the heavy quark/antiquark fields produced in the hard process. The following analysis is agnostic about what the hard process was. We allow the $c\bar{c}$ to radiate soft gluons. Soft gluons have a larger energy than the ultra-soft energy of the heavy quarks and will knock the quarks off-shell, producing a local operator in vNRQCD. This enables us to write down new production operators in the effective theory. These production operators are essential to consider because the emission of a soft gluon can change the angular momentum, spin, and color of the $c\bar{c}$ pair, and can potentially ``flip" a $c\bar{c}$ in a color-octet configuration to a $^3S_1^{[1]}$ state. It will be important to consider the order in the $\vv$ power-counting at which this process occurs. Our logic is similar to the theory of $J/\psi$ production in traditional NRQCD, in which a $c\bar{c}$ pair in a color-octet configuration must emit an ultra-soft gluon that flips its color, spin, or angular momentum to a $^3S_1^{[1]}$ (referred to as chomo-electric or chromomagnetic dipole emissions, see section \ref{sec: background}). The number of emissions gives the total $\vv$ power-counting suppression. 

%\spf{Not sure we need the following paragraph}
%We note that, because the soft gluons will knock the heavy quark propagator off-shell, our analysis is only valid in the limit that the $J/\psi$'s total transverse momentum is small and TMD factorization is used. This is because the small transverse momentum of the $J/\psi$ provides an additional scale for the matching coefficients of the operators, which are otherwise scaleless and power-divergent, and hence vanish in dimensional regularization \cite{Beneke:1997av}. We elaborate on this point in the coming sections. 

\begin{figure}[htbp]
\centering
\includegraphics[width=.4\textwidth]{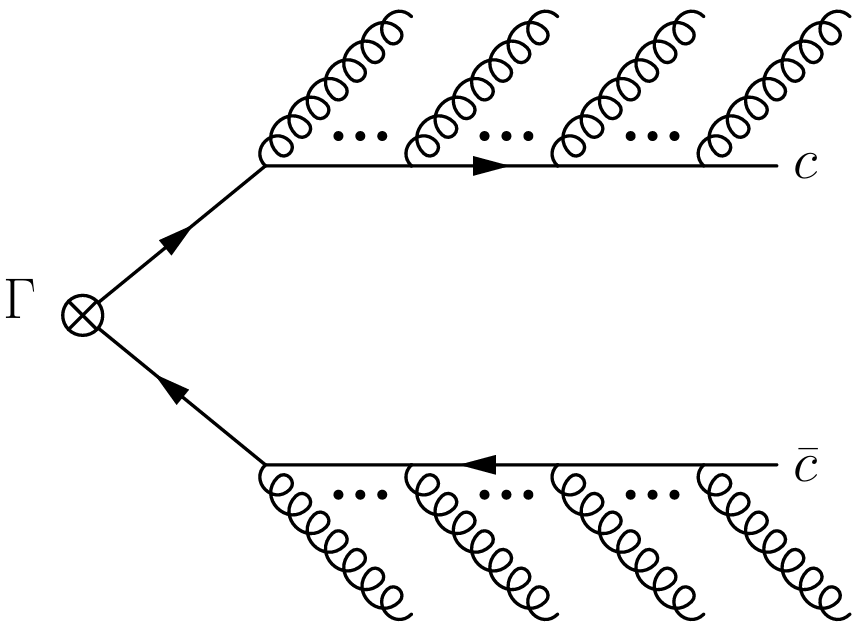}
\caption{Emission of an arbitrary number of soft gluons by a $c\bar{c}$ pair.\label{fig: arbitrary emissions}}
\end{figure}

We obtain our results by expanding all order tree-level QCD amplitudes in powers of $\vv$ and match the expressions onto vNRQCD operators. This means we consider an arbitrary number of soft gluon emissions from the $c\bar{c}$ pair, as shown in figure~\ref{fig: arbitrary emissions}.
Let the momentum of the $i$-th soft gluon emitted from a given quark or antiquark line be denoted by $p_i$. The ordering is such that $p_1$ is the gluon emitted furthest from $\Gamma$. In the following sections we use the shorthand notation
\begin{equation}
    p_{(\ell)} = \sum_{i=1}^\ell p_i 
\end{equation}
and write the QCD gluon field as
\begin{equation}
    A_i^\mu \equiv A^\mu(p_i).
\end{equation}
The $\ell$-th propagator for a quark/antiquark carrying momentum $p_{Q/\bar{Q}}+ p_1 + ... p_\ell$ is then compactly written as
\begin{equation}
    i\frac{\pm\slashed{p}_{Q/\bar{Q}}\pm \slashed{p}_{(\ell)} + m}{(p_{Q/\bar{Q}}+ p_{(\ell)})^2-m^2} \, ,
\label{eq: prop}
\end{equation}
where $p_{Q/\bar{Q}}$ is the momentum of the final state quark/antiquark.  Working in the quarkonium rest frame, we define a vector $v^\mu = (1,{\bf 0})$ so $p_{Q/\bar{Q}} = mv \pm q$, where $q = (0,{\bf q})$ is the soft relative three momentum of the charm quark pair which scales like $\qv  \sim m\vv$.

We expand the propagators and spinors in powers of $\vv$. Starting with the propagator in eq.~(\ref{eq: prop}) and using $\qv  \sim p_{(\ell)} \sim m\vv$ we find
\begin{equation}
    i\frac{\pm\slashed{p}_Q\pm \slashed{p}_{(\ell)} + m}{(p_Q + p_{(\ell)})^2-m^2} = i\bigg( \frac{1\pm\slashed{v}}{2} \pm \frac{\slashed{q}+\slashed{p}_{(\ell)}}{2m} \bigg) \frac{1}{p_{(\ell)}^0} \sum_{n=0}^\infty \bigg( \frac{-(q + p_{(\ell)})^2}{2mp_{(\ell)}^0} \bigg)^n \, .
\label{eq: expanded prop}
\end{equation}
Expanding the spinors in powers of $\vv$ gives
\begin{equation}
\begin{aligned}
    &u(p_Q) = u^{(0)}(p_Q) - \frac{{\boldsymbol{\gamma}}\cdot {\bf q}}{4m} u^{(0)}(p_Q)+ \cdots \\
    &v(p_{\bar{Q}}) = v^{(0)}(p_{\bar{Q}}) - \frac{{\boldsymbol{\gamma}}\cdot {\bf q}}{4m} v^{(0)}(p_{\bar{Q}})+ \cdots \\
\end{aligned}
\label{eq: expanded spinors}
\end{equation}
The leading order heavy quark spinors obey
\begin{equation}
    \slashed{v} u^{(0)}(p_Q) = u^{(0)}(p_Q) ~~~~  \slashed{v}v^{(0)}(p_{\bar{Q}}) = -v^{(0)}(p_{\bar{Q}}).
\label{eq: spinor EOM}
\end{equation}

\begin{figure}[htbp]
\centering
\includegraphics[width=.27\linewidth]{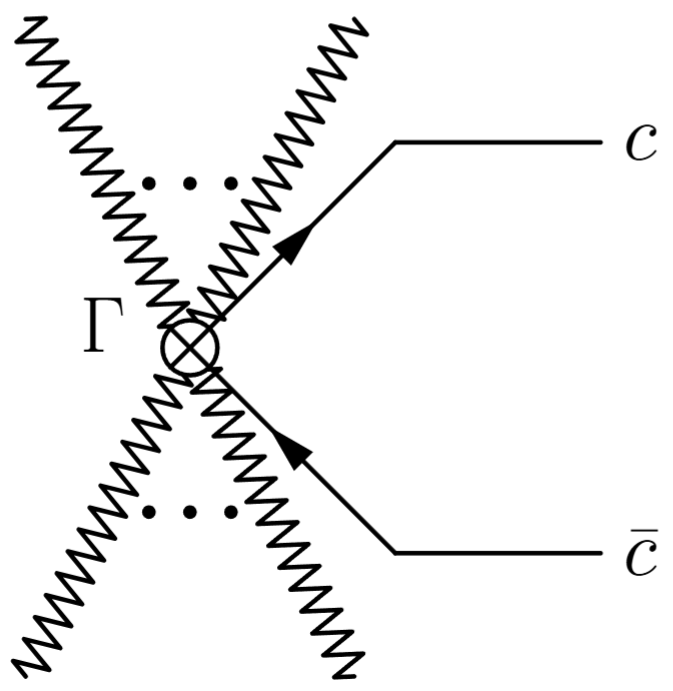}
\caption{vNRQCD diagram representing the emission of an arbitrary number of soft gluons by a $c\bar{c}$ pair with off-shell quark/anit-quark propagators contracted to a point.\label{fig: vNRQCD}}
\end{figure}

From an operator perspective, each emission is associated with the insertion of a soft gluon, $A^\mu_i$ and also introduces a propagator of the form of eq. (\ref{eq: prop}). Each soft gluon emission knocks the heavy quark off shell and contracts the propagator to a point, a process represented mathematically via the expansion in eq. (\ref{eq: expanded prop}). It was shown in ref. \cite{Fleming:2019pzj} that, after expanding to leading order in the $\vv$ expansion, the dominant term for each emission is a factor of $A_i^0/p_{(i)}^0$ which is not suppressed by $\vv$. This is why we must consider an arbitrary number of emissions. After expanding the various components of the QCD amplitude, we will match onto vNRQCD fields that reproduce the tree level results. We can represent the vNRQCD operators diagrammatically as shown in figure \ref{fig: vNRQCD}.

This procedure was used in ref. \cite{Fleming:2019pzj} to derive the color-octet S-wave and P-wave quarkonium production operators in vNRQCD at leading order in the $\vv$ expansion \footnote[1]{This terminology is confusing because, while directly power-counting the fields themselves makes the color-octet operators in ref. \cite{Fleming:2019pzj} leading in the $\vv$ expansion, matrix elements of these color-octet operators are subleading in practice because additional insertions of the vNRQCD Lagrangian are necessary to bring the color-octet charm quarks to the same quantum number configuration of the final state quarkonium. These insertions bring additional $\vv$ suppression to the matrix element.}. In the sections that follow, we extend the analysis to subleading order in the $\vv$ expansion. 

%At the order we consider in this paper, the quarks are still knocked off shell and the propagators contracted to points, however as we will see, new structures appear that modify the properties of the final state $c\bar{c}$ pair. 
%

%
\subsection{Soft gluon radiation at subleading power in $\vv$}
\label{sec: subleading charm}
The process shown in figure \ref{fig: arbitrary emissions} starts with the $c\bar{c}$ pair produced in a hard process with some quantum number configuration denoted by the vertex $\Gamma$, followed by radiation of an arbitrary number $n$ of on-shell soft gluons from the charm quark and an arbitrary number $n'$ of on-shell soft gluons from the anticharm quark. The soft gluon field scales like the soft gluon momentum: $A_s^\mu \sim p_s^\mu \sim m( \vv, {\bf v})$. The QCD amplitude for this contribution is
            \begin{equation}
            \begin{aligned}
               \mathcal{A}= &(-g)^{n+n'}u^\dagger(p_Q) \slashed{A}_1 \frac{\slashed{p}_Q + \slashed{p}_{(1)} + m_c}{(p_Q + p_{(1)})^2 -m_c^2 +i\epsilon}\cdots\slashed{A}_n\frac{\slashed{p}_Q + \slashed{p}_{(n)} + m_c}{(p_Q + p_{(n)})^2 -m_c^2 +i\epsilon}\Gamma\\
                &\times\frac{-\slashed{p}_{\bar{Q}} - \slashed{p}_{(n')} + m_c}{(p_{\bar{Q}} + p_{(n')})^2 -m_c^2 +i\epsilon} \slashed{A}_{n'} \cdots \frac{-\slashed{p}_{\bar{Q}} - \slashed{p}_{(1')} + m_c}{(p_{\bar{Q}} + p_{(1')})^2 -m_c^2 +i\epsilon} \slashed{A}_{1'} v(p_{\bar{Q}}),
            \label{eq: charm QCD amp}
            \end{aligned}
            \end{equation}
where we have replaced the polarization vectors with the gluon fields ($\epsilon_i^\mu \to A_i^\mu$) for clarity.
Each propagator comes with an $A^\mu_i$ insertion, so we pair the propagators and gluons together to power count each component of the diagram. From eq.~(\ref{eq: expanded prop}) we see the leading part of a propagator is ${\cal O}(\vv^{-1})$ and paired with a gluon field that scales like ${\cal O}(\vv)$, we get an ${\cal O}(1)$ contribution:
            \begin{equation}
                \slashed{A}_i \frac{1\pm\slashed{v}}{2p_{(i)}^0} \sim \frac{m\vv}{m\vv} \,.
             \label{eq: O(v-1) prop}
           \end{equation}
A propagator expanded to ${\cal O}(\vv^0)$ paired with a gluon is ${\cal O}(\vv)$: 
            \begin{equation}
                \slashed{A}_i\bigg(\frac{\slashed{p}_{(i)}}{2m_c p^0_{(i)}}- \frac{1+\slashed{v}}{2 p^0_{(i)}}\frac{p_{(i)}^2}{2m_c  p^0_{(i)}}\bigg) \sim \frac{m\vv}{m}.
            \label{eq: O(v0) prop}
           \end{equation}

\begin{figure}[htbp]
\centering
\begin{minipage}{0.39\textwidth}
\centering
\subfloat[]{\includegraphics[width=.8\textwidth]{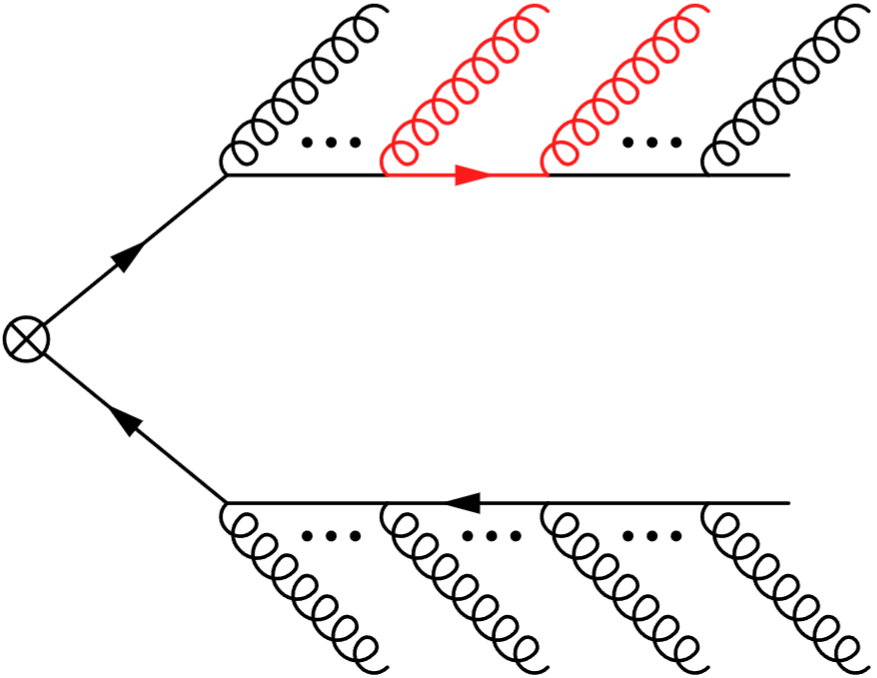}}
\end{minipage}%
\begin{minipage}{0.39\textwidth}
\centering
\subfloat[]{\includegraphics[width=.8\textwidth]{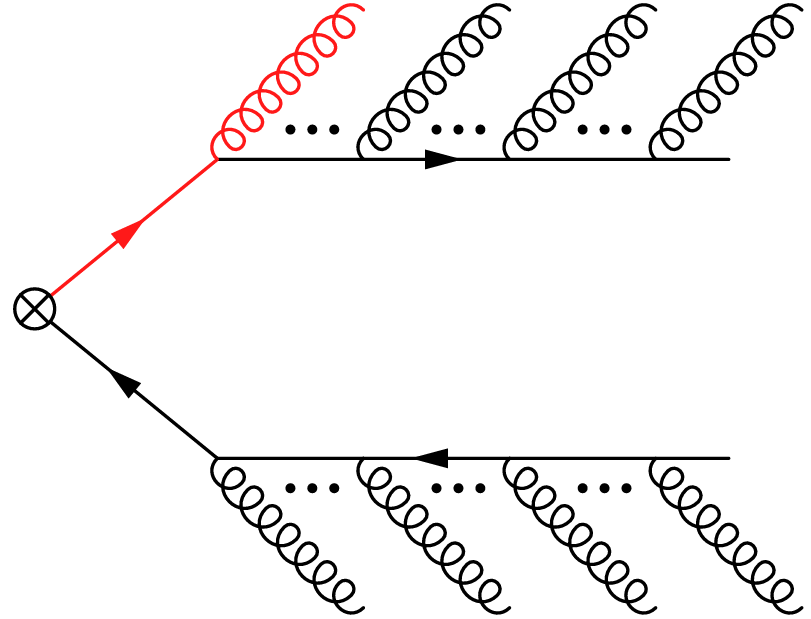}}
\end{minipage}%
\caption{Emission of an arbitrary number of soft gluons by a $c\bar{c}$ pair representing subleading corrections coming from the quark line. The diagram on the left corresponds to case (a) considered in the text and the diagram on the right to case (b) in the text.\label{fig: subleading quark 1}}
\end{figure}

Therefore, to obtain the subleading contributions, we can either: (a) expand the $\ell$-th quark propagator (for $\ell \neq n$) somewhere in the chain of gluons to ${\cal O}(\vv^0)$ and expand the remainder of the propagators to ${\cal O}(\vv^{-1})$, or (b) expand the $n$-th quark propagator closest to the vertex to ${\cal O}(\vv^0)$ and the rest of the propagators to ${\cal O}(\vv^{-1})$. Each case will produce a unique production operator at subleading order. These two scenarios are illustrated in figure~\ref{fig: subleading quark 1}, where the components that are highlighted red indicate the piece that will be expanded to ${\cal O}(\vv^{0})$.

First, consider case (a), where the $\ell$-th propagator is expanded to ${\cal O}(v^0)$ (for $\ell \neq n$) and the remaining propagators are expanded to leading order. This gives
        \begin{equation}
        \begin{aligned}
            \mathcal{A}_{3a} \approx &(-g)^{n+n'}u^{(0)\dagger}(p_Q) \slashed{A}_1 \frac{1+\slashed{v}}{2p_{(1)}^0}\cdots\slashed{A}_\ell\bigg(\frac{\slashed{p}_{(\ell)}}{2m_c p^0_{(\ell)}}- \frac{1+\slashed{v}}{2 p^0_{(\ell)}}\frac{p_{(\ell)}^2}{2m_c  p^0_{(\ell)}}\bigg)\cdots\slashed{A}_n \frac{1+\slashed{v}}{2p_{(n)}^0}\Gamma 
            \\
            &\times \frac{1-\slashed{v}}{2p_{(n')}^0} \slashed{A}_{n'} \cdots \frac{1-\slashed{v}}{2p_{(1')}^0} \slashed{A}_{1'} v(p_{\bar{Q}}).
        \end{aligned}
        \label{eq: expanded SL quark}
        \end{equation}
This is simplified by religiously applying $\slashed{A}_i (1+\slashed{v}) = (1-\slashed{v})\slashed{A}_i + 2A^0_i$ and $u^{(0)\dagger} (1-\slashed{v}) = 0$ on the quark leg, and $ (1-\slashed{v}) \slashed{A}_i = \slashed{A}_i(1+\slashed{v}) - 2A^0_i$ and $(1+\slashed{v})v^{(0)}  = 0$ on the antiquark leg. We also use
\begin{equation}
\begin{aligned}
    \slashed{A}_\ell ~&\slashed{p}_{(\ell)} \slashed{A}_{\ell +1} (1+\slashed{v}) = 2\bigg(A_\ell^0 A_{\ell+1}^{(0)}p_{(\ell)}^0-{\boldsymbol \gamma} \cdot {\bf A}_\ell {\boldsymbol \gamma} \cdot {\bf A}_{\ell +1} p_{(\ell)}^0 \\
             &+ {\boldsymbol \gamma}\cdot {\bf A}_\ell {\boldsymbol \gamma} \cdot {\bf p}_{(\ell)} A^0_{\ell +1} + A^0_{\ell}  {\boldsymbol \gamma} \cdot {\bf p}_{(\ell)} {\boldsymbol \gamma}\cdot {\bf A}_{\ell +1}\bigg)+(1-\slashed{v})\slashed{A}_\ell\slashed{p}_{(\ell)}\slashed{A}_{\ell+1}
\label{eq: ApA simplification}
\end{aligned}
\end{equation}
to obtain the ${\cal O}(v)$ approximation of eq.~(\ref{eq: expanded SL quark})
        \begin{equation}
        \begin{aligned}
            \mathcal{A}_{3a} \approx &\frac{(-g)^n(g)^{n'}}{2m} u^{(0) \dagger}(p_Q) \prod_{j = 1}^{\ell -1}  \frac{A^0_j}{p^0_{(j)}} \frac{1}{p_{(\ell)}^0} \bigg[A_\ell^0 A^0_{\ell +1} \frac{{\bf p}_s(\ell)^2}{p_{(\ell)}^0} -{\boldsymbol \gamma} \cdot {\bf A}_\ell {\boldsymbol \gamma} \cdot {\bf A}_{\ell +1} p_{(\ell)}^0 \\
             &+ {\boldsymbol \gamma}\cdot {\bf A}_\ell {\boldsymbol \gamma} \cdot {\bf p}_{(\ell)} A^0_{\ell +1} + A^0_{\ell}  {\boldsymbol \gamma} \cdot {\bf p}_{(\ell)} {\boldsymbol \gamma}\cdot {\bf A}_{\ell +1}\bigg] \frac{1}{p^0_{(\ell+1)}} \prod_{k=\ell+2}^n \frac{A^0_k}{p^0_{(k)}}\Gamma \prod_{w' = 1'}^{n'}  \frac{A^0_{w'}}{p^0_{(w')}} v^{(0)}(p_{\bar{Q}}).
        \label{eq: Simp SL quark}
        \end{aligned}
        \end{equation}
Now we apply what we call the ``magic formula''\footnote[2]{Of course there is no actual magic involved in eq. (\ref{eq: magic formula}) as it can be proved to be true.} \cite{Fleming:2019pzj}
%
%Magic formula
\begin{equation}
\begin{aligned}
    \frac{1}{p_{(i)}^0} \prod_{k=i+1}^n (-g)^{n-i}   \frac{A^0_k}{p^0_{(k)}} =& \sum_{\rho  = i}^n \frac{1}{p^0_{(\rho)} }\bigg[g^{\rho - i} \prod^{\rho}_{j = i + 1} \frac{A_j^0}{\sum_{k = i+1}^j p_{(\rho + i + 1 -k)}^0}\bigg]\\
    & \times\bigg[(-g)^{n-\rho} \prod^{n}_{j = \rho + 1} \frac{A_j^0}{\sum_{k = \rho+1}^j p_{(k)}^0}\bigg].
\label{eq: magic formula}
\end{aligned}
\end{equation}
Repeatedly applying the magic formula to eq.~(\ref{eq: Simp SL quark}) gives
        \begin{equation}
        \begin{aligned}
             \mathcal{A}_{3a} \approx&\frac{1}{2m} u^{(0) \dagger}(p_Q) \Bigg\{\bigg[ \bigg(\prod_{j = 1}^{\ell }  (-g)^\ell\frac{A^0_j}{p^0_{(j)}} \bigg) {\bf p}_s(\ell)^2-g\bigg( \prod_{j = 1}^{\ell -1}  (-g)^{\ell-1} \frac{A^0_j}{p^0_{(j)}} \bigg){\boldsymbol \gamma}\cdot {\bf A}_\ell {\boldsymbol \gamma} \cdot {\bf p}_{(\ell)} \bigg] \\
             & \times  \sum_{\rho  = \ell}^n \frac{1}{p^0_{(\rho)} }\bigg[g^{\rho - \ell} \prod^{\rho}_{j = \ell + 1} \frac{A_j^0}{\sum_{k = \ell+1}^j p_{(\rho + \ell + 1 -k)}^0}\bigg]\bigg[(-g)^{n-\rho} \prod^{n}_{j = \rho + 1} \frac{A_j^0}{\sum_{k = \rho+1}^j p_{(k)}^0}\bigg] \\      
             & -g\bigg[ \bigg(\prod_{j = 1}^{\ell }  (-g)^\ell\frac{A^0_j}{p^0_{(j)}} \bigg) {\boldsymbol \gamma} \cdot {\bf p}_{(\ell)} {\boldsymbol \gamma}\cdot {\bf A}_{\ell +1} + g\bigg( \prod_{j = 1}^{\ell -1}  (-g)^{\ell-1} \frac{A^0_j}{p^0_{(j)}} \bigg){\boldsymbol \gamma} \cdot {\bf A}_\ell {\boldsymbol \gamma} \cdot {\bf A}_{\ell +1} \bigg]\\
             & \times  \sum_{\rho  = \ell}^n \frac{1}{p^0_{(\rho)} }\bigg[g^{\rho - \ell-1} \prod^{\rho}_{j = \ell + 2} \frac{A_j^0}{\sum_{k = \ell+2}^j p_{(\rho + \ell + 2 -k)}^0}\bigg]\bigg[(-g)^{n-\rho} \prod^{n}_{j = \rho + 1} \frac{A_j^0}{\sum_{k = \rho+1}^j p_{(k)}^0}\bigg]\Bigg\}\\
             &\times \Gamma \prod_{w' = 1'}^{n'} g^{n'} \frac{A^0_{w'}}{p^0_{(w')}} v^{(0)}(p_{\bar{Q}}).
        \label{eq: magic simp SL quark}
        \end{aligned}
        \end{equation}
Although eq.~(\ref{eq: magic simp SL quark}) is a horrible mess, it can be simplified considerably. We sum over the number of gluons attached to the quark and antiquark legs, sum over all possible permutations, and normalize by the number of permutations. Then we identify
\begin{equation}
    S^\dagger_v = \sum_\ell \sum_{\rm perm} \frac{(-g)^\ell}{\ell!}\prod^n_{j=1}\frac{A^0_i}{p^0_{(j)}} 
\end{equation}
as a soft Wilson line in momentum space.
Now consider the combination 
\begin{equation}
\begin{aligned}
     &\sum_n \sum_{\rho  = i}^n \sum_{\rm perm} \frac{1}{p^0_{(\rho)}}\bigg[\frac{g^{\rho - i}}{(\rho-i)!}\prod^{\rho}_{j = i + 1} \frac{A_j^0}{\sum_{k = i+1}^j p_{(\rho + i + 1 -k)}^0}\bigg] \bigg[\frac{(-g)^{n-\rho}}{(n-\rho)!} \prod^{n}_{j = \rho + 1} \frac{A_j^0}{\sum_{k = \rho+1}^j p_{(k)}^0}\bigg].
\end{aligned}
\label{eq: magic gluons}
\end{equation}
For $i = 0$, this can be identified with
\begin{equation}
\begin{aligned}
     &\bigg[\frac{1}{v\cdot{\cal P}}S_v\bigg] S_v^\dagger \,,
\end{aligned}
\end{equation}
where we define a momentum projector operator that essentially acts like eq.~(\ref{eq: projector}) on the gluons: 
\begin{equation}
\begin{aligned}
    &{\cal P}^\mu A^\nu_i = p^\mu_i A_i \\
    {\cal P}^\mu A^{\mu_i}_i A^{\mu_{i+1}}_{i+1}  \cdots A^{\mu_m}_m = (&p_i + p_{i+1}+ \cdots + p_m)^\mu A_i^{\mu_i} A^{\mu_{i+1}}_{i+1}\cdots A^{\mu_m}_m.    
\end{aligned}
\label{eq: QCD proj}
\end{equation}
Our notation is such that a projector only acts on objects in the same square brackets as itself. For $i > 0$, the factor of $1/p_{(\rho)}^0$ contains momentum from the gluons in the first bracket of eq.~(\ref{eq: magic gluons}), but also from all gluons coming before. Thus the $1/p_{(\rho)}^0$ acts as a projector on the gluons from 1 to $\rho$, i.e.~the Wilson line that forms in the first bracket of eq.~(\ref{eq: magic gluons}) and the Wilson lines and gluons that come before it. This allows us to write eq.~(\ref{eq: magic simp SL quark}) in a compact form 
\begin{equation}
\begin{aligned}
   \mathcal{A}_{3a} \approx  \frac{-1}{2!}\frac{1}{2m}u^{(0) \dagger}(p_Q)\bigg[&\frac{1}{v\cdot{\cal P}}S_v^\dagger \big(-{\bfcal{P}}^{\dagger 2}-g   {\boldsymbol \gamma}\cdot {\bf A}_1 {\boldsymbol \gamma} \cdot {\bfcal P}^\dagger - g {\boldsymbol \gamma} \cdot {\bfcal P}^\dagger  {\boldsymbol \gamma}\cdot {\bf A}_2\\
    & + g^2 {\boldsymbol \gamma} \cdot {\bf A}_1  {\boldsymbol \gamma}\cdot {\bf A}_2 \big) S_v\bigg]S_v^\dagger \Gamma S_v v(p_{\bar{Q}}) + (1\leftrightarrow 2).
\end{aligned}
\label{eq: Wilson Line SL quark}
\end{equation}
Here ${\cal P}^\dagger$ is the same as in eq.~(\ref{eq: QCD proj}) except it acts to the left:
\begin{equation}
    A_1^{\mu_1} A_2^{\mu_2} .. A_n^{\mu_n} {\cal P}^{\dagger, \mu} =( -p_1 - p_2... - p_n)^\mu A_1^{\mu_1} .. A_n^{\mu_n} \,.
\end{equation}
This is similar to the usual convention in Soft-Collinear Effective Theory (SCET) \cite{Bauer:2000ew,Bauer:2000yr,Bauer:2001ct,Bauer:2001yt,Bauer:2002nz}. We have permuted the $A_1$ and $A_2$ gluons and divided by the number of gluons to account for the crossing diagrams. Finally, making use of ${\boldsymbol \gamma} \cdot {\bfcal P}^\dagger {\boldsymbol \gamma} \cdot {\bfcal P}^\dagger = - {\bfcal P}^{\dagger 2} $ and defining ${\bf D}_i = {\bfcal P} - g{\bf A}_i$ and ${\bf D}_i^\dagger = {\bfcal P}^\dagger - g{\bf A}_i$ we express the expanded amplitude as
\begin{equation}
   \mathcal{A}_{3a} \approx \frac{-1}{2!}\frac{1}{2m}u^{(0) \dagger}(p_Q)\bigg[\frac{1}{v\cdot{\cal P}}S_v^\dagger \big({\boldsymbol \gamma}\cdot {\bf D}_1^\dagger {\boldsymbol \gamma}\cdot {\bf D}_2^\dagger \big) S_v\bigg]S_v^\dagger \Gamma S_v v(p_{\bar{Q}})  + (1\leftrightarrow 2).
\label{eq: SL quark ans}
\end{equation}
The appearance of the covariant derivative is a check on the derivation as it must appear to preseve gauge invariance. This expression can be further reduced by taking advantage of the ability to relabel the gluons fields as each of the two gluon momenta are implicitly summed over. This give a factor of two canceling the two factorial in the denominator and removes the term with one and two interchanged. Furthermore, using $\gamma^i \gamma ^j = g^{ij}-i\sigma^{ij}$ the above expression can be written as:
\begin{equation}
   \mathcal{A}_{3a} \approx \frac{1}{2m}u^{(0) \dagger}(p_Q)\bigg[\frac{1}{v\cdot{\cal P}}S_v^\dagger \big({\bf D}_1^{\dagger}\cdot {\bf D}_2^\dagger +i \sigma_{ij}D_1^{i\dagger}D_2^{j\dagger} \big) S_v\bigg]S_v^\dagger \Gamma S_v v(p_{\bar{Q}}) \,.
\label{eq: SL quark ans1}
\end{equation}
This expression can be put in a more enlightening form.

In ref. \cite{Rothstein:2018dzq} the authors introduce a soft gluon gauge invariant building block in position space called $B^\mu(x)$, which is convenient for constructing operators. In this paper, for reason that will become apparent, we switch notation and call the soft gluon gauge invariant building block ${\bfcal E}^\mu(x)$. As in ref. \cite{Rothstein:2018dzq} the definition is
\begin{equation}
     g{\bfcal E}(x) = - \; S_v^\dagger ({{\bfcal P}} - g {\bf A}(x)) S_v  \,.
\label{eq: Bdagger}
\end{equation}
Using the standard relations between fundamental and adjoint Wilson lines this operator can be expressed as
\begin{equation}
     {\cal E}^i(x) =  \; \frac{1}{v\cdot{\cal P}} v^\nu G_{\nu i}^b S^{ba}_v T^a\,,
\label{eq: Bdagger}
\end{equation}
where $S^{ba}_v$ is the soft Wilson line in the adjoint representation. If we consider the rest frame of the quarkonium where $v^\mu = (1,0,0,0)$ then $v^\nu G_{\nu i}^b = G_{0 i}^b=E^b_i$ is the chromoelectric field, hence the use of the symbol ${\cal E}$. The hermitian conjugate of ${\bfcal E}^\mu(x)$ is
\begin{equation}
    g{\bfcal E}^\dagger =  \; -S_v^\dagger ({{\bfcal P}^\dagger} - g {\bf A}) S_v\,,
\label{eq: Bdagger}
\end{equation}
however the chromoelectric field is hermitian so ${\bfcal E}^\dagger = {\bfcal E}$.
Using this, the first term in the parenthesis in eq. (\ref{eq: SL quark ans1}) can be written in terms of ${\bfcal E}$:
\begin{equation}
    S^\dagger_v D_1^\dagger\cdot D_2^\dagger S_v = S^\dagger_v D_1^\dagger S_v\cdot S_v^\dagger D_2^\dagger S_v =g{\bfcal E}^\dagger_1\cdot g{\bfcal E}_2^\dagger\,, 
\end{equation}
which corresponds to a double electric transition.

The second term in parenthesis can be manipulated by once again using the property that allows us to relabel the gluons fields: 
\begin{equation}
    i \sigma_{ij}D_1^{i\dagger}D_2^{j\dagger}= \frac{i}{2}\sigma_{ij}[D_1^{i\dagger},D_2^{j\dagger}]=\frac{g}{2}\sigma_{ij}G^{ij}_{12}\,,
\end{equation}
with $G^{\mu \nu}$ the gluon field strength tensor. We recognize this term as the relativistic version of the chromomagnetic hyperfine splitting. 
Putting this together gives 
\begin{equation}
   \mathcal{A}_{3a} \approx \frac{1}{2m}u^{(0) \dagger}(p_Q)\bigg[\frac{1}{v\cdot{\cal P}} \big(g{\bfcal E}^\dagger_1 \cdot g{\bfcal E}^\dagger_2 +\frac{g}{2} \sigma_{ij}S_v^\dagger G^{ij}_{12} S_v \big) \bigg] \Gamma S_v v(p_{\bar{Q}}) \,.
\label{eq: SL quark ans final}
\end{equation}

Now consider case (b) shown in figure~\ref{fig: subleading quark 1}(b). Many of the steps are the same as above and, in fact, it is a much simpler derivation. We begin by expanding the ``$n$-th" propagator closest to the vertex, $\Gamma$, to ${\cal O}(\vv^0)$ and the rest of the propagators to ${\cal O}(\vv^{-1})$,
        \begin{equation}
        \begin{aligned}
           \mathcal{A}_{3b} \approx &(-g)^n (g)^n u^{(0)\dagger}(p_Q) \slashed{A}_1 \frac{1+\slashed{v}}{2p_{(1)}^0}\cdots\slashed{A}_n\bigg(\frac{\slashed{p}_{(n)}}{2m_c p^0_{(n)}}- \frac{1+\slashed{v}}{2 p^0_{(n)}}\frac{p_{(n)}^2}{2m_c  p^0_{(n)}}\bigg)\\
            & \times  \Gamma \frac{1+\slashed{v}}{2p_{(n')}^0} \slashed{A}_{n'} \cdots \frac{1+\slashed{v}}{2p_{(1')}^0} \slashed{A}_{1'} v^{(0}(p_{\bar{Q}}).
        \end{aligned}
        \label{eq: expanded end SL quark}
        \end{equation}
Again, we use $\slashed{A}_i (1+\slashed{v}) = (1-\slashed{v})\slashed{A}_i + 2A^0_i$ and $u^{(0)\dagger} (1-\slashed{v}) = 0$ on the quark leg, and $ (1-\slashed{v}) \slashed{A}_i = \slashed{A}_i(1+\slashed{v}) - 2A^0_i$ and $(1+\slashed{v})v^{(0)}  = 0$ on the antiquark leg to simplify. Expanding
\begin{equation}
    \slashed{A}_n {\slashed{p}_{(n)}} = A_n^0 p_{(n)}^0 + \slashed{v} {\bf \boldsymbol{\gamma}\cdot A}_n p_{(n)}^0 - \slashed{v} {\bf \boldsymbol{\gamma}\cdot p}_{(n)}A_n^0 + {\bf \boldsymbol{\gamma}\cdot A}_n  {\bf \boldsymbol{\gamma}\cdot p}_{(n)}\,
\label{eq: Ap simplification}
\end{equation}
allows us to write eq.~(\ref{eq: expanded end SL quark}) as 
\begin{equation}
        \begin{aligned}
             & \mathcal{A}_{3b} \approx \\
             &\frac{(-g)^n(g)^{n'}}{2m} u^{(0) \dagger}(p_Q) \prod_{j = 1}^{n-1}  \frac{A^0_j}{p^0_{(j)}}\bigg[ \frac{A_n^0}{p_{(n)}^0} \frac{{\bf p}_s(n)^2}{p_{(n)}^0} +{\bf \boldsymbol{\gamma}\cdot A}_n -  {\bf \boldsymbol{\gamma}\cdot p}_{(n)}\frac{A_n^0}{p_{(n)}} + {\bf \boldsymbol{\gamma}\cdot A}_n  {\bf \boldsymbol{\gamma}\cdot p}_{(n)}\frac{1}{p_{(n)}} \bigg] \\
             &\times \Gamma \prod_{w' = 1'}^{n'}  \frac{A^0_{w'}}{p^0_{(w')}} v^{(0)}(p_{\bar{Q}}).
        \label{eq: Simp SL end quark}
        \end{aligned}
        \end{equation}
Summing over all ``$n$" and permutations gives
        \begin{equation}
        \begin{aligned}
             \mathcal{A}_{3b} \approx&\frac{1}{2m} u^{(0) \dagger}(p_Q) \bigg[ S_v^\dagger\bigg(-\frac{{\bfcal P}^{\dagger ^2}}{v\cdot{\cal P}^\dagger} -g{\bf \boldsymbol{\gamma}\cdot A_1} +   \boldsymbol{\gamma}\cdot {\bfcal P}^{\dagger} + g{\bf \boldsymbol{\gamma}\cdot A_1} \frac{ \boldsymbol{\gamma}\cdot {\bfcal P}^{\dagger}}{v\cdot {\cal P}^\dagger}\bigg)\bigg] \Gamma S_v v^{(0)}(p_{\bar{Q}}).
        \label{eq: Wilson line SL end quark}
        \end{aligned}
        \end{equation}
%
%Notice the first and last operators in eq.~(\ref{eq: Wilson line SL end quark}) are redundant and reproduced by expanding the right most Wilson lines in eq.~(\ref{eq: Wilson Line SL quark}) to ${\cal O}(1)$. That leaves the middle two operators.
Which can be expressed in the compact form
\begin{equation}
        \begin{aligned}
              \mathcal{A}_{3b} \approx &-\frac{g}{2m} u^{(0) \dagger}(p_Q) \bigg[\boldsymbol{\gamma}\cdot {\bfcal E}_1^\dagger +  \boldsymbol{\gamma}\cdot {\bfcal E}_1^\dagger S_v^\dagger \frac{\boldsymbol{\gamma}\cdot {\bfcal P}^{\dagger} }{v\cdot{\cal P}^\dagger} S_v \bigg]S_v^\dagger \Gamma S_v v^{(0)}(p_{\bar{Q}}),
        \label{eq: SL end quark ans}
        \end{aligned}
        \end{equation}
where we have made use of the identity $S_v S_v^\dagger = 1$.

\begin{figure}[htbp]
\centering
\begin{minipage}{0.39\textwidth}
\centering
\subfloat[]{\includegraphics[width=.85\textwidth]{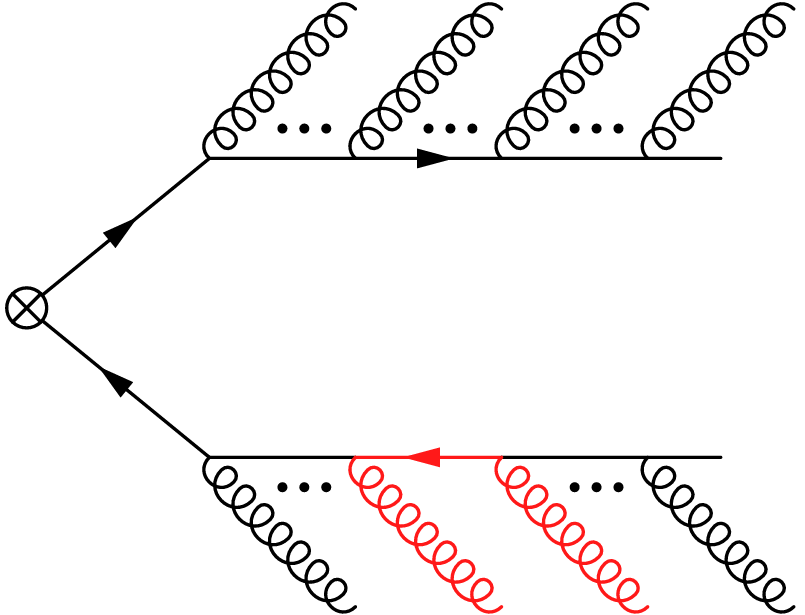}}
\end{minipage}%
\begin{minipage}{0.39\textwidth}
\centering
\subfloat[]{\includegraphics[width=.85\textwidth]{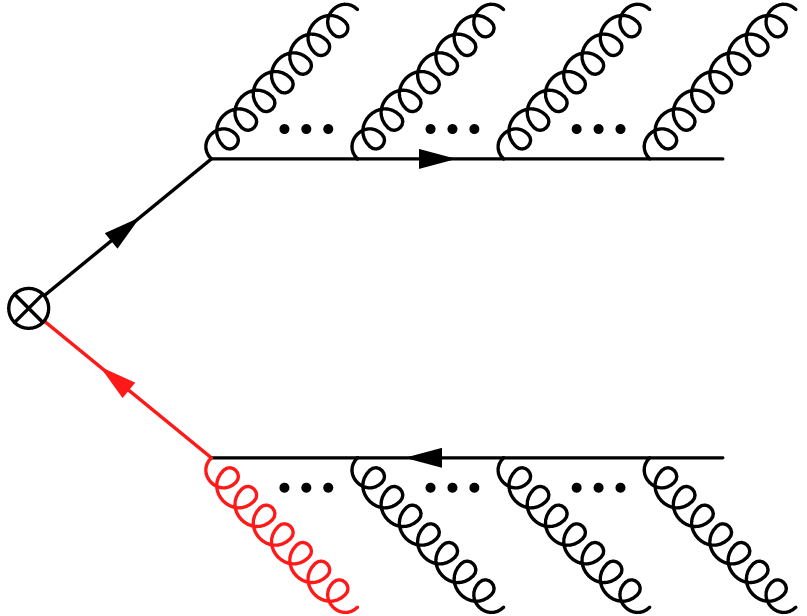}}
\end{minipage}%
\caption{Emission of an arbitrary number of soft gluons by a $c\bar{c}$ pair representing subleading corrections coming from the antiquark line. The diagram on
the left corresponds to case (a) considered in the text and the diagram on the right to case
(b) in the text.\label{fig: subleading antiquark}}
\end{figure}

Next we repeat the procedure for the antiquark line. Starting with eq. (\ref{eq: charm QCD amp}), where an arbitrary number of soft gluons are radiated from the quark and antiquark and we can either expand the $\ell$-th antiquark propagator to ${\cal O}(\vv^0)$ and the rest of the propagators to ${\cal O}(\vv^{-1})$ or the $n$-th antiquark propagator closest to the vertex to ${\cal O}(\vv^0)$ and the rest to ${\cal O}(\vv^{-1})$. These options are shown in figure~\ref{fig: subleading antiquark}.

The derivation of the antiquark operators is very similar to the derivation of the subleading quark operators in section \ref{sec: subleading charm}, so we relegate the details to appendix \ref{app: SL antiquark details}. We find that after expanding the $\ell$-th propagator, as in case (a) of figure \ref{fig: subleading antiquark}, eq.~(\ref{eq: charm QCD amp}) reduces to 
\begin{equation}
   \mathcal{A}_{4a} \approx - \frac{1}{2m}u^{(0) \dagger}(p_Q)S^\dagger \Gamma S_v \bigg[\frac{1}{v\cdot{\cal P}}(g{\bfcal E}_1\cdot g{\bfcal E}_2 + \frac{g}{2}S_v^\dagger G^{ij}_{12}S_v)\bigg]v(p_{\bar{Q}}).
\label{eq: SL antiquark ans 1}
\end{equation}

For case (b) of figure \ref{fig: subleading antiquark} we find that expanding the $n$-th propagator closest to the vertex gives us
\begin{equation}
        \begin{aligned}
           \mathcal{A}_{4b} \approx  &-\frac{g}{2m} u^{(0) \dagger}(p_Q)S_v^\dagger\Gamma S_v \bigg[ \boldsymbol{\gamma}\cdot {\bfcal E}_1 + S_v ^\dagger\frac{\boldsymbol{\gamma}\cdot {\bfcal P}}{v\cdot{\cal P}}S_v {\boldsymbol \gamma} \cdot {\bfcal E_1} \bigg]v^{(0)}(p_{\bar{Q}})\,.
        \label{eq: Wilson line SL end antiquark}
        \end{aligned}
        \end{equation}
This set of operators is the hermitian conjugate of the ones we found when expanding the quark line.

\subsection{Matching onto vNRQCD}
\label{sec: subleading vNRQCD}
%{\color{red} NEED TO DOUBLE CHECK/EXPLAIN NORMALIZATION OF THE SPINORS}

It is straightforward to construct vNRQCD operators that will reproduce the tree level results from section \ref{sec: subleading charm}. 
The expanded amplitudes for the quark and antiquark diagrams in figures \ref{fig: subleading quark 1} and \ref{fig: subleading antiquark} can be expressed as
\begin{equation}
\begin{aligned}
   \mathcal{A}_{3} &\approx \frac{-1}{2m} u^{(0)\dagger}(p_Q) \bigg[ \frac{1}{v\cdot{\cal P}} \big(g{\bfcal E}^\dagger_1 \cdot g{\bfcal E}^\dagger_2 +\frac{g}{2} \sigma_{ij}S_v^\dagger G^{ij}_{12} S_v \big) + g {\boldsymbol \gamma} \cdot  {\bfcal E}_1^\dagger  + g \boldsymbol{\gamma}\cdot {\bfcal E}_1^\dagger  S_v^\dagger  \frac{\boldsymbol{\gamma}\cdot {\bfcal P}^{\dagger} }{v\cdot{\cal P}^\dagger} S_v \bigg]\\ 
   &  \times S_v^\dagger\Gamma S_v v^{(0)}(p_{\bar{Q}})
\label{eq: QCD SL charm}
\end{aligned}
\end{equation}
and
\begin{equation}
\begin{aligned}
        \mathcal{A}_{4} &\approx  \frac{-1}{2m}  u^{(0) \dagger}(p_Q)  S_v^\dagger \Gamma S_v  \\
        &\times\bigg[ \frac{1}{v\cdot{\cal P}} \big(g{\bfcal E}_1 \cdot g{\bfcal E}_2 +\frac{g}{2} \sigma_{ij}S_v^\dagger G^{ij}_{12} S_v \big) +g{\boldsymbol \gamma} \cdot {\bfcal E}_1 + gS_v^\dagger \frac{\boldsymbol{\gamma}\cdot {\bfcal P}}{v\cdot{\cal P}}S_v {\boldsymbol \gamma} \cdot {\bfcal E}_1 \bigg]v^{(0)}(p_{\bar{Q}}) \, .
\label{eq: QCD SL anticharm}
\end{aligned}
\end{equation}
Next we simplify using the leading order spinors in the Dirac representation  \cite{Braaten:1996jt, Fleming:2019pzj},
\begin{equation}
\begin{aligned}
    u^{(0)}(p_Q) = \begin{pmatrix} \sqrt{2m} ~\xi \\ 0\end{pmatrix}\\
    v^{(0)}(p_{\bar{Q}}) = \begin{pmatrix} 0 \\ \sqrt{2m} ~\eta\end{pmatrix}.
\end{aligned}
\end{equation}
It is also useful to note that products of the Cartesian components of two gamma matrices in this basis are given by
\begin{equation}
    {\boldsymbol \gamma}^i {\boldsymbol \gamma}^j = \begin{pmatrix}
    -\sigma^i \sigma^j & 0\\
    0 & -\sigma^i \sigma^j\end{pmatrix}\,, 
\label{eq: cartesian double gamma}
\end{equation} 
and 
\begin{equation}
   \sigma^{ij} = \epsilon^{ijk}\begin{pmatrix}
    \sigma^k  & 0\\
    0 & \sigma^k\end{pmatrix} \,.
\label{eq: cartesian double gamma}
\end{equation}

Notice that the vertex containing information from the hard partonic cross section, $\Gamma$, must be a $4\times 4$ matrix for our expressions to make sense. Generically, for $2\times 2$ spin matrices, $\Gamma_a, \Gamma_b, \Gamma_c$ and $\Gamma_d$, $\Gamma$ has the form 
\begin{equation}
    \Gamma = \begin{pmatrix}
        ~~\Gamma_a~~ & \Gamma_b \\
        \Gamma_c & ~~\Gamma_d~~
    \end{pmatrix}.
\end{equation}
Therefore, the expressions in eq. (\ref{eq: QCD SL charm}) and eq. (\ref{eq: QCD SL anticharm}) reduce to 
\begin{equation}
\begin{aligned}
   \mathcal{A}_3 \approx & \, \frac{-1}{2m}\xi^\dagger \bigg[\frac{1}{v\cdot{\cal P}} \big(g{\bfcal E}^\dagger_1 \cdot g{\bfcal E}^\dagger_2 +\frac{g}{2} S_v^\dagger \epsilon_{ijk}G^{ij}_{12}\sigma^k S_v \big)   - g \boldsymbol{\sigma}\cdot {\bfcal E}_1^\dagger  S_v^\dagger  \frac{\boldsymbol{\sigma}\cdot {\bfcal P}^{\dagger} }{v\cdot{\cal P}^\dagger} S_v \bigg] S_v^\dagger \Gamma_b S_v \eta\, \\
    & + g \xi^\dagger  {\boldsymbol \sigma} \cdot  {\bfcal E}_1^\dagger S_v^\dagger \Gamma_d S_v \eta\,
\label{eq: expanded SL charm}
\end{aligned}
\end{equation}
and 
\begin{equation}
\begin{aligned}
   \mathcal{A}_4 \approx & \, \frac{-1}{2m}\xi^\dagger  S_v^\dagger \Gamma_b S_v  \bigg[ \frac{1}{v\cdot{\cal P}} \big(g{\bfcal E}_1 \cdot g{\bfcal E}_2 +\frac{g}{2} S_v^\dagger \epsilon_{ijk}G^{ij}_{12}\sigma^k S_v \big)   - g S_v^\dagger \frac{\boldsymbol{\sigma}\cdot {\bfcal P} }{v\cdot{\cal P}}S_v\boldsymbol{\sigma}\cdot {\bfcal E}_1    \bigg]\eta \, \\
    & + g\xi^\dagger S_v^\dagger \Gamma_a S_v{\boldsymbol \sigma} \cdot {\bfcal E}_1\eta 
\label{eq: expanded SL anticharm}
\end{aligned}
\end{equation}
respectively. We identify the combination ${\cal B}_{k,12}\equiv S_v^\dagger \epsilon_{ijk}G^{ij}_{12}S_v$ as the soft gauge invariant chromomagnetic field. 

We can match these objects onto the vNRQCD operators consisting of vNRQCD fields $\psi_{\pv_Q}$, $\chi_{\pv_{\bar{Q}}},$ and $A_q$ that reproduce the results at tree level. We find our set of operators is given by 
\begin{equation}
\begin{aligned}
   \mathcal{O}_3 = &\frac{-1}{2m}\psi^\dagger_{\pv_Q} \bigg[\frac{1}{v\cdot{\cal P}} \big(g{\bfcal E}^\dagger_q \cdot g{\bfcal E}^\dagger_{q'} +\frac{g}{2} {\bfcal B}_q\cdot\boldsymbol{\sigma} \big)   - g \boldsymbol{\sigma}\cdot {\bfcal E}^\dagger_q  S_v^\dagger  \frac{\boldsymbol{\sigma}\cdot {\bfcal P}^{\dagger} }{v\cdot{\cal P}^\dagger} S_v \bigg] S_v^\dagger \Gamma_b S_v\chi_{\pv_{\bar{Q}}} \, \\
    & - \frac{1}{2m}g\psi^\dagger_{\pv_Q}  {\boldsymbol \sigma} \cdot  {\bfcal E}_q^\dagger S_v^\dagger \Gamma_d S_v \chi_{\pv_{\bar{Q}}}\,,
\label{eq: vNRQCD SL charm}
\end{aligned}
\end{equation}
where we have used the hermiticity of the chromoelectric field, and 
\begin{equation}
\begin{aligned}
   \mathcal{O}_4 = &\frac{-1}{2m}\psi^\dagger_{\pv_Q}   S_v^\dagger \Gamma_b S_v  \bigg[\frac{1}{v\cdot{\cal P}} \big(g{\bfcal E}_q \cdot g{\bfcal E}_{q'} +\frac{g}{2} {\bfcal B}_q\cdot\boldsymbol{\sigma} \big)   - g S_v^\dagger \frac{\boldsymbol{\sigma}\cdot {\bfcal P} }{v\cdot{\cal P}}S_v \boldsymbol{\sigma}\cdot {\bfcal E}_q    \bigg]\chi_{\pv_{\bar{Q}}} \, \\
    & - \frac{1}{2m} \psi^\dagger_{\pv_Q}S_v^\dagger \Gamma_a S_v g{\boldsymbol \sigma} \cdot {\bfcal E}_q\chi_{\pv_{\bar{Q}}} .
\label{eq: vNRQCD SL anticharm}
\end{aligned}
\end{equation}
Some interesting features are apparent in these operators: the first term on the right of the expression in parenthesis involves two chromoelectric field operators. This term has no spin structure, but can be in either a color-singlet or color-octet configuration. The second term in parenthesis involves one chromomagnetic field operator in a color-octet configuration with a single Pauli matrix, meaning it will flip the quark spin. The final term is also in a color-octet configuration and can be expressed as a linear combination of the identity (in spin space) and a Pauli matrix. Thus depending on the structure of $\Gamma_{a,b}$ different parts of the operator will mediate transitions to various quarkonium states. In general the $\Gamma$ will have a spin structure (either $\mathbb{1}$ or $\sigma^i$), a color structure (either $\mathbb{1}$ or $T^A$) and possibly any number of derivatives. Since each derivative is suppressed by a power of $\vv$ or more these contributions are subleading and will not be considered. 

%When matching onto the leading production operators, only the off diagonal $\Gamma_b$ component of $\Gamma$ shows up in the vNRQCD operator vertex, since the vertex is between quark and antiquark spinors
%
%\begin{equation}
 %   \begin{pmatrix} \sqrt{2m} ~\xi^\dagger & 0\end{pmatrix} S_v^\dagger \Gamma S_v \begin{pmatrix}0 %\\ \sqrt{2m} ~\eta \end{pmatrix} = 2m \xi^\dagger \Gamma_b \eta.
%\end{equation}
%
Therefore, we can interpret operators with the $\Gamma_b$ structure as objects that will mediate the transition of charm quarks in S-wave color-octet configurations into some other quantum number configuration (we will focus on $^3S_1^{[1]}$ states) via soft gluon emission. In principle, the operators with the $\Gamma_a$ and $\Gamma_d$ vertices can also contribute to production physics. It is straightforward to show (using  $v\cdot A = 0$ gauge) that the $\Gamma_a$ and $\Gamma_d$ contributions in eq. (\ref{eq: vNRQCD SL charm}) and eq. (\ref{eq: vNRQCD SL anticharm}) cancel if the spin structure for the vertex from the hard process is $\Gamma \propto \gamma^\mu$. However, if $\Gamma$ is proportional to the identity or $\sigma^{\mu \nu}$ the $\Gamma_a$ and $\Gamma_d$ contributions are additive and do not vanish. 

\section{Projecting out the $^3S_1^{[1]}$ components}
\label{sec: projection}
%%%%%%%%%%%%%%%%%%%%%%%%%%%%%%%%%%%%%%%%%%%%%%%%%%%%%%%%%%%%%%%%%%%%%%%%%%%%%%%

\begin{figure}[htbp]
    \centering
\begin{tabular}{>{\centering\arraybackslash}m{1.4in}>{\centering\arraybackslash}m{2.2in}}
        
         {\Large $ \{^1S_0^{[8]}, ^3S_1^{[8]}, ^3P_J^{[8]}\}$} & 
       \includegraphics[width = 1.1\linewidth]{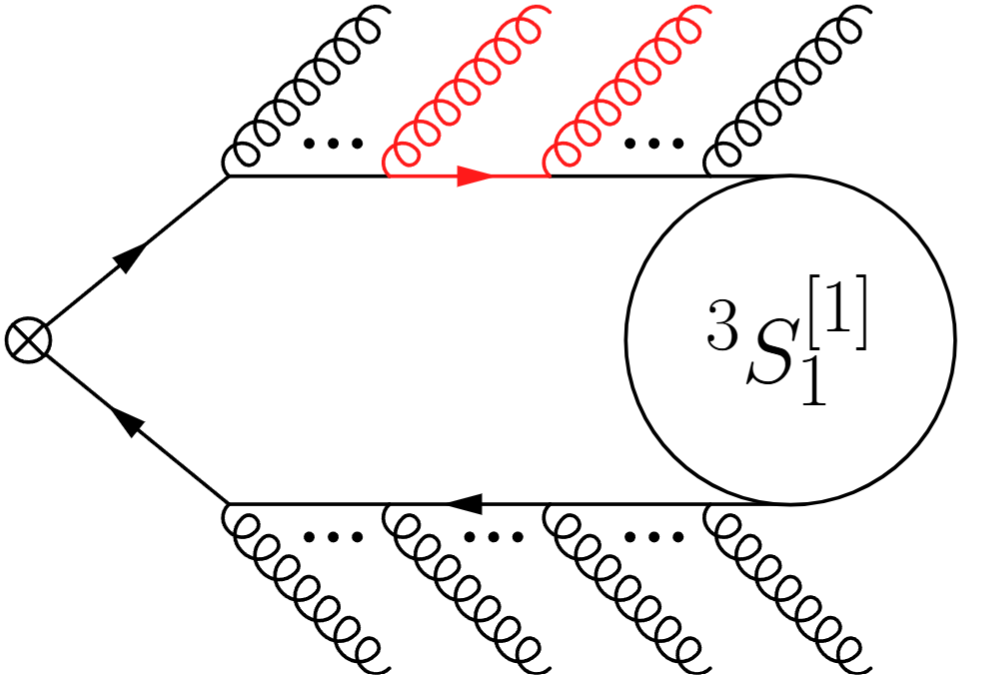} 
       \end{tabular}    \caption{Examples of the transition of a $c\bar{c}$ pair in a color-octet configuration to a $^3S_1^{[1]}$ state via soft gluon radiation at next-to-leading power in the $\vv$ expansion. Subleading operator insertion mediating the transition is indicated by the red propagator and gluons. The P-wave contribution is subleading to the two S-wave contributions.}
    \label{fig: subleading transition}
\end{figure}

In this paper we study the production of $J/\psi$ which is a $^3S_1^{[1]}$ state and therefore look for structures of $\Gamma$ that result in the charm anti-charm pair being in a $^3S_1^{[1]}$ configuration. In order to identify the dominant contributions, we must substitute in particular values for the vertex, $\Gamma$, and then pick out the $^3S_1^{[1]}$ component. An example of such a transition via a subleading operator is illustrated in figure \ref{fig: subleading transition}. The color-octet P-wave is subleading (by one power of $\vv$) relative to the other two operators and will be dropped from now on. We do not pick out other configurations, such as the $^1S_0^{[8]}$ or $^3S_1^{[8]}$, from the operator because charm-anticharm pairs in these configurations will need to transition to a $^3S_1^{[1]}$ state via additional ultrasoft or soft gluon emissions. This will induce additional $v$ suppression in the power-counting. 

As mentioned above, we find that if $\Gamma \propto \gamma^\mu$ the only operators in eqs. (\ref{eq: vNRQCD SL charm}) and (\ref{eq: vNRQCD SL anticharm}) that contribute to our (perturbative) analysis are those with the off diagonal $\Gamma_b$ vertices. 
%The $\Gamma_{a/d}$ operators are new objects that we will consider in a future section. 
To compute the transition, we need to insert values for $\Gamma_b$, which is the vertex that comes from the hard production of the $c\bar{c}$. In particular, we are interested in the two color-octet vertices: $^1S_0^{[8]}$ and $^3S_1^{[8]}$ denoted by their spin and color quantum numbers.
%and $^3P_J^{[8]}$, that define the LDMEs when inserted into the leading order operators \cite{Fleming:2019pzj}. 
% \\    &\Gamma^a_{^3P_0^{[8]}} = { \qv \cdot \boldsymbol \sigma } T^a.
These vertices are given by
\begin{equation}
\begin{aligned}
    &\Gamma^a_{^1S_0^{[8]}} = T^a \\ 
    &\Gamma^{\ell, a}_{^3S_1^{[8]}} = \sigma^\ell T^a \,.
\end{aligned}
\label{eq: production vertex}
\end{equation}
After substituting the vertices, we directly match onto a final $^3S_1^{[1]}$ configuration of the charm-anticharm pair by keeping only those terms that have no color matrix and a single Pauli matrix.   

One may be concerned about loop corrections to the leading order graphs, since these corrections could potentially be the same order in $\vv$ and $\alpha_s$ as our real emission diagrams. This, however is not the case. In the matching, loops will either arise from contractions of soft gluons in the Wilson lines or from a contraction of gluons from the subleading operators dressed by Wilson lines. In the former case the loops will just give an ${\cal O}(\alpha_s)$ correction to the operators being considered and in the later case the loops will give an ${\cal O}(\alpha_s)$ correction to a different operator that requires insertions of $\vv$ suppressed operators to have any overlap with a $^3S_1^{[1]}$ configuration. In either case the result is subleading to what we are considering.

\subsection{Subleading transitions to $^3S_1^{[1]}$}

We now consider how the operators in eqs. (\ref{eq: vNRQCD SL charm}) and (\ref{eq: vNRQCD SL anticharm}) can give rise to a $^3S_1^{[1]}$ configuration if the production vertex is either $^1S_0^{[8]}$ and $^3S_1^{[8]}$. Any of the three operators in the square brackets of eqs. (\ref{eq: vNRQCD SL charm}) and (\ref{eq: vNRQCD SL anticharm}) can mediate the transition from a color-octet state to a color singlet state, while only the double electric transition can mediate the transition from a color-singlet state to a color-singlet state (which we will not consider since it gives a subleading contribution). All of the terms can give a single Pauli matrix if $\Gamma_b \propto \sigma^i$, but only the second and third term can give a single Pauli matrix if $\Gamma_b \propto \mathbb{1}$. Thus for a $^1S_0^{[8]}$ production vertex only the second and third terms are important while for a $^3S_1^{[8]}$ production vertex we have to considere all of the terms.

The analysis is simplest in the axial gauge, $A^0_s = 0$ where the Wilson lines are set to one: $S_v = S_v^\dagger = 1$. First, we combine eqs. (\ref{eq: vNRQCD SL charm}) and (\ref{eq: vNRQCD SL anticharm}) and then simplify using 
\begin{equation}
    \sigma^i \sigma^j = 
    \delta^{ij} + i\epsilon^{ijk}\sigma^k. 
\label{eq: cartesian double gamma}
\end{equation} 
Some terms can be dropped because the onshell gauge condition $k_\mu A^\mu_k =0$ means that ${\bfcal P}\cdot {\bf A}_k = 0$ in $A^0_s = 0$ gauge. We can reduce the operator structure to
        \begin{equation}
        \begin{aligned}
            & \psi^\dagger_{\bf p_Q}  {\cal O}_{S.L.}(\Gamma) \chi_{\bf p_{\bar{Q}}}=  \\
            & \qquad \frac{-1}{2m}\psi^\dagger_{\bf p_Q}\bigg[ \frac{1}{v\cdot\Pc} \bigg(g^2 \big\{ {\bfcal E}_{q} \cdot {\bfcal E}_{q'}, \Gamma_b\big\} +\frac{g}{2}\big\{ {\bfcal B}_q\cdot \boldsymbol{\sigma},\Gamma_b\big\} -i g \epsilon_{ijk}[\Gamma_b,{\bfcal P}^i{\bfcal E}^j_q \sigma^k] \bigg) \bigg] \chi_{\bf p_{\bar{Q}}}\,.\\
        \end{aligned}
        \label{eq: combined subleading}
        \end{equation}

Next we will consider, in turn, the consequences of inserting a $^1S_0^{[8]}$ or $^3S_1^{[8]}$ production vertex. Since we are looking for a $^3S_1^{[1]}$ structure we will only keep the color-singlet terms that are proportional to a single Pauli matrix. For the $^1S_0^{[8]}$ vertex given in eq. (\ref{eq: production vertex}) the only color singlet term which survives is the anti-commutator with the chromomagnetic field. Letting ${\bfcal B} = {\bfcal B}^b T^b$ we obtain
%
%
%\begin{equation}
%    P_{^3S_1^{[1]}}\big[{\cal O}] = \frac{1}{2N_c}{\rm Tr}\big[{\cal O} \sigma^\ell \big] \sigma^\ell,
%\end{equation}
%
%where the trace is in color and spin space. This spits out the $^3S_1^{[1]}$ piece of ${\cal O}$ when it is sandwiched between $\psi^\dagger$ and $\chi$.
%
%Now, we insert particular values for the vertex. For a $c\bar{c}$ initially in a $^1S_0^{[8]},$ the vertex is $ \Gamma = T^a$.  Plugging this into eq. (\ref{eq: combined subleading}) and projecting out the $^3S_1^{[1]}$ components gives
%
        \begin{equation}
        \begin{aligned}
            & \psi^\dagger_{\bf p_Q} P_{^3S_1^{[1]}}\big[{\cal O}_{S.L.}(\Gamma^a_{^1S_0^{[8]}}) \big] \chi_{\bf p_{\bar{Q}}} =   \frac{-g}{12m} \psi^\dagger_{\bf p_Q} \bigg[ \frac{1}{v\cdot\Pc}{\bfcal B}^a\cdot\boldsymbol{\sigma}\bigg] \chi_{\bf p_{\bar{Q}}}\,,\\
        \label{eq: 1S0 transition 1}
        \end{aligned}
        \end{equation}
where $P_{^3S_1^{[1]}}$ indicates that we are projecting out the color-singlet ${}^3S_1$ state.
Although the dominant component of eq. (\ref{eq: 1S0 transition 1}) scale like $\vv^4$ in the power-counting, there is an additional power of the $g$ yielding a mild additional suppression because at the soft scale $\alpha_s(m\vv) \sim \vv$.

We now repeat this procedure for the $^3S_1^{[8]}$ production vertex. This time the commutator with the single electric field and the anti-commutator with the double electric field survives. Letting ${\bfcal E}= {\bfcal E}^a T^a$ and using the ability to interchange the labels $q$ and $q'$ we find
%In eq. (\ref{eq: combined subleading}) we insert the vertex for charm quarks initially in a $^3S_1^{[8]}$ configuration, $\Gamma^a_{^3S_1^{[8]}} = \sigma^\ell T^a$, and project out the $^3S_1^{[1]}$ component. This gives
%
\begin{equation}
        \begin{aligned}
            &\psi^\dagger_{\bf p_Q} P_{^3S_1^{[1]}}\big[{\cal O}_{S.L.}(\Gamma^{\ell,a}_{^3S_1^{[8]}}) \big]\chi_{\bf p_{\bar{Q}}} = \frac{-g}{12m} \psi^\dagger_{\bf p_Q} \bigg[\frac{1}{v\cdot\Pc}  \bigg(2\epsilon^{ijk}\epsilon^{\ell k m} {\bfcal P }_i{\bfcal E}^a_j \sigma_m+gd^{abc} {\bfcal E}_{q}^{ b} \cdot {\bfcal E}_{q'}^{c} \sigma^\ell \bigg) \bigg] \chi_{\bf p_{\bar{Q}}}\,.
        \end{aligned}
    \label{eq: 3S18 trans operator}
\end{equation}
The first term in eq. (\ref{eq: 3S18 trans operator}) is leading in the $\vv$ power-counting and scales like $g\times\vv^4$. Hence, it is the same order as eq. (\ref{eq: 1S0 transition 1}). The second term in eq. (\ref{eq: 3S18 trans operator}) scales like $\alpha_s(m\vv) \vv^4$ meaning that it is slightly suppressed (by a mere $\sqrt{\vv}$) relative to the operator appearing in eq. (\ref{eq: 1S0 transition 1}). If we take $\alpha_s(m\vv) \sim 1$, all operators in eq. (\ref{eq: 3S18 trans operator}) and eq. (\ref{eq: 1S0 transition 1}) are order $\vv^4$.
%approximately the same size as the second operator in eq. (\ref{eq: 1S0 transition}). Therefore, both of these operators are slightly subleading in the power-counting with respect to the first term in eq. (\ref{eq: 1S0 transition}). Note that these operators are only subleading due to the factor $\alpha_s(m\vv)$. If somehow $\alpha_s(m\vv) \gg v$ then these operators would suddenly be the same order as the dominant term in eq. (\ref{eq: 1S0 transition}). 

In this section we have deduced the dominant operators for S-wave color-octet state transitions to a $^3S_1^{[1]}$ configuration. We have not discussed P-wave color-octet transitions because they will be sub-leading in the power-counting. P-wave operators were considered in ref. \cite{Fleming:2019pzj} in which the authors wrote down the correct operators that arise from soft-gluon emissions transitioning $c\bar{c}$ pairs into a $^3P_J^{[8]}$ state. However, from here, the color-octet P-wave charm quarks are assumed to transition to a $^3S_1^{[1]}$ state via ultrasoft emission. 
%Using our approach, one would need to radiate more soft gluons off of the P-wave operator and expand one of the quark/antiquark propagators to ${\cal O}(\vv^0)$. Then, one would need to project out the leading $^3S_1^{[1]}$ component which should yield the final operator that arises from soft gluon transitions. This operator will be sub-sub-leading in the $\vv$ expansion and thus it's derivation is beyond the scope of this work.   

%{\color{red} Talk about how the $^3S_1^{[8]}$ and $^1S_0^{[8]}$ are the same size if $g \sim 1$. }
%
%which is a $^1S_0^{[1]}$. This is interesting because it shows that hard charm quarks produced in a $^3S_1^{[8]}$ configuration still require higher order operators in order to transition to the correct quantum numbers of the $J/\psi$. In other words, it is suppressed with respect to the $^1S_0^{[8]}$ mechanism.
%

%%%%%%%%%%%%%%%%%%%%%%%%%%%%%%%%%%%%%%%%%%%%%%%%%%%%%%%%%%%%%
\section{$J/\psi$ production in SIDIS in the TMD framework}
\label{sec: SIDIS}
%%%%%%%%%%%%%%%%%%%%%%%%%%%%%%%%%%%%%%%%%%%%%%%%%%%%%%%%%%%%%

Recently, factorization theorems for $J/\psi$ leptoproduction at small transverse momentum were derived using a combination of vNRQCD and SCET \cite{Echevarria:2024idp}. In this analysis, the authors match the hadronic currents onto a combination of SCET and vNRQCD operators which allows them to factorize the cross section in terms of the gluon TMDPDFs in the proton and so-called ``TMD shape functions" (TMDShFs) \cite{Fleming:2019pzj, Echevarria:2024idp, Boer:2023zit, Maxia:2025zee} describing the $J/\psi$'s hadronization. The TMDShFs incorporate the ultra-soft physics responsible for hadronization of the $c\bar{c}$ pair into the $J/\psi$ but do not include the soft physics we have been considering. In this section we, therefore, perform a similar analysis by matching the hadronic currents onto the soft operators we have identified as the dominant power corrections in the $\vv$ expansion.

We begin by introducing the kinematics of $J/\psi$ production SIDIS
\begin{equation}
    \ell(l) + N(P_N) \rightarrow \ell(l^\prime)+J/\psi(P_\psi) + X \, ,
\end{equation}
where $\ell$ is a lepton and $N$ is the initial nucleon. The momenta of the particles are indicated in parentheses. We use lightcone coordinates, where any vector can be written as $v^\mu \equiv (v^+, v^-, {\bf v}_T)$ and define light-like vectors, $n$ and $\bar{n}$ such that $n\cdot v=v^-$, $\bar{n}\cdot v=v^+$, and $n\cdot \bar{n} = 1$. We use the conventional kinematic variables \\
\begin{equation}
    Q^2 = -q^2 = -(l-l')^2, ~~x_B = \frac{Q^2}{2P_N\cdot q}, ~~y= \frac{P_N\cdot q}{P_N\cdot l}, ~~z = \frac{P_N\cdot P_\psi}{P_N\cdot q} 
\end{equation}
and work in the hadron frame, where the proton and $J/\psi$ have no transverse momentum, but the virtual photon does. In this frame, momenta are given by
\begin{equation}
\begin{aligned}
    q^\mu =& \; \left(-\frac{Q^2-\qv_T^2}{\sqrt{2}Q},\frac{Q}{\sqrt{2}},\qv_T \right) \, , \\
    P_N^\mu =& \; \left(\frac{Q}{\sqrt{2} x_B},0,{\bf 0} \right) \, , \\
    P_\psi^\mu =& \; \left(\frac{M^2}{\sqrt{2}zQ},\frac{zQ}{\sqrt{2}}, {\bf 0}\right) \, ,
\end{aligned}
\end{equation}
where we have neglected target mass corrections ($M_N \approx 0$). Note, $q^2 = -Q^2$, $P_N^2 = 0$, and $P_\psi^2 = M^2$. We can write the $J/\psi$'s momentum as $P_{\psi} = Mv^\mu$, where the velocity of the $J/\psi$ is defined in the boosted frame. The transverse momentum of the photon, $\qv_T$, is with respect to the $z$-axis defined by the direction of the nucleon's momentum. 

Using this notation, the differential cross section for SIDIS can be written in the standard form
\begin{equation}
    \frac{d\sigma}{dx_B \, dz \, dQ^2 \, d{\bf q}_T^2} = \frac{\alpha_{\rm em}^2 M}{2Q^2x_Bzs} L^{\mu\nu}W_{\mu\nu} \, ,
\end{equation}
where the leptonic and hadronic tensors are defined by
\begin{equation}
    L_{\mu \nu} = e^{-4} \bra{l'} J_\mu(0)\ket{l}\bra{l} J^\dagger_\nu (0) \ket{l'} \, 
\end{equation}
\begin{equation}
W_{\mu \nu} = \int \frac{d^4 b} {(2\pi)^4} e^{ib\cdot q} \sum_X \bra{p} J_\mu^\dagger(b) \ket{J/\psi,X}\bra{J/\psi,X} J_\nu(0) \ket{p} .
\label{eq: hadronic tensor}
\end{equation}
As in ref. \cite{Echevarria:2024idp}, we will use a combination of vNRQCD and SCET to factorize the hadronic tensor. In SCET, the relevant scales in our problem are the virtual momentum transfer of the photon ($Q$) and the transverse momentum of the final state ($\qv_T$). The power-counting parameter for TMD factorization in SCET is then given by $\lambda^2 = \qv_T^2/Q^2 \ll 1$. In vNRQCD, the scales were already discussed in section \ref{sec: background} - they are the $J/\psi$'s mass ($M$), the relative momentum of the heavy quarks ($m_c \vv \sim M \vv/2$, and the energy of the bound state ($m_c\vv^2 \sim M\vv^2/2$). We work in a regime where the SCET power-counting parameter is of the same order as the vNRQCD power-counting parameter, i.e., $\lambda \sim \vv$. This means that there is an overlap between the soft regions of SCET and vNRQCD and the two theories must be combined into a new EFT containing all relevant degrees of freedom. This new EFT has been referred to as SCET$_Q$ \cite{Fleming:2019pzj}. We will also work at leading order in $\alpha_s(M)$, where the hard production of a $c\bar{c}$ pair is given by the partonic cross section \cite{Fleming:1997fq, Copeland:2023qed, Echevarria:2024idp} 
\begin{equation}
    \gamma^* + g \rightarrow c\bar{c}(n)+g_s \, ,
\end{equation}
where the final state gluon $g_s$ is soft. Thus it does not contribute a factor of $\alpha_s(M)$ and should be paired with the heavy quark fields when constructing the low-energy operators. We will derive the factorization theorem by matching the hadronic tensor onto effective currents defined in terms of vNRQCD and SCET operators. The operators we need are the gauge invariant gluon building block from SCET
\begin{equation}
    B_{n\perp}^\mu(x) = \frac{1}{g}\bigg[W_n^\dagger (i\partial_{n\perp}^\mu + g A_{n\perp}^\mu)W_n\bigg](x)
\end{equation}
where $W_n$ is a $n$-collinear Wilson line in position space \cite{Echevarria:2024idp}, a soft Wilson line describing radiation in the $n$ direction from the initial state proton,
\begin{equation}
    S_n(x) = {\cal P} ~{\exp}\bigg[ig \int_{-\infty}^0 d\lambda n \cdot A_s(x + n \lambda)\bigg],
\end{equation}
and the operator derived in eq. (\ref{eq: 1S0 transition 1}).

\begin{figure}[htbp]
    \centering
\begin{tabular}{>{\centering\arraybackslash}m{1.5in}>{\centering\arraybackslash}m{.35in}>{\centering\arraybackslash}m{1.5in}>
{\centering\arraybackslash}m{.35in}>{\centering\arraybackslash}m{1.5in}}
        
       \hspace{-.25cm}\includegraphics[width = 1.1\linewidth]{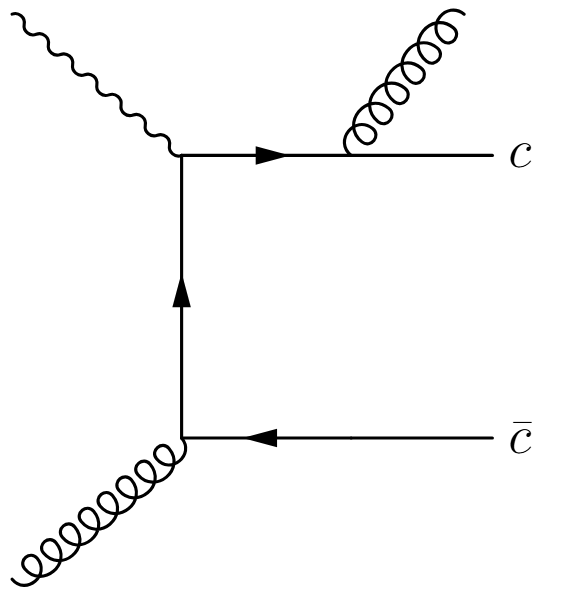}   &   {\huge $\longrightarrow$} & 
       \includegraphics[width = 1.1\linewidth]{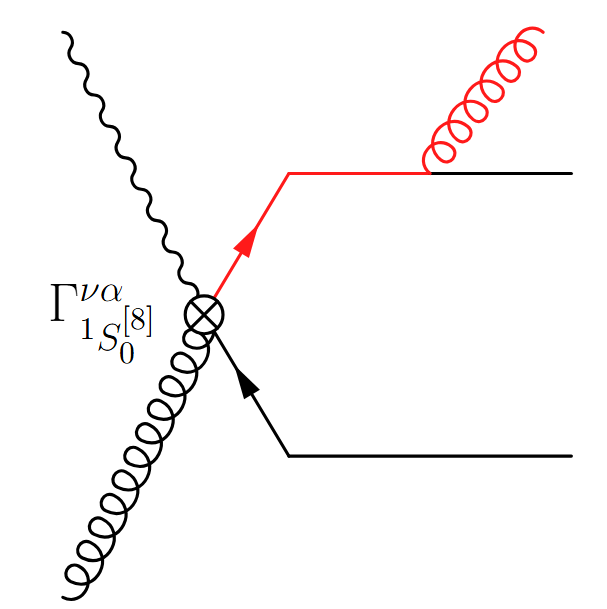} & {\huge $\longrightarrow$} &  
       \includegraphics[width = 1.1\linewidth]{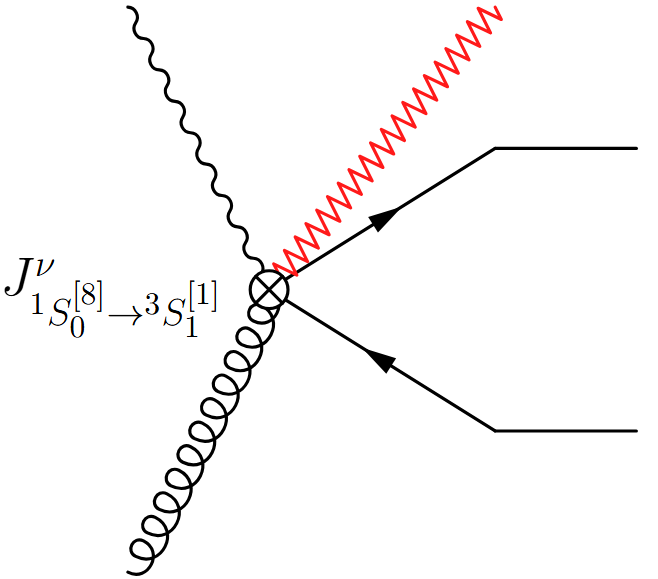}
       \end{tabular} 
    \caption{Feynman diagrams showing the sequential matching that occurs by integrating out the hard and soft quark propagators. The red gluon indicates the soft chromomagnetic field.}
    \label{fig: sequential matching}
\end{figure}

We find that the leading effective current in $\lambda$ and $\vv$ power-counting is given by
\begin{equation}
\begin{aligned}
    &J^\nu_{^1S_0^{[8]} \to ^3S_1^{[1]} }(x) = C_{^1S_0^{[8]}} (Q, M, \mu)\Gamma^{\nu \alpha}_{^1S_0^{[8]}} e^{i(Mv^+x^- + Mv^-x^+)}  \\
    &\times\bigg(\frac{g}{6M} \psi^\dagger_{\bf p_Q}\bigg[\frac{1}{v\cdot\Pc} {\boldsymbol{\sigma} \cdot {\bfcal B}}_s^a \bigg] \chi_{\bf p_{\bar{Q}}}\times{\cal S}_n^{ae} {\bf B}^e_{n,\perp, \alpha} \bigg)(x),
\label{eq: leading current}
\end{aligned}
\end{equation}
%
%{\color{red} Write down the operator for $^3S_1^{[8]}$ and argue that the matching coefficient is 0.}
%
%
where $C_{^1S_0^{[8]}}$ is the hard Wilson coefficient and $\Gamma^{\nu \alpha}_{^1S_0^{[8]}}$ is the same matching tensor obtained from calculating the hard partonic cross section, $\gamma^* + g \to c\bar{c}(^1S_0^{[8]})$ \cite{Echevarria:2024idp}. We follow the convention of ref. \cite{Echevarria:2024idp} so that the Wilson coefficient is a pure number and a function of $Q$ and $M$. The coefficients are related to the Hard function by
\begin{equation}
    H_{^1S_0^{[8]}} = \bigg|C_{^1S_0^{[8]}} (Q, M, \mu)\bigg|^2 = 1 ~+ ~{\cal O}(\alpha_s).
\end{equation}
where the $\alpha_s$ corrections to the coefficient come from virtual higher order QCD diagrams. Note these corrections can not change the structure of the matching tensor \cite{Echevarria:2024idp}. 
In eq. (\ref{eq: leading current}), the adjoint Wilson line is defined as ${\cal S}^{ba}T^a = S_n^\dagger T^bS_n$. 
The phase in the current arises because we have integrated out the heavy quark masses from the vNRQCD spinors and in the hadron frame. 

Formally, we derive this current by first considering the hard partonic cross section for $\gamma^* + g \to c\bar{c}+g_s$, where the final state gluon, $g_s,$ is soft. Then, we integrate out the hard scale, contracting the hard propagator to a point and subsequently matching onto the vertex from the partonic process $\gamma^* + g \to c\bar{c}(^1S_0^{[8]})$. Lastly, we expand the final quark propagator to subleading order in $\vv$, contracting it to a point and effectively integrating out the off-shell quark carrying soft momentum. The final current in eq. (\ref{eq: leading current}) is then written in terms of the initial collinear gluon, the soft chromomagnetic gluon, and the heavy quark fields. This sequential process is shown in figure  \ref{fig: sequential matching}. The diagram crossing the photon and collinear gluon should be considered too, as well as the diagrams containing a soft final state gluon on the antiquark propagator. Note the diagram with a soft gluon radiated off of the middle propagator between the initial collinear gluon and the photon will be power suppressed due to an additional hard propagator that is created. Therefore, we do not consider this contribution. 

We could also match onto another effective current using the operator from eq. (\ref{eq: 3S18 trans operator}), since it is the same order in $\vv$ and only down by one power of $g$
\begin{equation}
\begin{aligned}
    &J^\nu_{^3S_1^{[8]} \to ^3S_1^{[1]} }(x) = C_{^3S_1^{[8]}} (Q, M, \mu)\Gamma^{\nu \alpha}_{^3S_1^{[8]},\ell} e^{i(P^+_{c\bar{c}}x^- + P^-_{c\bar{c}}x^+)}  \\
    &\times\bigg( \frac{-g}{6M} \psi^\dagger_{\bf p_Q} \bigg[\frac{1}{v\cdot\Pc}  \bigg(2\epsilon^{ijk}\epsilon^{\ell k m} {\bfcal P }_i{\bfcal E}^a_j \sigma_m+gd^{bca} {\bfcal E}_{q}^{ b} \cdot {\bfcal E}_{q'}^{c} \sigma^\ell\bigg)  \bigg] \chi_{\bf p_{\bar{Q}}}\times{\cal S}_n^{ae} {\bf B}^e_{n,\perp, \alpha} \bigg)(x),
\label{eq: leading current 2}
\end{aligned}
\end{equation}
however, it's matching coefficient, $ C_{^3S_1^{[8]}} (Q, M, \mu)$, is zero at leading order in $\alpha_s(M_\psi)$.

From here, we match eq. (\ref{eq: leading current}) directly onto the the hadronic tensor in eq. (\ref{eq: hadronic tensor}). In SCET$_Q$, the Hilbert space factorizes into collinear and soft sectors because, after a BPS field redefinition, the effective Lagranian can be written in such a way that the soft and collinear sectors do not interact, so
\begin{equation}
    \ket{J/\psi, X} = \ket{X_n} \otimes \ket{J/\psi, X_s}.
\end{equation}
Therefore, at leading power, the hadronic tensor becomes
\begin{equation}
\begin{aligned}
    W^{\mu \nu} = &|C_{^1S_0^{[8]}}|^2 \Gamma^{\mu \alpha'}_{^1S_0^{[8]}}\Gamma^{\nu \alpha}_{^1S_0^{[8]}}\sumint_{X_n} \int \frac{d^4 b}{(2\pi)^4} e^{i q\cdot b} e^{-i P_\psi \cdot b} \bra{N} {\bf B}^{\dagger e'}_{n,\perp, \alpha'} (b) \ket{X_n}\bra{X_n} {\bf B}^e_{n,\perp, \alpha} (0) \ket{N}\\
    &\times \frac{1}{36M^2}\sumint_{X_s} \bra{0}\bigg( {\cal S}_n^{\dagger a'e'}  \chi^\dagger_{\bf p_{\bar{Q}}}\bigg[\frac{1}{v\cdot\Pc}   g{\boldsymbol{\sigma} \cdot {\bfcal B}}_s^{a'} \bigg] \psi_{\bf p_{Q}}\bigg)(b) \ket{J/\psi, X_s}\\
    &\times \bra{J/\psi, X_s}\bigg( {\cal S}_n^{ ae} \psi^\dagger_{\bf p_{Q}}\bigg[\frac{1}{v\cdot\Pc}   g{\boldsymbol{\sigma} \cdot {\bfcal B}}_s^a \bigg]  \psi_{\bf p_{Q}}\bigg) (0)\ket{0}.
\end{aligned}
\end{equation}
From here, we can simplify further by noting that, because $q \sim Q(1,1,\lambda)$, the spacial separation has to scale like $b\sim Q^{-1}(1,1,\lambda^{-1})$. Therefore, we can multipole expand the collinear $ {\bf B}^{\dagger e'}_{n,\perp, \alpha'}$ field, as well as the soft vNRQCD operators and Wilson lines. After the multipole expansion, the $b^+$ dependence of the collinear field is suppressed and the $b^+$ and $b^-$ dependence of the soft operator is suppressed. Now, we can evaluate the $b^+$ integral and get a delta function
\begin{equation}
    \delta(q^- - P_{J/\psi}^-) = \frac{\sqrt{2}}{Q}\delta(1-z).
\end{equation}
Define the partonic variable
\begin{equation}
    \xi = \frac{x_B}{Q^2}(Q^2-\qv_T^2 -\frac{1}{z}M^2)\,,
\end{equation}
so that $q^+-P_\psi^+ = -\xi P_N^+$. Completing the sum over collinear states, we are left with,
\begin{equation}
\begin{aligned}
    W^{\mu \nu} = &H^{\mu\alpha',\nu\alpha}_{^1S_0^{[8]}}\int\frac{d^2\bv_T}{(2\pi)^2} e^{-i \qv_T \cdot \bv_T} \int\frac{db^-}{2\pi} e^{-i \xi P^+_Nb^-} \bra{N} {\bf B}^{\dagger e'}_{n,\perp, \alpha'} (b^-,\bv_T) {\bf B}^e_{n,\perp, \alpha} (0) \ket{N}\\
    &\times \frac{1}{36M^2}\sumint_{X_s} \bra{0}\bigg( {\cal S}_n^{\dagger a'e'} \chi^\dagger_{\bf p_{\bar{Q}}}\bigg[\frac{1}{v\cdot\Pc}   g{\boldsymbol{\sigma} \cdot {\bfcal B}}_s^{a'} \bigg] \psi_{\bf p_{Q}}\bigg)(\bv_T) \ket{J/\psi, X_s}\\
    &\times \bra{J/\psi, X_s}\bigg( {\cal S}_n^{ a'e'}  \psi^\dagger_{\bf p_{Q}}\bigg[\frac{1}{v\cdot\Pc}   g{\boldsymbol{\sigma} \cdot {\bfcal B}}_s^a \bigg]  \psi_{\bf p_{Q}}\bigg) (0)\ket{0}\delta(1-z).
\end{aligned}
\end{equation}
where we have defined,
\begin{equation}
   H^{\mu\alpha',\nu\alpha}_{^1S_0^{[8]}} = \frac{\sqrt{2}}{Q}|C_{^1S_0^{[8]}}|^2 \Gamma^{\mu \alpha'}_{^1S_0^{[8]}} \Gamma^{\nu \alpha}_{^1S_0^{[8]}}
\end{equation}
The gluon TMDPDF is defined in terms of collinear matrix elements and a soft function \cite{Echevarria:2024idp}
\begin{equation}
    G_{g/N,\alpha \alpha'}(\xi, \bv_T) = \int\frac{db^-}{2\pi} e^{-i \xi P^+_Nb^-} \bra{N} {\bf B}^{\dagger e}_{n,\perp, \alpha'} (b^-,\bv_T) {\bf B}^e_{n,\perp, \alpha} (0) \ket{N}\sqrt{S(\bv_T)}\,,
\end{equation}
where the soft function is given by 
\begin{equation}
    S(\bv_T) = \frac{1}{N_c^2-1}{\rm Tr}\bra{0}\big[S_n^\dagger S_{\bar{n}}](\bv_T)[S_{\bar{n}}^\dagger S_{n}](0)]\ket{0}\,.
\end{equation}
The soft function is included to subtract rapidity divergences from the collinear matrix element. Using this, we can write the factorized hadronic tensor in terms of the gluon TMDPDF to arrive at the final expression:
\begin{equation}
\begin{aligned}
    W^{\mu \nu} = &H^{\mu\alpha',\nu\alpha}_{^1S_0^{[8]}}\int \frac{d^2 \bv_T}{(2\pi)^2} e^{-i \qv_T \cdot \bv_T} G_{g/N,\alpha \alpha'}(\xi,\bv_T) T_{^1S_0^{[8]} \to ^3S_1^{1}}(\bv_T)\delta(1-z)
\end{aligned}
\label{eq: factorized W}
\end{equation}
where $T_{^1S_0^{[8]} \to ^3S_1^{[1]}}(\bv_T)$ is a newly defined {\it TMD soft transition function} (TMDSTF)
\begin{equation}
\begin{aligned}
      &T_{^1S_0^{[8]} \to ^3S_1^{[1]}}(\bv_T)=\\
      &\frac{g^2  }{36M^2{(N_c^2-1)}\sqrt{S(\bv_T)}}\sumint_{X_s}{{\rm Tr}_c}\bra{0}\big[ {\cal S}_n^{\dagger ae}  \chi^\dagger_{\bf p_{\bar{Q}}}\bigg[\frac{1}{v\cdot\Pc}   g{\boldsymbol{\sigma} \cdot {\bfcal B}}_s^a \bigg] \psi_{\bf p_{Q}}\big]({\bf b}_T) \ket{J/\psi,X_s}\\
      &\times\bra{J/\psi, X_s}\big[ {\cal S}_n^{ a'e} \psi^\dagger_{\bf p_{Q}}\bigg[\frac{1}{v\cdot\Pc}   g{\boldsymbol{\sigma} \cdot {\bfcal B}}_s^a \bigg] \psi_{\bf p_{Q}} ](0)\ket{0}
\end{aligned}
\label{eq: TMDSTF}
\end{equation}
and is one of the key results in this paper. The TMDSTF is defined similarly to the TMDShFs defined in refs. \cite{Fleming:2019pzj, Echevarria:2024idp}, except it is strictly in a $^3S_1^{[1]}$ state and contains both the heavy quarks and the soft chromomagnetic gluon field. This object mediates the transition of charm quarks produced in a $^1S_0^{[8]}$ configuration during the hard process to a $^3S_1^{[1]}$ final state configuration via soft gluon radiation. 

Note, eq. (\ref{eq: TMDSTF}) is defined in the $v\cdot A_q = 0$ gauge, which we have chosen for clarity. However, gauge invariance can be restored by placing soft Wilson lines in in the $v$ direction in the appropriate places in the operator definition. This would produce operators like those we have derived in section \ref{sec: subleading vNRQCD}. 

\subsection{Matching $T_{^1S_0^{[8]} \to ^3S_1^{[1]}}(\bv_T)$ onto $\braket{{\cal O}^{J/\psi}\big(^3S_1^{[1]}\big)}$}

In the limit that the soft sector is treated perturbatively, i.e., $(m\vv)^2 \gg \Lambda_{QCD}^2$, one can evaluate the TMDSTF directly at leading order and match it onto the collinear $^3S_1^{[1]}$ LDME. This is effectively an operator product expansion, expanding the operator in powers of ($\bv_T \Lambda_{QCD}$). This evaluation is easiest to do in momentum space by Fourier transforming: 
\begin{equation}
    \tilde{T}_{^1S_0^{[8]} \to ^3S_1^{[1]}}(\kv_T) = \int \frac{d^2 \bv_T}{(2\pi)^2} e^{-i \kv_T \cdot \bv_T} T_{^1S_0^{[8]} \to ^3S_{1}^{[1]}}(\bv_T).
\end{equation}
At leading order in $\alpha_s$, the soft Wilson lines are set to unity and we only need the one gluon terms in the chromomagnetic field, so the evaluation of the matrix element becomes
\begin{equation}
\begin{aligned}
    \tilde{T}_{^1S_0^{[8]} \to ^3S_1^{[1]}}(\kv_T) = &\int \frac{d^2 \bv_T}{(2\pi)^2}e^{-i \kv_T \bv_T}\int \frac{d^3\qv}{(2\pi)^3}\frac{1}{2E_q} \frac{g^2\epsilon^{ijk}\epsilon^{\ell m n}}{9M^2} \\
    &\times\bra{0} \chi^\dagger_{\pv_{\bar{Q}}}\bigg[\frac{1}{v\cdot\Pc} {\bfcal P}_j {\bf A}^a_{q,i} \sigma_k \bigg] \psi_{\bf p_Q} (\bv_T)\ket{c\bar{c}(^3S_1^{[1]})g(q)}\\
    &\times\bra{c\bar{c}(^3S_1^{[1]})g(q)} \psi^\dagger_{\bf p_Q}\bigg[\frac{1}{v\cdot\Pc} {\bfcal P}_m {\bf A}^b_{q,\ell}  \sigma_n\bigg]\chi_{\pv_{\bar{Q}}}(0)\ket{0}.
\end{aligned}
\label{eq: leading order TMDSTF}
\end{equation}
It is easiest to evaluate this in the rest frame of the $J/\psi$, i.e., $P_\psi = (M,{\bf 0})$, $p_Q = P_\psi/2 + {\bf r}$, and $p_{\bar{Q}} = P_\psi/2 - {\bf r}$, where ${\bf r}$ is the relative momentum of the $c\bar{c}$. We get
\begin{equation}
\begin{aligned}
    \tilde{T}_{^1S_0^{[8]} \to ^3S_1^{[1]}}(\kv_T) =&\sum_\lambda \int \frac{d^3\qv}{(2\pi)^3}\frac{1}{2E_q}\frac{4\pi \alpha_s(\mu_s)}{9M^2} \epsilon^{ijk}\epsilon^{\ell m n} \frac{\epsilon^a_{i,\lambda}(q) \qv_j}{E_q}\frac{\epsilon^b_{\ell,\lambda}(q) \qv_m}{E_q}\delta^{(2)}(\qv_T-\kv_T)\\
    &\times\big[M^2 \eta^\dagger \sigma^k\xi \xi^\dagger\sigma^n \eta\big] \delta^{ab}\,,
\end{aligned}
\label{eq: pertur 1S0 op}
\end{equation}
where $\mu_s \sim m\vv$ is the soft scale. The spinor structure can be matched onto the $^3S_1^{[1]}$ LDME at leading order using
\begin{equation}
    M^2 \xi^\dagger \sigma^k\eta \eta^\dagger\sigma^n \xi = \frac{1}{3} \delta^{kn} \braket{{\cal O}^{J/\psi}\big(^3S_1^{[1]}\big)},
\label{eq: 3S11 matching}
\end{equation}
by spin and rotational symmetry \cite{Braaten:1996jt}. In the axial gauge, summing over gluon polarizations lets us replace the polarization vectors with
\begin{equation}
    \sum_\lambda \epsilon^a_{i,\lambda}(q) \epsilon^a_{\ell,\lambda}(q) = \delta_{i\ell} - \frac{\qv_i \qv_\ell}{(q^0)^2}. 
\label{eq: axial pol}
\end{equation}
Eq. (\ref{eq: pertur 1S0 op}) simplifies to 
\begin{equation}
    \tilde{T}_{^1S_0^{[8]} \to ^3S_1^{1}}(\kv_T) = \frac{2\alpha_s(\mu_s)}{27M^2}  \int \frac{d^{D-2}\qv_T dq_3 }{(2\pi)^{D-2}}\frac{1}{\sqrt{q_3^2 + (\qv_T\big|_{D-2})^2 }} \delta^{(2)}(\kv_T - \qv_T)\braket{{\cal O}^{J/\psi}\big(^3S_1^{[1]}\big)}
\end{equation}
and we evaluate the integral using dimension regularization. In $D = 4-2\epsilon$ dimensions, $\qv_T\big|_{D-2} = \qv_T + \qv_{\epsilon}$, where $\qv_T$ is a usual two dimensional vector and $\qv_{\epsilon}$ is the portion of the vector in the $\epsilon$ dimensions which is orthogonal to $\qv_T$. Using this prescription we can evaluate the two dimensional delta function and are left with  
\begin{equation}
    \tilde{T}_{^1S_0^{[8]} \to ^3S_1^{[1]}}(\kv_T,\mu_s, \zeta) = \frac{\alpha_s(\mu_s)}{27M^2} \mu_s^{2\epsilon}i_s^\epsilon\int \frac{d^{-2\epsilon}\qv_{\epsilon} dq_3 }{2\pi^2(2\pi)^{-2\epsilon}}\frac{1}{\sqrt{q_3^2+ \kv_T^2 + \qv^2_{\epsilon }}} \braket{{\cal O}^{J/\psi}\big(^3S_1^{[1]}\big)}.
\end{equation}
where $i_s = e^{\gamma_E}/4\pi$. Evaluating the $d^{-2\epsilon}\qv_{\epsilon}$ integral first, we find in the $\overline{MS}$ scheme
\begin{equation}
    \tilde{T}_{^1S_0^{[8]} \to ^3S_1^{[1]}}(\kv_T;\mu_s, \zeta) = \frac{1}{2}\frac{\alpha_s(\mu_s)}{27M^2\pi^2}\bigg[\frac{1}{\epsilon_{UV}}-\log\bigg(\frac{\kv_T^2}{\mu_s^2}\bigg)\bigg] \braket{{\cal O}^{J/\psi}\big(^3S_1^{[1]}\big)}.
\label{eq: pert TMDSTF}
\end{equation}
where the $1/{\epsilon}$ pole can be canceled by introducing the operator
\begin{equation}
    \sumint_{X_s}\bra{0}\chi^\dagger_{\pv_{\bar{Q}}}\sigma^i  \psi_{\bf p_Q} \Theta[\kv_T-{\bfcal{P}}_T]\ket{J/\psi, X_S}\bra{J/\psi,X_S} \psi^\dagger_{\bf p_Q}\sigma^i\chi_{\pv_{\bar{Q}}}\ket{0}.
\end{equation}

In $\bv_T$ space, the TMDSTF is determined by taking the Fourier transform of eq. (\ref{eq: pert TMDSTF}):
\begin{equation}
     T_{^1S_0^{[8]} \to ^3S_{1}^{[1]}}(\bv_T;\mu_s,\zeta) =  \frac{2\alpha_s(\mu_s)}{27M^2\pi}\frac{1}{\bv_T^2} \braket{{\cal O}^{J/\psi}\big(^3S_1^{[1]}\big)}\,.
\label{eq: TMDSTF bT}
\end{equation}

If it is indeed valid to treat the soft scale in quarkonium production as perturbative, then the expression in eq. (\ref{eq: TMDSTF bT}) is attractive from a phenomenological point of view. This is because it only contains one free parameter, $\Otsosing$, which is well constrained by experiment (see table \ref{tab: LDMEs}). Therefore, eq. (\ref{eq: TMDSTF bT}) enables a straightforward extraction of the gluon TMDPDFs in the proton because the factorized hadronic tensor in eq. (\ref{eq: factorized W}) contains no other free parameters. This is fortunate because, unlike the LDMEs, the TMDSTF as defined in eq. (\ref{eq: TMDSTF}) depends on the initial state soft radiation due to the soft Wilson lines in the $n$ direction. As a result it is process dependent and consequently, not  universal. However, the perturbative matching in this section shows that the TMDSTF can be written in terms of the universal LDME, $\Otsosing$, at leading order and we expect that it can be matched onto other subleading LDMEs at higher orders as well.

The perturbative result for the TMDSTF in eq. (\ref{eq: TMDSTF bT}) scales like $\alpha_s(m\vv)\vv^5 \sim \vv^6$ in the vNRQCD power-counting. This is because it matches onto the $^3S_1^{[1]}$ LDME which scales like $\vv^3$ and the factor of $1/\bv_T^2$ scale as $\sim \vv^2$. Our result should be compared with the scaling for the color-octet TMDShFs which have been used for quarkonium production at small transverse momentum in previous studies \cite{Boer:2023zit, Echevarria:2024idp}. When evaluated perturbatively, the TMDShFs are matched onto the sub-leading color-octet LDMEs, such as $\Oosz$ and $\Otpz$. We list the matching results from ref. \cite{Echevarria:2024idp} up to ${\cal O}(\alpha_s(\mu_s)\vv^7)$ for convenience
\begin{equation}
\begin{aligned}
    S_{^1S_0^{[8]} \to J/\psi}(\bv_T;\mu_s, \zeta) &= \Oosz[1+\frac{\alpha_s(\mu_s)C_A}{2\pi} \log\bigg(\frac{\bv_T^2 \mu_s^2 e^{2\gamma_E}}{4}\bigg)(1-\log(\zeta))]\\
    S_{^3P_0^{[8]} \to J/\psi}(\bv_T;\mu_s, \zeta) &= \Otpz[1+\frac{\alpha_s(\mu_s)C_A}{2\pi} \log\bigg(\frac{\bv_T^2 \mu_s^2 e^{2\gamma_E}}{4}\bigg)(1-\log(\zeta))]\,.
\end{aligned}
\label{eq: TMD ShF}
\end{equation}
Notice that because color-octet LDMEs scale like $\vv^7$ in the NRQCD power-counting, the TMDShFs are suppressed by a factor of $\vv$ with respect to the TMDSTF.
%and therefore will not give the dominant contribution to $J/\psi$ production in SIDIS, or other processes, in the TMD framework. 
The $\vv$ scaling for both the TMDSTF and the TMDShFs remains the same if the soft sector is non-perturbative. 

The key difference between color-octet TMDShFs and TMDSTFs is that TMDShFs give the probability for color-octet operators to transition to a $J/\psi$ state via insertions of the vNRQCD Lagrangian (for example via ultrasoft chromomagnetic or chromoelectric emissions). These insertions give additional $\vv$ suppression, leading to the $\vv^7$ scaling of the color-octet LDMEs. On the other hand, our TMDSTFs describe operators that have already transitioned to a $^3S_1^{[1]}$ state via soft gluon radiation, meaning there is no additional suppression from insertions of the vNRQCD Lagrangian. 

\begin{figure}[tp]
    \centering
    \includegraphics[width=.9\linewidth]{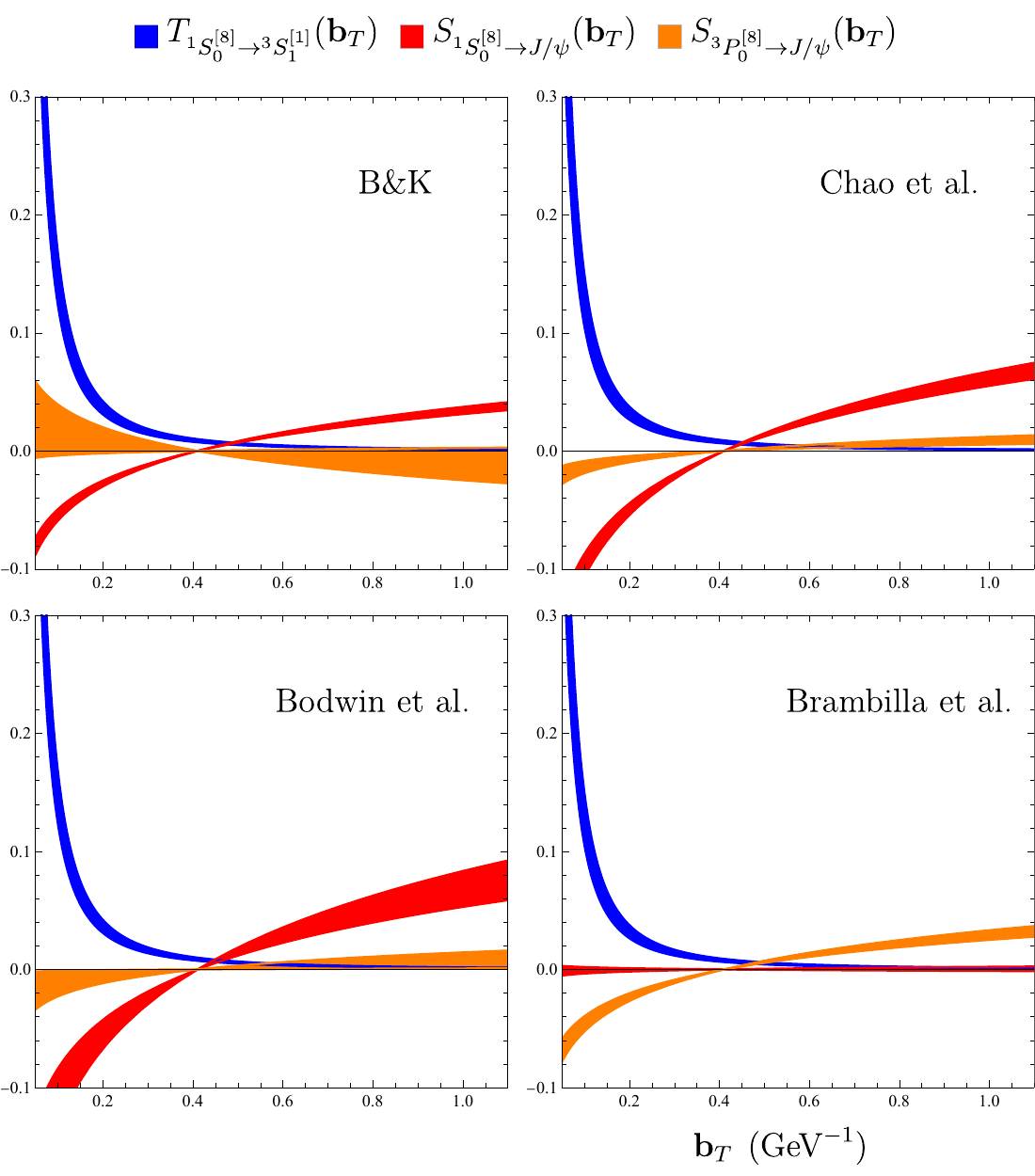}
    \caption{Comparison of fixed order calculations for the TMDSTF and the TMDShFs as a function of the transverse separation, $\bv_T$. The soft scale is arbitrarily chosen to be $\mu_s = 750$ MeV for all curves. Predicted results are shown using the different values of the LDMEs from table \ref{tab: LDMEs}.}
    \label{fig: TMDSTF vs ShF}
\end{figure}

Interestingly, while the TMDSTF is enhanced in pure $\vv$ power-counting, it appears to be sub-leading in the TMD power-counting. This is obvious when comparing the expressions in eqs. (\ref{eq: TMDSTF bT}) and (\ref{eq: TMD ShF}). In $\bv_T$ space, the TMDSTF goes like $1/\bv_T^2$ while the TMDShFs are a constant at leading order. For large $\bv_T$ (or equivalently small $\kv_T$), the TMDShFs should begin to dominate, even though they are subleading in the $\vv$ counting. This creates a nuanced picture of the competing effects between the TMDSTF and TMDShFs. To compare the relative importance of each operator, it is interesting to plot the perturbative results in eqs. (\ref{eq: TMDSTF bT}) and (\ref{eq: TMD ShF}) as a function on the transverse separation, $\bv_T$. 

In figure \ref{fig: TMDSTF vs ShF}, we compare the $\bv_T$ dependence of the $^1S_0^{[8]} \to {}^3S_1^{[1]}$ TMDSTF (shown in blue) against the color-octet TMDShFs from ref. \cite{Echevarria:2024idp} (shown in red and orange). As shown in table \ref{tab: LDMEs}, there are many different extractions for the color-octet LDMEs, each of which produces significantly different values for each parameter. Since the size of $\Oosz$ and $\Otpz$ directly impacts the overall magnitude of the TMDShFs in eq. (\ref{eq: TMD ShF}), we plot the TMDSTFs and TMDShFs using each set of LDMEs presented in table \ref{tab: LDMEs}. The panels are labeled according to the extraction used. In figure \ref{fig: TMDSTF vs ShF}, we plot the operators for a moderate range of $\bv_T$, between $0.1$ and $1.1$ GeV$^{-1}$ where we expect our perturbative results are most valid. It appears that for $\bv_T \lesssim 0.5$ GeV$^{-1}$, the TMDSTF becomes one of the most significant contributions, regardless of the LDME set used. In particular, for the B\&K, Chao et al., and Bodwin et al. LDME sets, the TMDSTF is either around the same size as or much greater than the $^3P_0^{[8]}$ TMDShF for all values of $\bv_T$ shown. For these same sets, the TMDSTF becomes larger in magnitude than the $^1S_0^{[8]}$ TMDShF once $\bv_T \lesssim 0.5$ GeV$^{-1}$. At larger values of $\bv_T$, the TMDSTF becomes quickly suppressed because, as discussed, it is subleading in the TMD power-counting. The same narrative is true for the Brambilla et al. LDME set, except with the roles of the $^1S_0^{[8]}$ and $^3P_0^{[8]}$ TMDShFs reversed.  For $\bv_T \gg 1$ GeV$^{-1}$ non-perturbative effects will begin to play a role because the operator product expansion no longer holds. Usually these non-perturbative effects provide additional suppression, so we'd expect the TMDShFs to begin to drop at long distances as well. Also, we note that for $\bv_T  \lesssim 0.4$ GeV$^{-1}$ logarithms of $\bv_T$ will begin to dominate and resummation using the TMD evolution framework will become necessary. This should suppress the divergences observed as $\bv_T \to 0$ in figure \ref{fig: TMDSTF vs ShF}, but we leave an analysis of the TMD evolution of the TMDSTF for future work. 

%Since the TMDShFs are suppressed by $\vv^3$, we argue that the TMD soft transition functions are the most important objects to match onto when studying quarkonium production at small transverse momentum. 

%

It is interesting to discuss our TMDSTF in the collinear limit where transverse momentum is integrated over. In principle there is a collinear soft transition function which also contributes to quarkonium production in the collinear limit. However, as pointed out in ref. \cite{Beneke:1997av} such a contribution in perturbation theory is pure cutoff and should not be taken into account to estimate the scaling of low-energy matrix elements. For example, consider the $^1S_0^{[8]} \to ^3S_1^{[1]}$ transition via a soft gluon at large transverse momentum scales. For such a process, we can write down the amplitude
\begin{equation}
\begin{aligned}
    \braket{{\cal O}({^1S_0^{[8]} \to ^3S_1^{[1]}})} = &\int \frac{d^3q}{(2\pi)^3}\frac{1}{\sqrt{2E_q}} \epsilon^{ijk}\epsilon^{\ell m n}\frac{g^2}{9M^2} \\
    &\times\bra{0} \chi^\dagger_{\pv_{\bar{Q}}}\bigg[\frac{1}{v\cdot\Pc} {\bfcal P}_j {\bf A}^a_{q,i}  \sigma_k  \bigg]\psi_{\bf p_Q}(0)\ket{c\bar{c}(^3S_1^{[1]})g(q)}\\
    &\times\bra{c\bar{c}(^3S_1^{[1]})g(q)} \psi^\dagger_{\bf p_Q}\bigg[\frac{1}{v\cdot\Pc} {\bfcal P}_m {\bf A}^a_{q,\ell}  \sigma_n\bigg]\chi_{\pv_{\bar{Q}}}(0)\ket{0}.
\end{aligned}
\label{eq: collinear transition function}
\end{equation}
Roughly, one can think of this as eq. (\ref{eq: leading order TMDSTF}) in the limit that $\kv_T^2 \gg (m\vv)^2$. This operator can be evaluated using the same steps as above and we find it reduces to
\begin{equation}
   \frac{\alpha_s(m\vv)}{27M^2} \int \frac{d^3\qv}{2\pi^2}\frac{1}{|\qv|} \braket{{\cal O}^{J/\psi}\big(^3S_1^{[1]}\big)}.
\label{eq: ui LO 1S0}
\end{equation}
This is a scaleless integral that now diverges quadratically since there's no delta function constraining the $\qv_T$ integral. Therefore, it vanishes identically in dimensional regularization. This result could have been expected, as the operator in eq. (\ref{eq: collinear transition function}) is what one would get from an insertion of the magnetic dipole operator on a $^1S_0^{[8]}$ $c\bar{c}$ pair when the gluon is taken to be soft. 

Before concluding, we would like to comment on the similarities between eq. (\ref{eq: collinear transition function}) and the calculation of the $^1S_0^{[8]}$ LDME using pNRQCD in refs. \cite{Brambilla:2022ayc,Brambilla:2022rjd}. In both this work and the pNRQCD approach, charm quarks in a $^1S_0^{[8]}$ configuration are allowed to transition to a $^3S_1^{[1]}$ state via the radiation of a chromomagnetic soft gluon, enabling the color-octet LDME to be matched directly onto $\Otsosing$ \footnote[3]{In the pNRQCD approach, $\Oosz$ is actually matched onto the wave function at the origin, but this is related to $\Otsosing$ at leading order in $\vv$ \cite{Bodwin:1994jh, Braaten:1996jt}.}. In the pNRQCD approach the matching coefficient is written in terms of a chromomagnetic field correlator and the authors expect their expression to be valid to all orders, whereas in our analysis, we have only considered matching onto the $\Otsosing$ at leading order in perturbation theory (given in eq. (\ref{eq: ui LO 1S0})). However, we can take similar steps to what was done in their analysis and ``integrate out" the heavy quark fields by approximating $\ket{J/\psi} \approx \ket{c\bar{c}}$ at leading order, which allows us to contract the heavy quark operators with the Fock state, yielding a factor of $M^2\eta^\dagger \sigma^i \xi \xi^\dagger \sigma^i \eta$. This can then be matched onto $\Otsosing$, allowing us to write eq. (\ref{eq: collinear transition function}) in a form that mimics their approach 
\begin{equation}
\begin{aligned}
    \braket{{\cal O}({^1S_0^{[8]} \to ^3S_1^{[1]}})} = &\frac{1}{3N_c^2M^2} \braket{{\bf O}_{\bfcal B}}\Otsosing.
\end{aligned}
\label{eq: 1S08 collinear matched}
\end{equation}
where 
\begin{equation}
    \braket{{\bf O}_{\bfcal B}} = \bra{0} \bigg(\frac{1}{v\cdot\Pc} g {\bfcal B}_s\bigg)^2 \ket{0}
\label{eq: Bs}
\end{equation}
is a magnetic field correlator that corresponds to their $c_F^2{\cal B}_{00}$ parameter since $c_F = 1$ at leading order in $\alpha_s$. At higher orders, eq. (\ref{eq: 1S08 collinear matched}) should contain a factor of $c_F^2$ as well. Since we are working in the $v\cdot A=0$ gauge, the soft Wilson lines in the $v$ direction are absent in our correlator. They can be restored to insure gauge invariance. The matching coefficient of $1/(3N_c^2 M^2)$ in eq. (\ref{eq: 1S08 collinear matched}) agrees with the matching coefficient of eq. (3.48c) in ref. \cite{Brambilla:2022ayc} since $\Otsosing = 3N_c/(2\pi) |R^{(0)}_{J/\psi}(0)|^2$, where $R^{(0)}_{J/\psi}(0)$ is the radial wavefunction at the origin. Additionally, when calculating the parameter ${\cal B}_{00}$ in perturbation theory, the authors in ref. \cite{Brambilla:2022ayc} noted that the correlator diverges quadratically and vanishes identically in dimensional regularization, which is also  what we found in eq. (\ref{eq: ui LO 1S0}). 

In order to get a nonzero result for ${\cal B}_{00}$ and calculate the scale dependence of this parameter, the authors in ref. \cite{Brambilla:2022ayc} consider the interaction of a ``non-perturbative" gluon with the chromomagnetic operator, which is a model for what may happen if the soft scale is actually non-perturbative. It is certainly interesting to consider the effects of a non-perturbative soft scale and it is not an unreasonable assumption given that empirically $m\vv \sim 750$ MeV for charmonium. In the scenario of a non-perturbative soft scale, operators like eq. (\ref{eq: collinear transition function}) do not necessarily vanish because non-perturbative scales arise dynamically at all orders, preventing soft gluon emissions from producing scaleless integrals. In this case, operators like eq. (\ref{eq: collinear transition function}) cannot be evaluated analytically and must be treated as a non-perturbative parameters to be extracted from experiment. This is in essence the prescription taken by refs. \cite{Brambilla:2022rjd, Brambilla:2022ayc}. In these analyses, the chromomagnetic correlator is treated as a universal parameter to be determined by fitting to experiment (so are the chromoelectric correlators $\varepsilon_{00}$ and $\varepsilon_{10,10}$ which appear in the matching of the $\Otsooct$ and $\Otpz$ onto the $\Otsosing$). It is not clear whether the soft scale is perturbative or not, but the effects of a non-perturbative soft scale in vNRQCD should be explored further in future work.

Finally, we would like to point out that our analysis of eq. (\ref{eq: 1S08 collinear matched}) is purely a leading order perturbative statement and not a true factorization theorem. This is because the approximation $\ket{J/\psi} \approx \ket{c\bar{c}}$ is not valid at higher orders since the $J/\psi$ Fock state in vNRQCD must include soft and ultrasoft gluons as well. Since soft gluons can not be decoupled from the heavy quarks in the vNRQCD Lagrangian (for example, by a BPS field redefinition \cite{Bauer:2001yt}), soft gluon fields can not be factorized from quarkonium matrix elements. It becomes obvious that eq. (\ref{eq: 1S08 collinear matched}) can not be completely correct in vNRQCD when one considers that $\Otsosing$ contains soft states
\begin{equation}
\begin{aligned}
    \Otsosing& =\sum_{X_s} \bra{0} \chi^\dagger \sigma^i \psi \ket{J/\psi + X_s}\bra{J/\psi +X_s} \psi \sigma^i \chi\ket{0} \,.
\end{aligned}
\end{equation}
Thus the soft chromomagnetic operator and implicit sum over soft states in eq. (\ref{eq: Bs}) should not have been separated from $\Otsosing$, which also contains soft states. This is markedly different from the pNRQCD approach in ref. \cite{Brambilla:2022ayc}. In the pNRQCD Lagrangian, the soft scale is completely integrated out from the theory so there are no interactions between heavy quark fields and soft gluons. It seems essential to fully understand the form this factorization takes in vNRQCD because the factorized LDMEs in refs. \cite{Brambilla:2022ayc,Brambilla:2022rjd} enable powerful relationships between the LDMEs of different S-wave quarkonium states and reduce the number of universal parameters in the theory from 12 to 3. We leave a detailed analysis of the factorization of the LDMEs in vNRQCD to future work.

%%%%%%%%%%%%%%%%%%%%%%%%%%%%%%%%%%%%%%%%%%%%%
\section{Conclusion}
\label{sec: conclusion}
%%%%%%%%%%%%%%%%%%%%%%%%%%%%%%%%%%%%%%%%%%%%%%%

In this paper we have derived new operators that will contribute to the production of $J/\psi$ mesons in the transverse momentum dependent framework. These operators arise from the emission of soft gluons from a $c\bar{c}$ pair that has been produced in some quantum number configuration via a hard process, and enable the transition of color-octet charm quark pairs to charm quarks in a $^3S_1^{[1]}$ configuration. We derived a new factorization theorem for $J/\psi$ production in SIDIS and showed that, at leading power in vNRQCD and SCET power-counting, the hadronic tensor can be written as a convolution of the gluon TMDPDF in the proton with new objects, which we dub TMD soft transition functions. We evaluate the TMDSTF at leading order and demonstrate that it is larger by a factor of $1/\vv$ than the color-octet TMDShFs, which have been introduced in the literature to study quarkonium production in the TMD framework so far. We note that, while the TMDSTF is enhanced in the $\vv$ power-counting, it is actually sub-leading in the TMD power-counting, meaning that there are different regimes where each operator may dominate. Lastly, we demonstrate that, when the soft scale is perturbative, the vacuum production matrix elements of these operators vanish in the collinear framework. We then commented on the similarities between our approach and that of ref. \cite{Brambilla:2022ayc, Brambilla:2022rjd}, which uses pNRQCD to match the color-octet LDMES onto the wave functions at the origin. We show that we can reproduce their expression for the matching of the $\Oosz$ which is a check on our calculation. We then discuss the validity of this matching and the implications of the the soft scale turning non-perturbative for quarkonium production physics. 

 This paper demonstrates the importance of the soft scale, $m\vv$, in studies of quarkonium production at small transverse momentum scales and there are many ways to extend this work in the future. For example, a number of new factorization theorems can be derived for the production of the $J/\psi$ meson in various experiments by matching onto the operators presented in this paper. In general, it will be worthwhile to categorize a variety of different TMD soft transition functions that arise from the transitions of charm quarks in color-octet configurations to a $^3S_1^{[1]}$ state. In order to explore the validity of a perturbative evaluation of the TMDSTF, one could possibly compare with $J/\psi$ production from $e^+e^-$ collisions at small transverse momentum, in which the cross section is only sensitive to the hadronization dynamics of the meson. Equivalent studies should be carried out for bottomonium as well, particularly because the soft scale for bottomonium is larger than the soft scale for charmonium due to the larger mass of the bottom quark, $m_b \vv \sim 1.5 $ GeV. 

Also, as already alluded to, the production of quarkonium at small transverse momentum promises to be an excellent probe of the gluon TMDPDFs in the proton, which are poorly constrained quantities. In this analysis we have laid the necessary theoretical foundations for this endeavor by identifying one of the most essential production operators in the TMD framework. From here, it is necessary to compare theoretical calculations of quarkonium production in SIDIS with experimental measurements in order to extract the gluon TMDs. To improve the accuracy of the calculations, logarithms of $\bv_T$ that  appear in the TMDSTF should be resummed to all orders using the TMD evolution framework. Naturally, our calculations can also be improved to higher accuracy in $\alpha_s$, as well as in the $\vv$ and $\lambda$ expansions. Finally, it will be necessary to continue to study the effects of the soft scale in quarkonium production physics, not just in the TMD framework, but in the collinear factorization framework as well using vNRQCD. Particularly if the soft scale is non-perturbative.

\acknowledgments

This research is dedicated to the memory of Tom Mehen, who provided insightful feedback during the completion of this work. We also thank Iain Stewart for helpful suggestions. M.~C. and R.H. are supported by the U.S. Department of Energy, Office of Science, Office of Nuclear Physics under grant Contract No.~DE-FG02-05ER41367. M.C. is also supported by the National Science Foundation Graduate Research Fellowship under Grant No.~DGE 2139754. R.H. was also supported by the U.S. Department of Energy through the Topical Collaboration in Nuclear Theory on Heavy-Flavor Theory (HEFTY) for QCD Matter under award no. DESC0023547. S.~F. is supported by the U.S. Department of Energy, Office of Science, Office of Nuclear Physics,  under award No.~DE-FG02-04ER41338. 

\appendix

\section{Single and double soft gluon emissions}
\begin{figure}[htbp]
\centering
\begin{minipage}{0.25\textwidth}
\centering
\subfloat[]{\includegraphics[width=.8\textwidth]{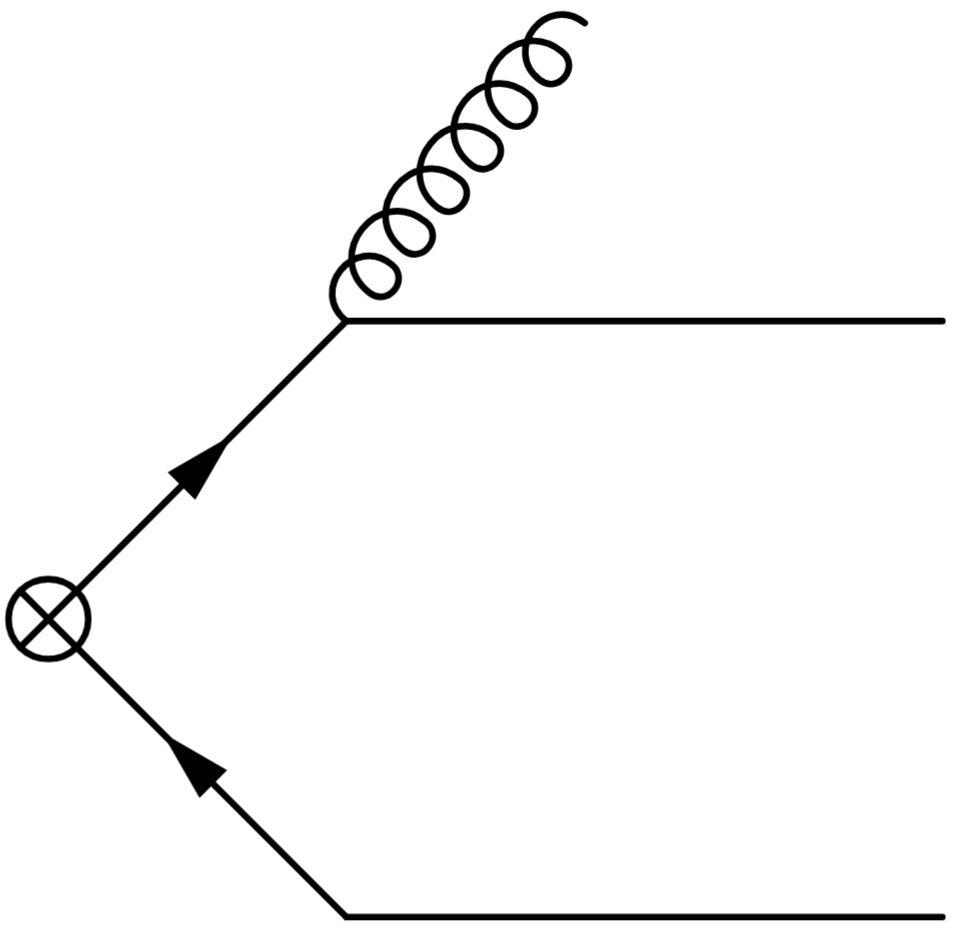}}
\end{minipage}%
\begin{minipage}{0.25\textwidth}
\centering
\subfloat[]{\includegraphics[width=.8\textwidth]{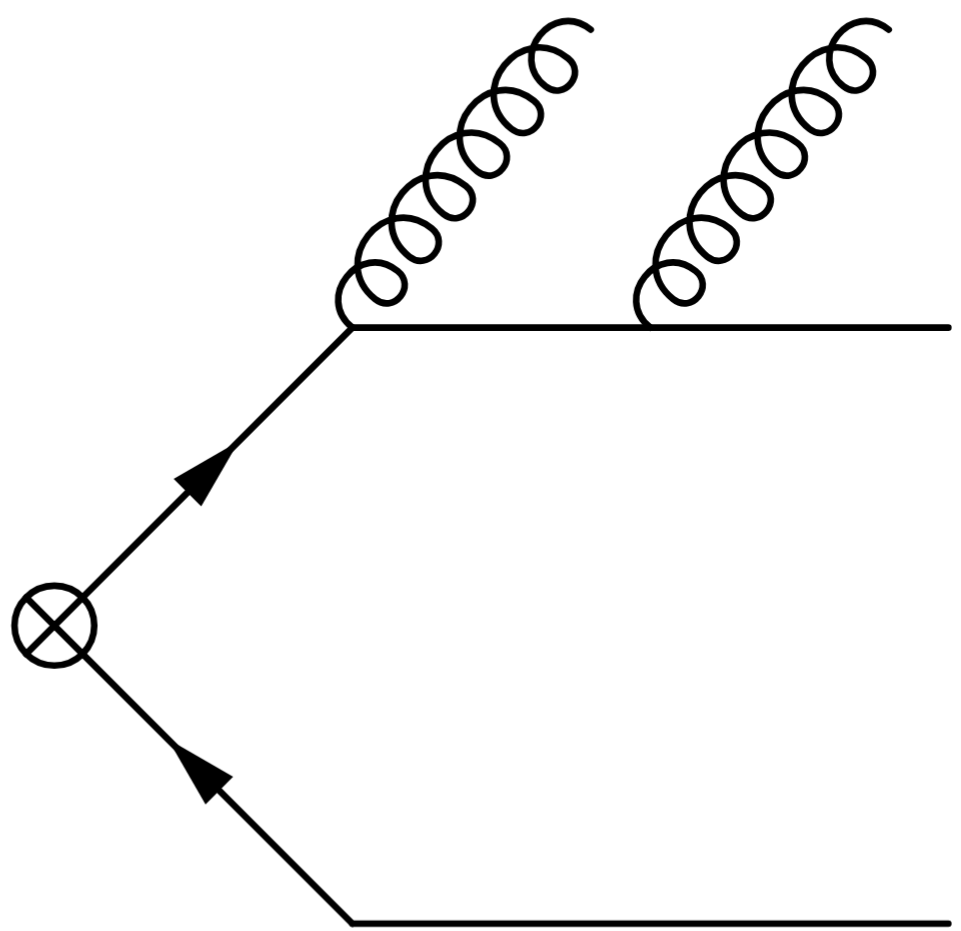}}
\end{minipage}
\begin{minipage}{0.25\textwidth}
\centering
\subfloat[]{\includegraphics[width=.8\textwidth]{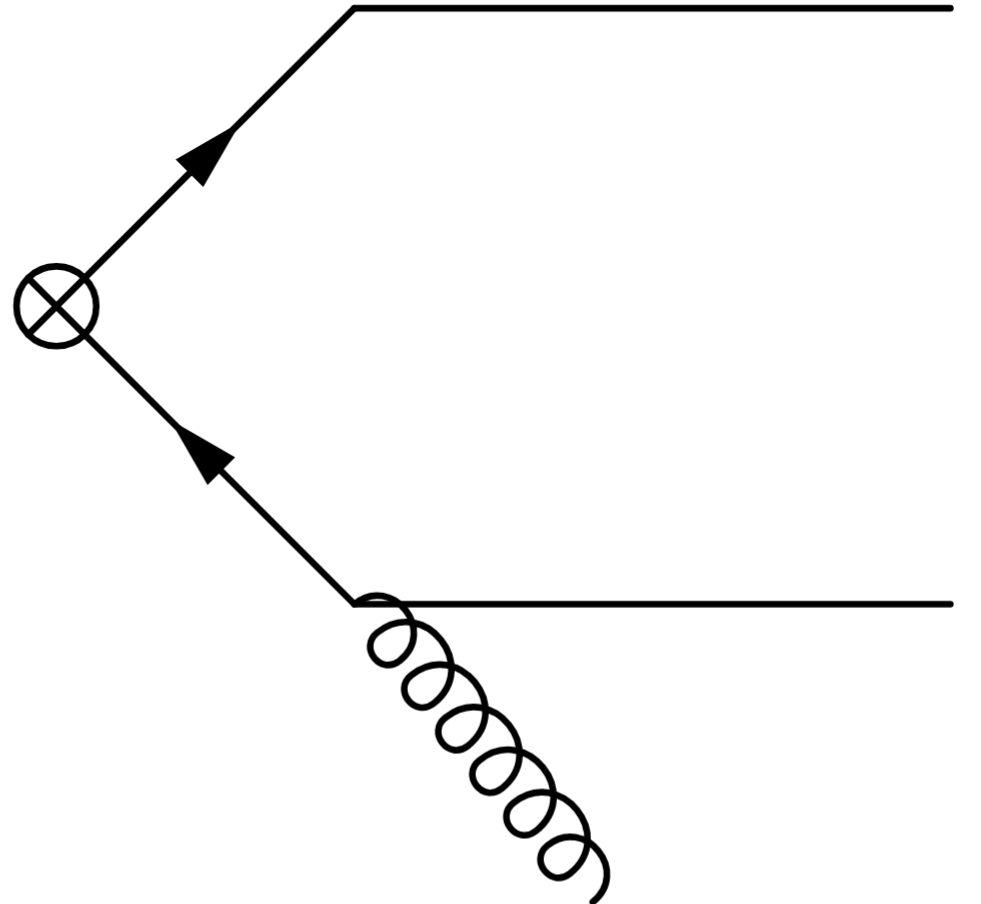}}
\end{minipage}%
\begin{minipage}{0.25\textwidth}
\centering
\subfloat[]{\includegraphics[width=.8\textwidth]{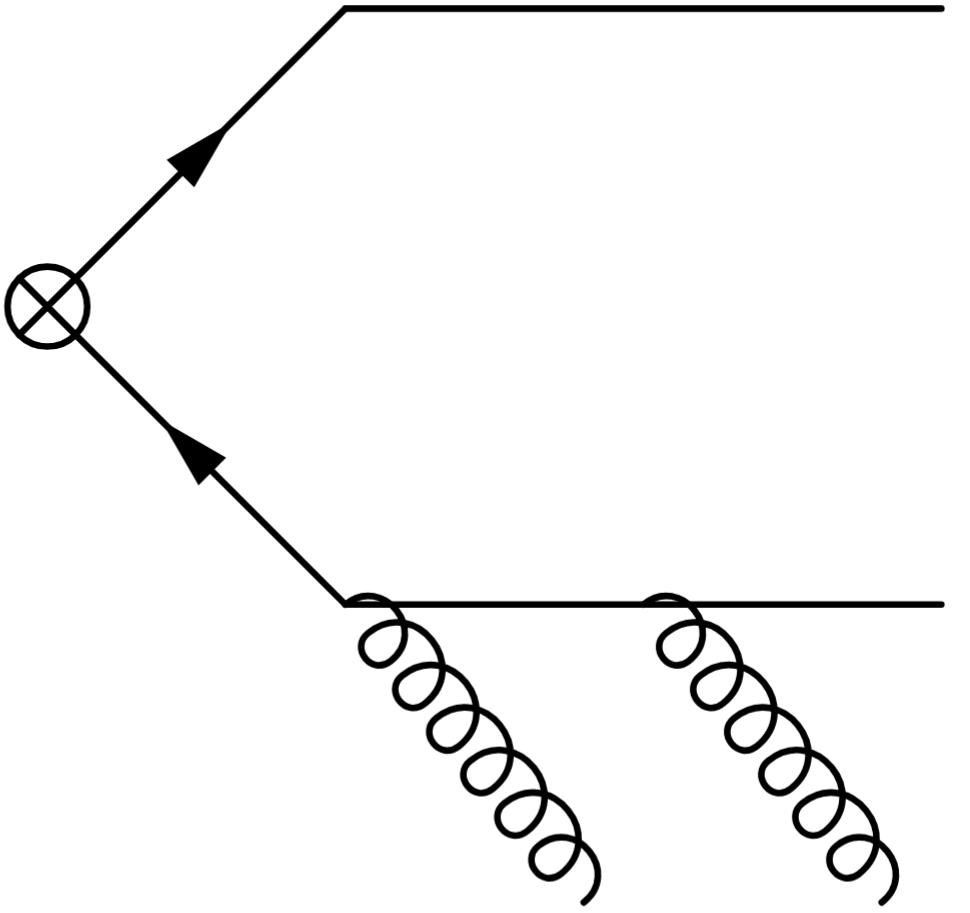}}
\end{minipage}%
\caption{Single and double soft gluon emissions from a $c\bar{c}$ pair.\label{fig: 1 and 2 gluons}}
\end{figure}

In order to verify our results, we have checked that our operators from section \ref{sec: subleading} reproduce the tree level results for single and double gluon emissions from quark and antiquark lines after expanding out the soft Wilson lines to ${\cal O}(\alpha_s(m\vv)^2)$, shown in figure \ref{fig: 1 and 2 gluons}.

For completeness, we list the tree level results for these diagrams here. If you first consider the emission of a single soft gluon from a quark propagator, as well as an emission from an antiquark propagator, then we find, after expanding the propagator to ${\cal O}(\vv)$, the amplitude is given by 
\begin{equation}
\begin{aligned}
    \mathcal{A}_{(a)+(c)} -\frac{g}{2mp_s^0} \bar{u} \bigg[ & ( 2mA^{a0} + A^a \cdot p_s ) [t^a, \Gamma] -(A^{a0}\mathbf{p}_s^i - p_s^0 \mathbf{A}^{ai})\{ t^a \boldsymbol{\gamma}^i, \Gamma\} 
    \\
    & - i(\mathbf{A}^a \times \mathbf{p}_s)^i \{ t^a \boldsymbol{\sigma}^i , \Gamma \} \bigg] v \, .
\end{aligned}
\end{equation}
Likewise, for two gluon emissions off of the quark or antiquark propagators, we have
\begin{equation}
\begin{aligned}
    \mathcal{A}_{(b)+(d)} &\frac{g^2}{4mp_2^0} \frac{1}{\frac{1}{m}p_1\cdot p_2 + p_s^0}  \bar{u}(p_Q)\\
    &\times \bigg[  (2m A_2^{a0}A_1^{b0} + 2 A_2^{a0}A_1^b\cdot p_2 + A_1^{b0}A_2^a\cdot p_2 )
    \bigg( \frac{2}{3}\delta^{ab}\Gamma + d^{abc}\{t^c,\Gamma\} + i f^{abc}[t^c,\Gamma]\bigg) 
    \\& + ( A_2^{a0} p_s^0 \mathbf{A}_1^{bi} - A_1^{b0}A_2^{a0} \mathbf{p}_s^i )  \bigg( \frac{1}{3}\delta^{ab}[\boldsymbol{\gamma}^i,\Gamma] + d^{abc}[t^c\boldsymbol{\gamma}^i,\Gamma] + i f^{abc}\{t^c\boldsymbol{\gamma}^i,\Gamma\}\bigg) \\
    & -i\bigg( A_1^{b0}(\mathbf{A}_2^a\times \mathbf{p}_2)^i + A_2^{a0}(\mathbf{A}_1^b \times \mathbf{p}_1)^i \bigg) \bigg( \frac{1}{3}\delta^{ab}[\boldsymbol{\sigma}^i,\Gamma] + d^{abc}[t^c\boldsymbol{\sigma}^i,\Gamma] + i f^{abc}\{t^c\boldsymbol{\sigma}^i,\Gamma\}\bigg) \\
    & - p_2^0 (\mathbf{A}_2^a \times \mathbf{A}_1^b)^i f^{abc}\{ t^c \boldsymbol{\sigma}^i,\Gamma \}   -p_2^0 A_2^a \cdot A_1^b \bigg( \frac{2}{3}\delta^{ab}\Gamma + d^{abc}\{t^c, \Gamma\} \bigg) \bigg] v(p_{\bar{Q}}) + (1 \leftrightarrow 2) \,.
\end{aligned}
\end{equation}
It can be verified that, by expanding the soft gluon Wilson lines in eqs. (\ref{eq: vNRQCD SL charm}) and (\ref{eq: vNRQCD SL anticharm}) to ${\cal O}(g)$ and ${\cal O}(g^2)$, you can reproduce these explicit one and two gluon emission results, respectively. 

\section{Expanding the soft gluon radiation from an anticharm quark}
%%%%%%%%%%%%%%%%
\label{app: SL antiquark details}
%%%%%%%%%%%%%%%%
In section \ref{sec: subleading charm} we deduced the dominant contributions that come from expanding QCD amplitudes that come from an arbitrary number emissions off of quark and antiquark propagators. In this appendix we derive the terms that come from expanding antiquark propagators to ${\cal O}(\vv^0)$. We again begin with the general expression that comes from radiating off $n'$ soft gluons from the charm propagator and $n$ soft gluons from the anticharm propagator
            \begin{equation}
            \begin{aligned}
              \mathcal{A}=  &(-g)^{n+n'}u^\dagger(p_Q) \slashed{A}_{1'} \frac{\slashed{p}_Q + \slashed{p}_{(1')} + m_c}{(p_Q + p_{(1')})^2 -m_c^2 +i\epsilon}\cdots\slashed{A}_{n'}\frac{\slashed{p}_Q + \slashed{p}_{(n')} + m_c}{(p_Q + p_{(n')}^2 -m_c^2 +i\epsilon}\Gamma\\
                &\times\frac{-\slashed{p}_{\bar{Q}} - \slashed{p}_{(n)} + m_c}{(p_{\bar{Q}} + p_{(n)})^2 -m_c^2 +i\epsilon} \slashed{A}_{n} \cdots \frac{-\slashed{p}_{\bar{Q}} - \slashed{p}_{(1)} + m_c}{(p_{\bar{Q}} + p_{(1)})^2 -m_c^2 +i\epsilon} \slashed{A}_{1} v(p_{\bar{Q}}),
            \label{eq: anticharm QCD amp}
            \end{aligned}
            \end{equation}
As explained above, we can either (a) expand the $\ell$-th antiquark propagator to ${\cal O}(\vv^0)$ and expand the remaining propagators to ${\cal O}(\vv^{-1})$, or we can (b) expand the $n$-th antiquark propagator closest to the vertex to ${\cal O}(\vv^0)$ and the rest of the propagators to ${\cal O}(\vv^{-1})$. These two scenarios are illustrated in figure~\ref{fig: subleading antiquark}, where the components that are highlighted red indicate the piece that will be expanded to ${\cal O}(\vv^{0})$.

In this appendix, we only list the steps for case (a), where the $\ell'$-th propagator is expanded to ${\cal O}(v^0)$ (for $\ell \neq n$) and the remaining propagators are expanded to leading order. This gives
        \begin{equation}
        \begin{aligned}
           \mathcal{A}_{4(a)} &(-1)^n(g)^{n+n'}u^{(0)\dagger}(p_Q) \slashed{A}_{1'} \frac{1+\slashed{v}}{2p_{(1')}^0}\cdots\slashed{A}_{n'} \frac{1+\slashed{v}}{2p_{(n')}^0}\Gamma 
            \\
            &\times \frac{-(1-\slashed{v})}{2p_{(n)}^0} \slashed{A}_{n} \cdots\slashed{A}_{(\ell+1)}\bigg(\frac{\slashed{p}_{(\ell)}}{2m_c p^0_{(\ell)}}+ \frac{1-\slashed{v}}{2 p^0_{(\ell)}}\frac{p_{(\ell)}^2}{2m_c  p^0_{(\ell)}}\bigg)\slashed{A}_{\ell} \cdots \frac{-(1-\slashed{v})}{2p_{(1)}^0} \slashed{A}_{1} v(p_{\bar{Q}}).
        \end{aligned}
        \label{eq: expanded SL antiquark}
        \end{equation}
From here, simplify by using the same tricks, i.e., use $\slashed{A}_i (1+\slashed{v}) = (1-\slashed{v})\slashed{A}_i + 2A^0_i$ and $u^{(0)\dagger} (1-\slashed{v}) = 0$ on the quark leg, and $ -(1-\slashed{v}) \slashed{A}_i = -\slashed{A}_i(1+\slashed{v}) + 2A^0_i$ and $(1+\slashed{v})v^{(0)}  = 0$ on the antiquark leg. Also use
\begin{equation}
\begin{aligned}
    -(1-\slashed{v})\slashed{A}_{\ell+1} &~\slashed{p}_{(\ell)} \slashed{A}_{\ell}  =2\bigg(A_{\ell+1}^0 A_{\ell}^{(0)}p_{(\ell)}^0-{\boldsymbol \gamma} \cdot {\bf A}_{(\ell+1)}{\boldsymbol \gamma} \cdot {\bf A}_{\ell} p_{(\ell)}^0 \\
             &+ {\boldsymbol \gamma}\cdot {\bf A}_{\ell+1} {\boldsymbol \gamma} \cdot {\bf p}_{(\ell)} A^0_{\ell} + A^0_{(\ell+1)}  {\boldsymbol \gamma} \cdot {\bf p}_{(\ell)} {\boldsymbol \gamma}\cdot {\bf A}_{\ell}\bigg) - \slashed{A}_{\ell+1}\slashed{p}_{(\ell)}\slashed{A}_{\ell}(1+\slashed{v})
\label{eq: ApA simplification}
\end{aligned}
\end{equation}
to write eq.~(\ref{eq: expanded SL quark}) as
        \begin{equation}
        \begin{aligned}
            \mathcal{A}_{4(a)}\approx &\frac{(-g)^{n'}(g)^{n}}{2m} u^{(0) \dagger}(p_Q) \prod_{j' = 1'}^{n'}  \frac{A^0_{j'}}{p^0_{(j')}}\Gamma\prod_{j = \ell+2}^{n}  \frac{A^0_{n+\ell+2-j}}{p^0_{(n+\ell+2-j)}} \frac{1}{p_{(\ell+1)}^0} \bigg[A_{\ell+1}^0 A_{\ell}^{(0)}p_{(\ell)}^0\\
            &-{\boldsymbol \gamma} \cdot {\bf A}_{\ell+1}{\boldsymbol \gamma} \cdot {\bf A}_{\ell} p_{(\ell)}^0 -A_{\ell+1}^0 A_{\ell}^0 p_{(\ell)}^2+  {\boldsymbol \gamma}\cdot {\bf A}_{(\ell+1)'} {\boldsymbol \gamma} \cdot {\bf p}_{(\ell')} A^0_{\ell'} \\
            &+ A^0_{\ell+1}  {\boldsymbol \gamma} \cdot {\bf p}_{(\ell)} {\boldsymbol \gamma}\cdot {\bf A}_{\ell}\bigg] \frac{1}{p^0_{(\ell)}} \prod_{k=1}^{\ell-1} \frac{A^0_{\ell-j}}{p^0_{(\ell-j)}}v^{(0)}(p_{\bar{Q}}).
        \label{eq: Simp SL antiquark}
        \end{aligned}
        \end{equation}
Now make use of
\begin{equation}
    \prod_{k = \ell + 2}^n \frac{A_{n+\ell+2-k}^0}{p^0_{(n + \ell + 2 -k)}} = \prod_{k = \ell + 2}^n  \frac{A_{n+\ell+2-k}^0}{p^0_{(k)}}
\end{equation}
and 
\begin{equation}
    \prod_{j = 1}^{\ell-1} \frac{A_{\ell-j}^0}{p^0_{(\ell - j)}} =  \prod_{j = 1}^{\ell-1} \frac{A_{\ell-j}^0}{p^0_{(j)}}
\end{equation}
to rearrange eq. (\ref{eq: Simp SL antiquark})
\begin{equation}
\begin{aligned}
\mathcal{A}_{4(a)} \approx \frac{(-g)^{n'}(g)^n}{2 m}& u^{(0)\dagger}(p_Q) \prod_{j' = 1'}^{n'}  \frac{A^0_{j'}}{p^0_{(j')}}\Gamma \Bigg[\bigg\{ \frac{1}{ p^0_{(\ell)} }
\left( \prod_{k=\ell+1}^{n} \frac{A^{0}_{n+\ell+1-k}}{p^0_{(k)}} \right)
((p^0_{(\ell)})^2 -p_{(\ell)}^2)
\\
&+\frac{1}{p_{(\ell+1)}^0 }
\left( \prod_{k=\ell+2}^{n} \frac{A^{0}_{n+\ell+2-k}}{p_{(n+\ell+2-k)}^0} \right)
 \gamma\cdot{\bf A}_{\ell+1} \gamma\cdot{\bf p}_{(\ell)}  \bigg\}\left( \prod_{j=1}^{\ell} \frac{A^{0}_{\ell+1-j}}{ p^0_{(j)}} \right) 
\\
&+ \bigg\{ \frac{1}{p^0_{(\ell)}} 
\left( \prod_{k=\ell +1 }^{n} \frac{A^{0}_{n+\ell+1-k}}{p^0_{(k)}} \right) 
 \gamma\cdot{\bf p}_{(\ell+1)} \gamma\cdot {\bf A}_\ell\\ 
&+ \frac{1}{p_{(\ell+1)}^0 }
\left( \prod_{k=\ell+2}^{n} \frac{A^{0}_{n+\ell+2-k}}{p_{(n+\ell+2-k)}^0} \right) \gamma\cdot{\bf A}_{\ell+1} \gamma\cdot {\bf A}_\ell
\bigg\}\prod_{j=1}^{\ell-1} \frac{A^{\circ}_{\ell-j}}{p_{(\ell-j)}^0 }
\Bigg] v^{(0)}(p_{\bar{Q}}).
\end{aligned}
\label{eq: rearranged SL antiquark}
\end{equation}

For the antiquark algebra, the magical formula takes the form of \cite{Fleming:2019pzj}
%
%Magic formula
\begin{equation}
\begin{aligned}
    \frac{1}{p_{(i)}^0} \prod_{k=i}^n (-g) \frac{A^0_{n+i+1-k}}{p^0_{(k)}} =& \sum_{\rho  = i}^n \frac{1}{p^0_{(\rho)} }\bigg[g^{n-\rho} \prod^{\rho}_{j = \rho+ 1} \frac{A_{n+\rho+1-j}^0}{\sum_{k = \rho+1}^j p_{(k)}^0}\bigg]\\
    & \times\bigg[(-g)^{\rho - i} \prod^{\rho}_{j = i+ 1} \frac{A_{\rho +i+1-j}^0}{\sum_{k = \rho+1}^j p_{(\rho +i +1-k)}^0}\bigg].
\label{eq: magic formula}
\end{aligned}
\end{equation}
After applying the magic formula to eq.~(\ref{eq: rearranged SL antiquark}) and making use of 
\begin{equation}
     \prod_{j = 1}^{\ell-1} \frac{A_{\ell-j}^0}{p^0_{(j)}} \equiv \prod_{j =1}^\ell g \frac{A^0_{\ell+1-j}}{p_{(j)}^0} \equiv S_v
\end{equation}
we are left with the following set of structures
\begin{equation}
\begin{aligned}
   \mathcal{A}_{4(a)}\approx \frac{1}{2m}u^\dagger(p_Q) S^\dagger_v \Gamma S_v \bigg[&\frac{1}{v\cdot \Pc}S_v^\dagger \bigg\{-(\gamma \cdot {\bfcal P} \gamma \cdot {\bfcal P}) S_v + g(\gamma\cdot {\bfcal P} \gamma \cdot {\bf A} \\
    &+ \gamma \cdot {\bf A} \gamma \cdot {\bfcal P})S_v - g^2 \gamma\cdot {\bf A}\gamma \cdot {\bf A} S_v\bigg\}\bigg]v^{(0)}(p_{\bar{Q}}).
\end{aligned}
\end{equation}
Using the definition of the covariant derivative ${\bf D}_i = {\bfcal P} - g{\bf A}_i$, we can quickly see that this reduces to the structure given in eq. (\ref{eq: SL antiquark ans 1}). A similar procedure can be used to derive the operators in case (b) of figure \ref{fig: subleading antiquark}.

%%%%%%%%%%%%%%%%%%%%%%%%%%%%%%%%%%%%%%%%
\bibliographystyle{JHEP}
\bibliography{biblio.bib}

\providecommand{\noopsort}[1]{}\providecommand{\singleletter}[1]{#1}%

\providecommand{\href}[2]{#2}\begingroup\raggedright\begin{thebibliography}{10}

\bibitem{LRP2023}
U.D.~of~Energy and N.S.~Foundation, \emph{A new era of discovery: The 2023 long range plan for nuclear science},  October, 2023.

\bibitem{Butterworth:2015oua}
J.~Butterworth, S.~Carrazza, A.~De~Roeck et~al., \emph{Pdf4lhc recommendations for lhc run ii}, \href{https://doi.org/10.1088/0954-3899/43/2/023001}{\emph{J. Phys. G} {\bfseries 43} (2016) 023001} [\href{https://arxiv.org/abs/1510.03865}{{\ttfamily 1510.03865}}].

\bibitem{Boussarie:2023izj}
R.~Boussarie et~al., \emph{{TMD Handbook}},  \href{https://arxiv.org/abs/2304.03302}{{\ttfamily 2304.03302}}.

\bibitem{Bodwin:1994jh}
G.T.~Bodwin, E.~Braaten and G.P.~Lepage, \emph{{Rigorous QCD analysis of inclusive annihilation and production of heavy quarkonium}}, \href{https://doi.org/10.1103/PhysRevD.55.5853}{\emph{Phys. Rev. D} {\bfseries 51} (1995) 1125} [\href{https://arxiv.org/abs/hep-ph/9407339}{{\ttfamily hep-ph/9407339}}].

\bibitem{Fleming:1997fq}
S.~Fleming and T.~Mehen, \emph{{Leptoproduction of J / psi}}, \href{https://doi.org/10.1103/PhysRevD.57.1846}{\emph{Phys. Rev. D} {\bfseries 57} (1998) 1846} [\href{https://arxiv.org/abs/hep-ph/9707365}{{\ttfamily hep-ph/9707365}}].

\bibitem{Yuan:2000cn}
F.~Yuan and K.-T.~Chao, \emph{{Polarized $J/\psi$ production in deep inelastic scattering at HERA}}, \href{https://doi.org/10.1103/PhysRevD.63.034017}{\emph{Phys. Rev. D} {\bfseries 63} (2001) 034017} [\href{https://arxiv.org/abs/hep-ph/0008301}{{\ttfamily hep-ph/0008301}}].

\bibitem{Beneke:1998re}
M.~Beneke, M.~Kramer and M.~Vanttinen, \emph{{Inelastic photoproduction of polarized J / psi}}, \href{https://doi.org/10.1103/PhysRevD.57.4258}{\emph{Phys. Rev. D} {\bfseries 57} (1998) 4258} [\href{https://arxiv.org/abs/hep-ph/9709376}{{\ttfamily hep-ph/9709376}}].

\bibitem{Chu:2024fpo}
Z.~Chu, J.~Chen, X.-P.~Wang and H.~Xing, \emph{{On the role of $J/{\psi}$ production in electron-ion collisions}},  \href{https://arxiv.org/abs/2406.01406}{{\ttfamily 2406.01406}}.

\bibitem{Maxia:2024cjh}
L.~Maxia and F.~Yuan, \emph{{Azimuthal Angular Correlation of $J/\psi$ Plus Jet Production at the EIC}},  \href{https://arxiv.org/abs/2403.02097}{{\ttfamily 2403.02097}}.

\bibitem{Bodwin:2010fi}
G.T.~Bodwin, X.~Garcia~i Tormo and J.~Lee, \emph{{Factorization in exclusive quarkonium production}}, \href{https://doi.org/10.1103/PhysRevD.81.114014}{\emph{Phys. Rev. D} {\bfseries 81} (2010) 114014} [\href{https://arxiv.org/abs/1003.0061}{{\ttfamily 1003.0061}}].

\bibitem{Braaten:2002fi}
E.~Braaten and J.~Lee, \emph{{Exclusive Double Charmonium Production from $e^+ e^-$ Annihilation into a Virtual Photon}}, \href{https://doi.org/10.1103/PhysRevD.72.099901}{\emph{Phys. Rev. D} {\bfseries 67} (2003) 054007} [\href{https://arxiv.org/abs/hep-ph/0211085}{{\ttfamily hep-ph/0211085}}].

\bibitem{Hagiwara:2003cw}
K.~Hagiwara, E.~Kou and C.-F.~Qiao, \emph{{Exclusive $J/\psi$ productions at $e^{+} e^{-}$ colliders}}, \href{https://doi.org/10.1016/j.physletb.2003.07.006}{\emph{Phys. Lett. B} {\bfseries 570} (2003) 39} [\href{https://arxiv.org/abs/hep-ph/0305102}{{\ttfamily hep-ph/0305102}}].

\bibitem{Bodwin:2008nf}
G.T.~Bodwin, X.~Garcia~i Tormo and J.~Lee, \emph{{Factorization theorems for exclusive heavy-quarkonium production}}, \href{https://doi.org/10.1103/PhysRevLett.101.102002}{\emph{Phys. Rev. Lett.} {\bfseries 101} (2008) 102002} [\href{https://arxiv.org/abs/0805.3876}{{\ttfamily 0805.3876}}].

\bibitem{Braaten:1994xb}
E.~Braaten, M.A.~Doncheski, S.~Fleming and M.L.~Mangano, \emph{{Fragmentation production of $J/\psi$ and $\psi^\prime$ at the Tevatron}}, \href{https://doi.org/10.1016/0370-2693(94)90182-1}{\emph{Phys. Lett. B} {\bfseries 333} (1994) 548} [\href{https://arxiv.org/abs/hep-ph/9405407}{{\ttfamily hep-ph/9405407}}].

\bibitem{Braaten:1994vv}
E.~Braaten and S.~Fleming, \emph{{Color octet fragmentation and the psi-prime surplus at the Tevatron}}, \href{https://doi.org/10.1103/PhysRevLett.74.3327}{\emph{Phys. Rev. Lett.} {\bfseries 74} (1995) 3327} [\href{https://arxiv.org/abs/hep-ph/9411365}{{\ttfamily hep-ph/9411365}}].

\bibitem{Baumgart:2014upa}
M.~Baumgart, A.K.~Leibovich, T.~Mehen and I.Z.~Rothstein, \emph{{Probing Quarkonium Production Mechanisms with Jet Substructure}}, \href{https://doi.org/10.1007/JHEP11(2014)003}{\emph{JHEP} {\bfseries 11} (2014) 003} [\href{https://arxiv.org/abs/1406.2295}{{\ttfamily 1406.2295}}].

\bibitem{Bain:2016clc}
R.~Bain, L.~Dai, A.~Hornig, A.K.~Leibovich, Y.~Makris and T.~Mehen, \emph{{Analytic and Monte Carlo Studies of Jets with Heavy Mesons and Quarkonia}}, \href{https://doi.org/10.1007/JHEP06(2016)121}{\emph{JHEP} {\bfseries 06} (2016) 121} [\href{https://arxiv.org/abs/1603.06981}{{\ttfamily 1603.06981}}].

\bibitem{Bain:2017wvk}
R.~Bain, L.~Dai, A.~Leibovich, Y.~Makris and T.~Mehen, \emph{{NRQCD Confronts LHCb Data on Quarkonium Production within Jets}}, \href{https://doi.org/10.1103/PhysRevLett.119.032002}{\emph{Phys. Rev. Lett.} {\bfseries 119} (2017) 032002} [\href{https://arxiv.org/abs/1702.05525}{{\ttfamily 1702.05525}}].

\bibitem{Kang:2017yde}
Z.-B.~Kang, J.-W.~Qiu, F.~Ringer, H.~Xing and H.~Zhang, \emph{{$J/\psi$ production and polarization within a jet}}, \href{https://doi.org/10.1103/PhysRevLett.119.032001}{\emph{Phys. Rev. Lett.} {\bfseries 119} (2017) 032001} [\href{https://arxiv.org/abs/1702.03287}{{\ttfamily 1702.03287}}].

\bibitem{Dai:2017cjq}
L.~Dai and P.~Shrivastava, \emph{{Quarkonium Polarization and the Long Distance Matrix Elements Hierarchies using Jet Substructure}}, \href{https://doi.org/10.1103/PhysRevD.96.036020}{\emph{Phys. Rev. D} {\bfseries 96} (2017) 036020} [\href{https://arxiv.org/abs/1707.08629}{{\ttfamily 1707.08629}}].

\bibitem{Wang:2025drz}
Y.~Wang, D.~Kang and H.S.~Chung, \emph{{NRQCD Re-Confronts LHCb Data on Quarkonium Production within Jets}},  \href{https://arxiv.org/abs/2507.19022}{{\ttfamily 2507.19022}}.

\bibitem{Copeland:2025osx}
M.~Copeland, L.~Dai, Y.~Fu and J.~Roy, \emph{{$\psi(2S)$ production in jets using NRQCD}},  \href{https://arxiv.org/abs/2508.00814}{{\ttfamily 2508.00814}}.

\bibitem{Flett:2024htj}
C.A.~Flett, J.P.~Lansberg, S.~Nabeebaccus, M.~Nefedov, P.~Sznajder and J.~Wagner, \emph{{Exclusive vector-quarkonium photoproduction at NLO in {\ensuremath{\alpha}}s in collinear factorisation with evolution of the generalised parton distributions and high-energy resummation}}, \href{https://doi.org/10.1016/j.physletb.2024.139117}{\emph{Phys. Lett. B} {\bfseries 859} (2024) 139117} [\href{https://arxiv.org/abs/2409.05738}{{\ttfamily 2409.05738}}].

\bibitem{Ivanov:2004vd}
D.Y.~Ivanov, A.~Schafer, L.~Szymanowski and G.~Krasnikov, \emph{{Exclusive photoproduction of a heavy vector meson in QCD}}, \href{https://doi.org/10.1140/epjc/s2004-01712-x}{\emph{Eur. Phys. J. C} {\bfseries 34} (2004) 297} [\href{https://arxiv.org/abs/hep-ph/0401131}{{\ttfamily hep-ph/0401131}}].

\bibitem{Chen:2019uit}
Z.-Q.~Chen and C.-F.~Qiao, \emph{{NLO QCD corrections to exclusive electroproduction of quarkonium}}, \href{https://doi.org/10.1016/j.physletb.2019.134816}{\emph{Phys. Lett. B} {\bfseries 797} (2019) 134816} [\href{https://arxiv.org/abs/1903.00171}{{\ttfamily 1903.00171}}].

\bibitem{Blask:2025jua}
S.K.~Blask, S.~Fleming, T.~Mehen, J.~Roy, I.W.~Stewart and F.~Zhao, \emph{{Relativistic corrections to exclusive photoproduction of Quarkonia near-threshold}},  \href{https://arxiv.org/abs/2506.18905}{{\ttfamily 2506.18905}}.

\bibitem{Sharma:2012dy}
R.~Sharma and I.~Vitev, \emph{{High transverse momentum quarkonium production and dissociation in heavy ion collisions}}, \href{https://doi.org/10.1103/PhysRevC.87.044905}{\emph{Phys. Rev. C} {\bfseries 87} (2013) 044905} [\href{https://arxiv.org/abs/1203.0329}{{\ttfamily 1203.0329}}].

\bibitem{Yao:2018nmy}
X.~Yao and T.~Mehen, \emph{{Quarkonium in-medium transport equation derived from first principles}}, \href{https://doi.org/10.1103/PhysRevD.99.096028}{\emph{Phys. Rev. D} {\bfseries 99} (2019) 096028} [\href{https://arxiv.org/abs/1811.07027}{{\ttfamily 1811.07027}}].

\bibitem{Yao:2020kqy}
X.~Yao, W.~Ke, Y.~Xu, S.A.~Bass, T.~Mehen and B.~M\"uller, \emph{{Quarkonium Production in Heavy Ion Collisions: From Open Quantum System to Transport Equation}}, \href{https://doi.org/10.1016/j.nuclphysa.2020.121854}{\emph{Nucl. Phys. A} {\bfseries 1005} (2021) 121854} [\href{https://arxiv.org/abs/2002.04079}{{\ttfamily 2002.04079}}].

\bibitem{Yang:2024ejk}
D.-L.~Yang and X.~Yao, \emph{{Quarkonium Polarization in Medium from Open Quantum Systems and Chromomagnetic Correlators}},  \href{https://arxiv.org/abs/2405.20280}{{\ttfamily 2405.20280}}.

\bibitem{Hoang:2002ae}
A.H.~Hoang, \emph{{Heavy quarkonium dynamics}},  \href{https://arxiv.org/abs/hep-ph/0204299}{{\ttfamily hep-ph/0204299}}.

\bibitem{Brambilla:2010cs}
N.~Brambilla et~al., \emph{{Heavy Quarkonium: Progress, Puzzles, and Opportunities}}, \href{https://doi.org/10.1140/epjc/s10052-010-1534-9}{\emph{Eur. Phys. J. C} {\bfseries 71} (2011) 1534} [\href{https://arxiv.org/abs/1010.5827}{{\ttfamily 1010.5827}}].

\bibitem{Nayak:2005rw}
G.C.~Nayak, J.-W.~Qiu and G.F.~Sterman, \emph{{Fragmentation, factorization and infrared poles in heavy quarkonium production}}, \href{https://doi.org/10.1016/j.physletb.2005.03.031}{\emph{Phys. Lett. B} {\bfseries 613} (2005) 45} [\href{https://arxiv.org/abs/hep-ph/0501235}{{\ttfamily hep-ph/0501235}}].

\bibitem{Nayak:2005rt}
G.C.~Nayak, J.-W.~Qiu and G.F.~Sterman, \emph{{Fragmentation, NRQCD and NNLO factorization analysis in heavy quarkonium production}}, \href{https://doi.org/10.1103/PhysRevD.72.114012}{\emph{Phys. Rev. D} {\bfseries 72} (2005) 114012} [\href{https://arxiv.org/abs/hep-ph/0509021}{{\ttfamily hep-ph/0509021}}].

\bibitem{Nayak:2006fm}
G.C.~Nayak, J.-W.~Qiu and G.F.~Sterman, \emph{{NRQCD Factorization and Velocity-dependence of NNLO Poles in Heavy Quarkonium Production}}, \href{https://doi.org/10.1103/PhysRevD.74.074007}{\emph{Phys. Rev. D} {\bfseries 74} (2006) 074007} [\href{https://arxiv.org/abs/hep-ph/0608066}{{\ttfamily hep-ph/0608066}}].

\bibitem{Butenschoen:2011yh}
M.~Butenschoen and B.A.~Kniehl, \emph{{World data of J/psi production consolidate NRQCD factorization at NLO}}, \href{https://doi.org/10.1103/PhysRevD.84.051501}{\emph{Phys. Rev. D} {\bfseries 84} (2011) 051501} [\href{https://arxiv.org/abs/1105.0820}{{\ttfamily 1105.0820}}].

\bibitem{Butenschoen:2012qr}
M.~Butenschoen and B.A.~Kniehl, \emph{{Next-to-leading-order tests of NRQCD factorization with $J/\psi$ yield and polarization}}, \href{https://doi.org/10.1142/S0217732313500272}{\emph{Mod. Phys. Lett. A} {\bfseries 28} (2013) 1350027} [\href{https://arxiv.org/abs/1212.2037}{{\ttfamily 1212.2037}}].

\bibitem{Chao:2012iv}
K.-T.~Chao, Y.-Q.~Ma, H.-S.~Shao, K.~Wang and Y.-J.~Zhang, \emph{{$J/\psi$ Polarization at Hadron Colliders in Nonrelativistic QCD}}, \href{https://doi.org/10.1103/PhysRevLett.108.242004}{\emph{Phys. Rev. Lett.} {\bfseries 108} (2012) 242004} [\href{https://arxiv.org/abs/1201.2675}{{\ttfamily 1201.2675}}].

\bibitem{Bodwin:2014gia}
G.T.~Bodwin, H.S.~Chung, U.-R.~Kim and J.~Lee, \emph{{Fragmentation contributions to $J/\psi$ production at the Tevatron and the LHC}}, \href{https://doi.org/10.1103/PhysRevLett.113.022001}{\emph{Phys. Rev. Lett.} {\bfseries 113} (2014) 022001} [\href{https://arxiv.org/abs/1403.3612}{{\ttfamily 1403.3612}}].

\bibitem{Brambilla:2024iqg}
N.~Brambilla, M.~Butenschoen and X.-P.~Wang, \emph{{How well does nonrelativistic QCD factorization work at next-to-leading order?}}, \href{https://doi.org/10.1103/8xqs-45pz}{\emph{Phys. Rev. D} {\bfseries 112} (2025) L011902} [\href{https://arxiv.org/abs/2411.16384}{{\ttfamily 2411.16384}}].

\bibitem{LHCb:2013izl}
{\scshape LHCb} collaboration, \emph{{Measurement of $J/\psi$ polarization in $pp$ collisions at $\sqrt{s}=7$ TeV}}, \href{https://doi.org/10.1140/epjc/s10052-013-2631-3}{\emph{Eur. Phys. J. C} {\bfseries 73} (2013) 2631} [\href{https://arxiv.org/abs/1307.6379}{{\ttfamily 1307.6379}}].

\bibitem{Brambilla:1999xf}
N.~Brambilla, A.~Pineda, J.~Soto and A.~Vairo, \emph{{Potential NRQCD: An Effective theory for heavy quarkonium}}, \href{https://doi.org/10.1016/S0550-3213(99)00693-8}{\emph{Nucl. Phys. B} {\bfseries 566} (2000) 275} [\href{https://arxiv.org/abs/hep-ph/9907240}{{\ttfamily hep-ph/9907240}}].

\bibitem{Brambilla:2004jw}
N.~Brambilla, A.~Pineda, J.~Soto and A.~Vairo, \emph{{Effective Field Theories for Heavy Quarkonium}}, \href{https://doi.org/10.1103/RevModPhys.77.1423}{\emph{Rev. Mod. Phys.} {\bfseries 77} (2005) 1423} [\href{https://arxiv.org/abs/hep-ph/0410047}{{\ttfamily hep-ph/0410047}}].

\bibitem{Luke:1999kz}
M.E.~Luke, A.V.~Manohar and I.Z.~Rothstein, \emph{{Renormalization group scaling in nonrelativistic QCD}}, \href{https://doi.org/10.1103/PhysRevD.61.074025}{\emph{Phys. Rev. D} {\bfseries 61} (2000) 074025} [\href{https://arxiv.org/abs/hep-ph/9910209}{{\ttfamily hep-ph/9910209}}].

\bibitem{Rothstein:2018dzq}
I.Z.~Rothstein, P.~Shrivastava and I.W.~Stewart, \emph{{Manifestly Soft Gauge Invariant Formulation of vNRQCD}}, \href{https://doi.org/10.1016/j.nuclphysb.2018.12.027}{\emph{Nucl. Phys. B} {\bfseries 939} (2019) 405} [\href{https://arxiv.org/abs/1806.07398}{{\ttfamily 1806.07398}}].

\bibitem{Brambilla:2022rjd}
N.~Brambilla, H.S.~Chung, A.~Vairo and X.-P.~Wang, \emph{{Production and polarization of S-wave quarkonia in potential nonrelativistic QCD}}, \href{https://doi.org/10.1103/PhysRevD.105.L111503}{\emph{Phys. Rev. D} {\bfseries 105} (2022) L111503} [\href{https://arxiv.org/abs/2203.07778}{{\ttfamily 2203.07778}}].

\bibitem{Brambilla:2022ayc}
N.~Brambilla, H.S.~Chung, A.~Vairo and X.-P.~Wang, \emph{{Inclusive production of J/{\ensuremath{\psi}}, {\ensuremath{\psi}}(2S), and {\ensuremath{\Upsilon}} states in pNRQCD}}, \href{https://doi.org/10.1007/JHEP03(2023)242}{\emph{JHEP} {\bfseries 03} (2023) 242} [\href{https://arxiv.org/abs/2210.17345}{{\ttfamily 2210.17345}}].

\bibitem{Fleming:2019pzj}
S.~Fleming, Y.~Makris and T.~Mehen, \emph{{An effective field theory approach to quarkonium at small transverse momentum}}, \href{https://doi.org/10.1007/JHEP04(2020)122}{\emph{JHEP} {\bfseries 04} (2020) 122} [\href{https://arxiv.org/abs/1910.03586}{{\ttfamily 1910.03586}}].

\bibitem{Echevarria:2011epo}
M.G.~Echevarria, A.~Idilbi and I.~Scimemi, \emph{{Factorization Theorem For Drell-Yan At Low $q_T$ And Transverse Momentum Distributions On-The-Light-Cone}}, \href{https://doi.org/10.1007/JHEP07(2012)002}{\emph{JHEP} {\bfseries 07} (2012) 002} [\href{https://arxiv.org/abs/1111.4996}{{\ttfamily 1111.4996}}].

\bibitem{Echevarria:2012js}
M.G.~Echevarr\'\i{}a, A.~Idilbi and I.~Scimemi, \emph{{Soft and Collinear Factorization and Transverse Momentum Dependent Parton Distribution Functions}}, \href{https://doi.org/10.1016/j.physletb.2013.09.003}{\emph{Phys. Lett. B} {\bfseries 726} (2013) 795} [\href{https://arxiv.org/abs/1211.1947}{{\ttfamily 1211.1947}}].

\bibitem{vonKuk:2023jfd}
R.~von Kuk, J.K.L.~Michel and Z.~Sun, \emph{{Transverse momentum distributions of heavy hadrons and polarized heavy quarks}}, \href{https://doi.org/10.1007/JHEP09(2023)205}{\emph{JHEP} {\bfseries 09} (2023) 205} [\href{https://arxiv.org/abs/2305.15461}{{\ttfamily 2305.15461}}].

\bibitem{vonKuk:2024uxe}
R.~von Kuk, J.K.L.~Michel and Z.~Sun, \emph{{Transverse momentum-dependent heavy-quark fragmentation at next-to-leading order}}, \href{https://doi.org/10.1007/JHEP07(2024)129}{\emph{JHEP} {\bfseries 07} (2024) 129} [\href{https://arxiv.org/abs/2404.08622}{{\ttfamily 2404.08622}}].

\bibitem{vonKuk:2025hdv}
R.O.~von Kuk, \emph{{Heavy-quark Effects in Factorization and Resummation}}, Ph.D. thesis, U. Hamburg (main), 2025.

\bibitem{Dai:2023rvd}
L.~Dai, C.~Kim and A.K.~Leibovich, \emph{{Heavy quark transverse momentum dependent fragmentation}},  \href{https://arxiv.org/abs/2310.19207}{{\ttfamily 2310.19207}}.

\bibitem{Copeland:2024wwm}
M.~Copeland and T.~Mehen, \emph{{Transverse momentum dependent PDFs in chiral effective theory}}, \href{https://doi.org/10.1103/PhysRevD.110.114026}{\emph{Phys. Rev. D} {\bfseries 110} (2024) 114026} [\href{https://arxiv.org/abs/2405.14965}{{\ttfamily 2405.14965}}].

\bibitem{Copeland:2024cgq}
M.~Copeland and T.~Mehen, \emph{{Probing nonperturbative transverse momentum dependent PDFs with chiral perturbation theory: the $\bar{d}-\bar{u}$ asymmetry}},  \href{https://arxiv.org/abs/2412.07717}{{\ttfamily 2412.07717}}.

\bibitem{Ke:2024ytw}
W.~Ke, J.~Terry and I.~Vitev, \emph{{Toward a first-principles description of transverse momentum dependent Drell-Yan production in proton-nucleus collisions}}, \href{https://doi.org/10.1007/JHEP02(2025)102}{\emph{JHEP} {\bfseries 02} (2025) 102} [\href{https://arxiv.org/abs/2408.10310}{{\ttfamily 2408.10310}}].

\bibitem{Catani:2014qha}
S.~Catani, M.~Grazzini and A.~Torre, \emph{{Transverse-momentum resummation for heavy-quark hadroproduction}}, \href{https://doi.org/10.1016/j.nuclphysb.2014.11.019}{\emph{Nucl. Phys. B} {\bfseries 890} (2014) 518} [\href{https://arxiv.org/abs/1408.4564}{{\ttfamily 1408.4564}}].

\bibitem{Kang:2014tta}
Z.-B.~Kang, Y.-Q.~Ma, J.-W.~Qiu and G.~Sterman, \emph{{Heavy Quarkonium Production at Collider Energies: Factorization and Evolution}}, \href{https://doi.org/10.1103/PhysRevD.90.034006}{\emph{Phys. Rev. D} {\bfseries 90} (2014) 034006} [\href{https://arxiv.org/abs/1401.0923}{{\ttfamily 1401.0923}}].

\bibitem{Sun:2012vc}
P.~Sun, C.P.~Yuan and F.~Yuan, \emph{{Heavy Quarkonium Production at Low Pt in NRQCD with Soft Gluon Resummation}}, \href{https://doi.org/10.1103/PhysRevD.88.054008}{\emph{Phys. Rev. D} {\bfseries 88} (2013) 054008} [\href{https://arxiv.org/abs/1210.3432}{{\ttfamily 1210.3432}}].

\bibitem{Catani:2010pd}
S.~Catani and M.~Grazzini, \emph{{QCD transverse-momentum resummation in gluon fusion processes}}, \href{https://doi.org/10.1016/j.nuclphysb.2010.12.007}{\emph{Nucl. Phys. B} {\bfseries 845} (2011) 297} [\href{https://arxiv.org/abs/1011.3918}{{\ttfamily 1011.3918}}].

\bibitem{Mukherjee:2016cjw}
A.~Mukherjee and S.~Rajesh, \emph{{Linearly polarized gluons in charmonium and bottomonium production in color octet model}}, \href{https://doi.org/10.1103/PhysRevD.95.034039}{\emph{Phys. Rev. D} {\bfseries 95} (2017) 034039} [\href{https://arxiv.org/abs/1611.05974}{{\ttfamily 1611.05974}}].

\bibitem{Mukherjee:2015smo}
A.~Mukherjee and S.~Rajesh, \emph{{Probing Transverse Momentum Dependent Parton Distributions in Charmonium and Bottomonium Production}}, \href{https://doi.org/10.1103/PhysRevD.93.054018}{\emph{Phys. Rev. D} {\bfseries 93} (2016) 054018} [\href{https://arxiv.org/abs/1511.04319}{{\ttfamily 1511.04319}}].

\bibitem{Boer:2012bt}
D.~Boer and C.~Pisano, \emph{{Polarized gluon studies with charmonium and bottomonium at LHCb and AFTER}}, \href{https://doi.org/10.1103/PhysRevD.86.094007}{\emph{Phys. Rev. D} {\bfseries 86} (2012) 094007} [\href{https://arxiv.org/abs/1208.3642}{{\ttfamily 1208.3642}}].

\bibitem{Echevarria:2019ynx}
M.G.~Echevarria, \emph{{Proper TMD factorization for quarkonia production: $pp\to\eta_{c,b}$ as a study case}}, \href{https://doi.org/10.1007/JHEP10(2019)144}{\emph{JHEP} {\bfseries 10} (2019) 144} [\href{https://arxiv.org/abs/1907.06494}{{\ttfamily 1907.06494}}].

\bibitem{DAlesio:2021yws}
U.~D'Alesio, L.~Maxia, F.~Murgia, C.~Pisano and S.~Rajesh, \emph{{J/\ensuremath{\psi} polarization in semi-inclusive DIS at low and high transverse momentum}}, \href{https://doi.org/10.1007/JHEP03(2022)037}{\emph{JHEP} {\bfseries 03} (2022) 037} [\href{https://arxiv.org/abs/2110.07529}{{\ttfamily 2110.07529}}].

\bibitem{Boer:2020bbd}
D.~Boer, U.~D'Alesio, F.~Murgia, C.~Pisano and P.~Taels, \emph{{J/\ensuremath{\psi} meson production in SIDIS: matching high and low transverse momentum}}, \href{https://doi.org/10.1007/JHEP09(2020)040}{\emph{JHEP} {\bfseries 09} (2020) 040} [\href{https://arxiv.org/abs/2004.06740}{{\ttfamily 2004.06740}}].

\bibitem{Bor:2022fga}
J.~Bor and D.~Boer, \emph{{TMD evolution study of the cos2\ensuremath{\phi} azimuthal asymmetry in unpolarized J/\ensuremath{\psi} production at EIC}}, \href{https://doi.org/10.1103/PhysRevD.106.014030}{\emph{Phys. Rev. D} {\bfseries 106} (2022) 014030} [\href{https://arxiv.org/abs/2204.01527}{{\ttfamily 2204.01527}}].

\bibitem{Kishore:2021vsm}
R.~Kishore, A.~Mukherjee and M.~Siddiqah, \emph{{Cos(2$\phi_h$) asymmetry in J/$\psi$ production in unpolarized $ep$ collision}}, \href{https://doi.org/10.1103/PhysRevD.104.094015}{\emph{Phys. Rev. D} {\bfseries 104} (2021) 094015} [\href{https://arxiv.org/abs/2103.09070}{{\ttfamily 2103.09070}}].

\bibitem{Scarpa:2019fol}
F.~Scarpa, D.~Boer, M.G.~Echevarria, J.-P.~Lansberg, C.~Pisano and M.~Schlegel, \emph{{Studies of gluon TMDs and their evolution using quarkonium-pair production at the LHC}}, \href{https://doi.org/10.1140/epjc/s10052-020-7619-1}{\emph{Eur. Phys. J. C} {\bfseries 80} (2020) 87} [\href{https://arxiv.org/abs/1909.05769}{{\ttfamily 1909.05769}}].

\bibitem{DAlesio:2019qpk}
U.~D'Alesio, F.~Murgia, C.~Pisano and P.~Taels, \emph{{Azimuthal asymmetries in semi-inclusive $J/\psi\,+\,\mathrm{jet}$ production at an EIC}}, \href{https://doi.org/10.1103/PhysRevD.100.094016}{\emph{Phys. Rev. D} {\bfseries 100} (2019) 094016} [\href{https://arxiv.org/abs/1908.00446}{{\ttfamily 1908.00446}}].

\bibitem{Bacchetta:2018ivt}
A.~Bacchetta, D.~Boer, C.~Pisano and P.~Taels, \emph{{Gluon TMDs and NRQCD matrix elements in $J/\psi$ production at an EIC}}, \href{https://doi.org/10.1140/epjc/s10052-020-7620-8}{\emph{Eur. Phys. J. C} {\bfseries 80} (2020) 72} [\href{https://arxiv.org/abs/1809.02056}{{\ttfamily 1809.02056}}].

\bibitem{Mukherjee:2016qxa}
A.~Mukherjee and S.~Rajesh, \emph{{$J/\psi $ production in polarized and unpolarized ep collision and Sivers and $\cos 2\phi $ asymmetries}}, \href{https://doi.org/10.1140/epjc/s10052-017-5406-4}{\emph{Eur. Phys. J. C} {\bfseries 77} (2017) 854} [\href{https://arxiv.org/abs/1609.05596}{{\ttfamily 1609.05596}}].

\bibitem{Rajesh:2018qks}
S.~Rajesh, R.~Kishore and A.~Mukherjee, \emph{{Sivers effect in Inelastic $J/\psi$ Photoproduction in $ep^\uparrow$ Collision in Color Octet Model}}, \href{https://doi.org/10.1103/PhysRevD.98.014007}{\emph{Phys. Rev. D} {\bfseries 98} (2018) 014007} [\href{https://arxiv.org/abs/1802.10359}{{\ttfamily 1802.10359}}].

\bibitem{Godbole:2013bca}
R.M.~Godbole, A.~Misra, A.~Mukherjee and V.S.~Rawoot, \emph{{Transverse Single Spin Asymmetry in $e+p^\uparrow \to e+J/\psi +X $ and Transverse Momentum Dependent Evolution of the Sivers Function}}, \href{https://doi.org/10.1103/PhysRevD.88.014029}{\emph{Phys. Rev. D} {\bfseries 88} (2013) 014029} [\href{https://arxiv.org/abs/1304.2584}{{\ttfamily 1304.2584}}].

\bibitem{Godbole:2012bx}
R.M.~Godbole, A.~Misra, A.~Mukherjee and V.S.~Rawoot, \emph{{Sivers Effect and Transverse Single Spin Asymmetry in $e+p^\uparrow \to e+J/\psi+X$}}, \href{https://doi.org/10.1103/PhysRevD.85.094013}{\emph{Phys. Rev. D} {\bfseries 85} (2012) 094013} [\href{https://arxiv.org/abs/1201.1066}{{\ttfamily 1201.1066}}].

\bibitem{denDunnen:2014kjo}
W.J.~den Dunnen, J.P.~Lansberg, C.~Pisano and M.~Schlegel, \emph{{Accessing the Transverse Dynamics and Polarization of Gluons inside the Proton at the LHC}}, \href{https://doi.org/10.1103/PhysRevLett.112.212001}{\emph{Phys. Rev. Lett.} {\bfseries 112} (2014) 212001} [\href{https://arxiv.org/abs/1401.7611}{{\ttfamily 1401.7611}}].

\bibitem{Kang:2014pya}
Z.-B.~Kang, Y.-Q.~Ma, J.-W.~Qiu and G.~Sterman, \emph{{Heavy Quarkonium Production at Collider Energies: Partonic Cross Section and Polarization}}, \href{https://doi.org/10.1103/PhysRevD.91.014030}{\emph{Phys. Rev. D} {\bfseries 91} (2015) 014030} [\href{https://arxiv.org/abs/1411.2456}{{\ttfamily 1411.2456}}].

\bibitem{Zhu:2013yxa}
R.~Zhu, P.~Sun and F.~Yuan, \emph{{Low Transverse Momentum Heavy Quark Pair Production to Probe Gluon Tomography}}, \href{https://doi.org/10.1016/j.physletb.2013.11.002}{\emph{Phys. Lett. B} {\bfseries 727} (2013) 474} [\href{https://arxiv.org/abs/1309.0780}{{\ttfamily 1309.0780}}].

\bibitem{Copeland:2023qed}
M.~Copeland, S.~Fleming, R.~Gupta, R.~Hodges and T.~Mehen, \emph{{Polarized $J/\psi$ production in semi-inclusive DIS at large $Q^2$: Comparing quark fragmentation and photon-gluon fusion}},  \href{https://arxiv.org/abs/2310.13737}{{\ttfamily 2310.13737}}.

\bibitem{Copeland:2023wbu}
M.~Copeland, S.~Fleming, R.~Gupta, R.~Hodges and T.~Mehen, \emph{{Polarized TMD fragmentation functions for $J/\psi$ production}},  \href{https://arxiv.org/abs/2308.08605}{{\ttfamily 2308.08605}}.

\bibitem{Echevarria:2023dme}
M.G.~Echevarria, S.F.~Romera and I.~Scimemi, \emph{{Gluon TMD fragmentation function into quarkonium}},  \href{https://arxiv.org/abs/2308.12356}{{\ttfamily 2308.12356}}.

\bibitem{Echevarria:2024idp}
M.G.~Echevarria, S.F.~Romera and P.~Taels, \emph{{Factorization for $J/\psi$ leptoproduction at small transverse momentum}},  \href{https://arxiv.org/abs/2407.04793}{{\ttfamily 2407.04793}}.

\bibitem{Maxia:2025zee}
L.~Maxia, D.~Boer and J.~Bor, \emph{{The impact of the TMD shape function on matching the transverse momentum spectrum in $J/\psi$ production at the EIC}},  \href{https://arxiv.org/abs/2504.19617}{{\ttfamily 2504.19617}}.

\bibitem{Boer:2023zit}
D.~Boer, J.~Bor, L.~Maxia, C.~Pisano and F.~Yuan, \emph{{Transverse momentum dependent shape function for $J/\psi$ production in SIDIS}},  \href{https://arxiv.org/abs/2304.09473}{{\ttfamily 2304.09473}}.

\bibitem{Georgi:1990um}
H.~Georgi, \emph{{An Effective Field Theory for Heavy Quarks at Low-energies}}, \href{https://doi.org/10.1016/0370-2693(90)91128-X}{\emph{Phys. Lett. B} {\bfseries 240} (1990) 447}.

\bibitem{Manohar:1999xd}
A.V.~Manohar and I.W.~Stewart, \emph{{Renormalization group analysis of the QCD quark potential to order v**2}}, \href{https://doi.org/10.1103/PhysRevD.62.014033}{\emph{Phys. Rev. D} {\bfseries 62} (2000) 014033} [\href{https://arxiv.org/abs/hep-ph/9912226}{{\ttfamily hep-ph/9912226}}].

\bibitem{Bauer:2000ew}
C.W.~Bauer, S.~Fleming and M.E.~Luke, \emph{{Summing Sudakov logarithms in $B \to X_s \gamma $in effective field theory.}}, \href{https://doi.org/10.1103/PhysRevD.63.014006}{\emph{Phys. Rev. D} {\bfseries 63} (2000) 014006} [\href{https://arxiv.org/abs/hep-ph/0005275}{{\ttfamily hep-ph/0005275}}].

\bibitem{Bauer:2000yr}
C.W.~Bauer, S.~Fleming, D.~Pirjol and I.W.~Stewart, \emph{{An Effective field theory for collinear and soft gluons: Heavy to light decays}}, \href{https://doi.org/10.1103/PhysRevD.63.114020}{\emph{Phys. Rev. D} {\bfseries 63} (2001) 114020} [\href{https://arxiv.org/abs/hep-ph/0011336}{{\ttfamily hep-ph/0011336}}].

\bibitem{Bauer:2001ct}
C.W.~Bauer and I.W.~Stewart, \emph{{Invariant operators in collinear effective theory}}, \href{https://doi.org/10.1016/S0370-2693(01)00902-9}{\emph{Phys. Lett. B} {\bfseries 516} (2001) 134} [\href{https://arxiv.org/abs/hep-ph/0107001}{{\ttfamily hep-ph/0107001}}].

\bibitem{Bauer:2001yt}
C.W.~Bauer, D.~Pirjol and I.W.~Stewart, \emph{{Soft collinear factorization in effective field theory}}, \href{https://doi.org/10.1103/PhysRevD.65.054022}{\emph{Phys. Rev. D} {\bfseries 65} (2002) 054022} [\href{https://arxiv.org/abs/hep-ph/0109045}{{\ttfamily hep-ph/0109045}}].

\bibitem{Bauer:2002nz}
C.W.~Bauer, S.~Fleming, D.~Pirjol, I.Z.~Rothstein and I.W.~Stewart, \emph{{Hard scattering factorization from effective field theory}}, \href{https://doi.org/10.1103/PhysRevD.66.014017}{\emph{Phys. Rev. D} {\bfseries 66} (2002) 014017} [\href{https://arxiv.org/abs/hep-ph/0202088}{{\ttfamily hep-ph/0202088}}].

\bibitem{Braaten:1996jt}
E.~Braaten and Y.-Q.~Chen, \emph{{Helicity decomposition for inclusive J / psi production}}, \href{https://doi.org/10.1103/PhysRevD.54.3216}{\emph{Phys. Rev. D} {\bfseries 54} (1996) 3216} [\href{https://arxiv.org/abs/hep-ph/9604237}{{\ttfamily hep-ph/9604237}}].

\bibitem{Beneke:1997av}
M.~Beneke, \emph{{Nonrelativistic effective theory for quarkonium production in hadron collisions}},  in \emph{{24th Annual SLAC Summer Institute on Particle Physics}: {The Strong Interaction, From Hadrons to Protons}}, pp.~549--574, 3, 1997 [\href{https://arxiv.org/abs/hep-ph/9703429}{{\ttfamily hep-ph/9703429}}].

\end{thebibliography}\endgroup

\end{document}